%% file: VertexPaper.tex
\newcommand*{\ATLASLATEXPATH}{latex/}
\documentclass[cernpreprint,txfonts,UKenglish,texlive=2016]{\ATLASLATEXPATH atlasdoc}
\pdfoutput=1
\usepackage{\ATLASLATEXPATH atlaspackage}
\usepackage{\ATLASLATEXPATH atlasbiblatex}

\usepackage{\ATLASLATEXPATH atlascontribute}

\usepackage{\ATLASLATEXPATH atlasphysics}
\graphicspath{{logos/}{figures/}}

\RequirePackage{xspace}

\def\um         {\ensuremath{\mathrm{\mu m}}\xspace}

\def\GeV        {\ensuremath{\mathrm{GeV}}\xspace}
\def\TeV        {\ensuremath{\mathrm{TeV}}\xspace}

\def\lumposx    {\ensuremath{\overline{x}_{\mathcal{L}}}\xspace}
\def\lumposy    {\ensuremath{\overline{y}_{\mathcal{L}}}\xspace}
\def\lumposz    {\ensuremath{\overline{z}_{\mathcal{L}}}\xspace}

\def\lumsigx    {\ensuremath{\sigma_{x{\mathcal L}}}\xspace}
\def\lumsigy    {\ensuremath{\sigma_{y{\mathcal L}}}\xspace}
\def\lumsigz    {\ensuremath{\sigma_{z{\mathcal L}}}\xspace}
\def\lumtiltx   {\ensuremath{\overline{x}_{\mathcal{L}}'}\xspace}
\def\lumtilty   {\ensuremath{\overline{y}_{\mathcal{L}}'}\xspace}

\def\xang       {\ensuremath{\phi}\xspace}           % full crossing angle                                                                                    
\def\Ebeam      {\ensuremath{E_{\mathrm{beam}}}\xspace}

\def\bst        {\ensuremath{\beta^*}\xspace}
\newcommand{\btajb}[1]{\ensuremath{\beta_{#1}}\xspace}

\newcommand{\emitjb}[1]{\ensuremath{\varepsilon_{#1}}\xspace}
\newcommand{\emitnjb}[1]{\ensuremath{\varepsilon_{\textrm N #1}}\xspace}
\newcommand{\bsigx}[1]{\ensuremath{\sigma_{x#1}}\xspace}
\newcommand{\bsigy}[1]{\ensuremath{\sigma_{y#1}}\xspace}
\newcommand{\bsigz}[1]{\ensuremath{\sigma_{z#1}}\xspace}
\newcommand{\bsigj}[1]{\ensuremath{\sigma_{#1}}\xspace}
\def\xang       {\ensuremath{\phi}\xspace}           % full crossing angle          

\AtlasTitle{Reconstruction of primary vertices at the ATLAS experiment in Run 1 proton--proton collisions at the LHC}
\AtlasRefCode{PERF-2015-01}
\PreprintIdNumber{CERN-EP-2016-150}

\AtlasJournal{EPJC}
\AtlasJournalRef{Eur. Phys. J. C 77 (2017) 332}
\AtlasDOI{10.1140/epjc/s10052-017-4887-5}

\AtlasAbstract{
\input{abstract}
}
\makeatletter
\newcommand{\vo}{\vec{o}\@ifnextchar{^}{\,}{}}
\makeatother

\begin{document}

\maketitle

\tableofcontents

% List of contributors - print here or after the Bibliography.
%\PrintAtlasContribute{0.30}
\clearpage

%-------------------------------------------------------------------------------
\section{Introduction}
\label{sec:intro}
\input{Introduction}
%-------------------------------------------------------------------------------

%-------------------------------------------------------------------------------
\section{The ATLAS detector and LHC beam parameters}
\label{sec::Atlas}
\input{Atlas}

%-------------------------------------------------------------------------------

%The ATLAS detector~\cite{PERF-2007-01} ...
% \input{atlas-detector}

%-------------------------------------------------------------------------------
\section{Data and Monte Carlo samples}
\label{sec:datasets}
\input{Datasets}

\section{Primary vertex reconstruction}
\label{sec::Reco}
\input{Reconstruction}

\section{Hard-scatter interaction vertices}
\label{sec::HardScatter}
\input{HardScatter}

\section{Primary vertices in minimum-bias data}
\label{sec::Perf}
\input{SingleVertexPerformance}

\section{Performance in the high pile-up regime}
\label{sec::HighPileUp}
\input{PileUp}

\section{Efficiency of vertex reconstruction as a function of pile-up}
\label{sec::Projection}
\input{VertexProjection}

\section{Conclusion}
\label{sec::Conclusion}
\input{Conclusion}
%-------------------------------------------------------------------------------

%-------------------------------------------------------------------------------
\section*{Acknowledgements}
\input{Acknowledgements}

%-------------------------------------------------------------------------------

%\input{acknowledgements/Acknowledgements}

%The \texttt{atlaslatex} package contains the acknowledgements that were valid 
%at the time of the release you are using.
%These can be found in the \texttt{acknowledgements} subdirectory.
%When your ATLAS paper or PUB/CONF note is ready to be published,
%download the latest set of acknowledgements from:\\
%\url{https://twiki.cern.ch/twiki/bin/view/AtlasProtected/PubComAcknowledgements}

%The supporting notes for the analysis should also contain a list of contributors.
%This information should usually be included in \texttt{mydocument-metadata.tex}.
%The list should be printed either here or before the table of contents.

%-------------------------------------------------------------------------------
%\clearpage
%\appendix
%\part*{Appendix}
%\addcontentsline{toc}{part}{Appendix}
%%-------------------------------------------------------------------------------

%-------------------------------------------------------------------------------
% If you use biblatex and either biber or bibtex to process the bibliography
% just say \printbibliography here
\printbibliography
% If you want to use the traditional BibTeX you need to use the syntax below.
%\bibliographystyle{bibtex/bst/atlasBibStyleWoTitle}
%\bibliography{atlas-document,bibtex/bib/ATLAS}
%-------------------------------------------------------------------------------

\newpage \input{atlas_authlist}

%-------------------------------------------------------------------------------
% Print the list of contributors to the analysis
% The argument gives the fraction of the text width used for the names
%-------------------------------------------------------------------------------
%\clearpage
%\PrintAtlasContribute{0.30}

%%-------------------------------------------------------------------------------
%\clearpage
%\appendix
%\part*{Auxiliary material}
%\addcontentsline{toc}{part}{Auxiliary material}
%%-------------------------------------------------------------------------------
%
%In an ATLAS paper, auxiliary plots and tables that are supposed to be made public 
%should be collected in an appendix that has the title \enquote{Auxiliary material}.
%This appendix should be printed after the Bibliography.
%At the end of the paper approval procedure, this information can be split into a separate document
%-- see \texttt{atlas-auxmat.tex}.
%
%In an ATLAS note, use the appendices to include all the technical details of your work
%that are relevant for the ATLAS Collaboration only (e.g.\ dataset details, software release used).
%This information should be printed after the Bibliography.

\end{document}

%% file: abstract.tex
 This paper presents the method and performance of primary vertex reconstruction in
 proton--proton collision data recorded by the ATLAS experiment during Run 1 of the LHC.
The studies presented focus on data taken during 2012 at a centre-of-mass energy of $\sqrt{s} = 8$~TeV. 
The performance has been measured as a function of the number of interactions per bunch crossing over a wide range, from one to seventy.
The measurement of the position and size of the luminous region
and its use as a constraint to improve the 
primary vertex resolution are discussed. 
A longitudinal vertex position  resolution of about $30\;\mu{\textrm m}$ is achieved for 
events with high  multiplicity of reconstructed tracks. The transverse position resolution is better than
$20\;\mu{\textrm m}$ and is dominated by the  precision on the size of the
luminous region.
An analytical model is proposed to describe the primary vertex reconstruction efficiency
as a function of the number of interactions per bunch crossing and of the longitudinal size of the luminous region.
Agreement between the data and the predictions of this model is better than~3\% up to seventy interactions per bunch crossing.

%% file: Introduction.tex
Efficient and precise reconstruction of primary vertices, defined as the points in space where
proton--proton ($pp$) interactions have occurred, is an important element of data analysis
at the LHC. It is of direct relevance to the reconstruction of hard-scatter interactions,
in which the correct assignment of charged-particle trajectories to the
hard-scatter primary vertex is essential in reconstructing the full kinematic properties of
the event. An aspect of primary vertex reconstruction requiring special attention  is the
superposition of multiple inelastic  $pp$ interactions reconstructed as a single physics event with many primary vertices.
These additional primary vertices, which are usually soft-QCD
interactions related to the dominant components of the total cross section, are referred
to as pile-up.
The average number of inelastic $pp$ interactions per bunch crossing 
under constant beam conditions is denoted as $\mu$ and 
is directly related to the instantaneous luminosity~\cite{Lumi8}.
The primary vertex reconstruction is also important for the determination of the luminous region, or beam spot,
where collisions take place within the ATLAS detector. 

This paper describes the performance of primary vertex reconstruction with the ATLAS detector,  
during Run 1 of the LHC from 2010 to 2012. The studies presented here  are based on the data collected in 2012 at 
a proton--proton centre-of-mass energy $\sqrt{s} = 8$~TeV.  
Averaged over the 2012 dataset, $\mu$ was approximately 20.
The 2012 data are representative of the full set of data taken from 2010 to 2012 in terms of the primary vertex 
performance.
Studies in this paper make use of dedicated datasets  recorded at very low values of
  $\mu$  ($\mu$ = 0.01), thereby providing a measurement of the performance in the absence of pile-up. Data 
recorded with the highest number of interactions per bunch crossing, leading to values of  $\mu$
  up to 72, are used to study the various mechanisms that lead to a degradation of the primary vertex
 reconstruction as pile-up increases.

The paper is organised  as follows: Section~\ref{sec::Atlas}  provides a brief description of the ATLAS detector,
a description of pile-up determination  and a discussion of the parameters of the  LHC accelerator  that determine the size of the
luminous region. Section~\ref{sec:datasets} describes the data and Monte Carlo (MC) simulation samples used.
Section~\ref{sec::Reco} presents the algorithms for primary vertex reconstruction in ATLAS.
The measurement and stability of the beam-spot parameters and their use as a constraint in primary vertex reconstruction are discussed.
The predicted impact of pile-up contamination on the reconstruction and selection of primary vertices from hard-scatter processes is
 discussed in Section~\ref{sec::HardScatter}.
Studies of single vertex reconstruction in minimum-bias data and the related comparisons to MC simulation
 are presented in Section~\ref{sec::Perf}.
Section~\ref{sec::HighPileUp}  describes the performance of vertex reconstruction in  high pile-up conditions.
In Section~\ref{sec::Projection}, the results of studies presented in  Sections~\ref{sec::HardScatter} through \ref{sec::HighPileUp} are
used to model the efficiency of primary vertex reconstruction in simulation, to predict its behaviour at high pile-up, and to compare the predictions to data.
Summary and conclusions are presented in Section~\ref{sec::Conclusion}.

%% file: Atlas.tex
The ATLAS detector \cite{DetectorPaper:2008} is a multi-purpose detector with a cylindrical geometry.
 It is comprised of an inner detector (ID)
surrounded by a thin superconducting solenoid, a calorimeter
 system and a muon spectrometer embedded in a toroidal magnetic field. The ID is the primary
 detector used for vertex reconstruction and it is described in further detail below in Section~\ref{sec::Detector}.
  Outside of the ID and the solenoid are electromagnetic sampling calorimeters made of
 liquid argon as the active material and lead as an absorber.   Surrounding the electromagnetic calorimeter is the iron and scintillator tile calorimeter for hadronic 
 energy measurements.  In the forward regions
it is complemented  by two end-cap calorimeters made of liquid argon  and copper or tungsten. The muon
 spectrometer surrounds the calorimeters and consists of three large superconducting eight-coil toroids, a 
 system of tracking chambers, and detectors for triggering. 

\subsection{\label{sec::Detector}The ATLAS inner detector}                
\begin{figure}[tp]
   \centering
   \begin{minipage}[h]{0.59\linewidth}
   \subfloat[~]{\includegraphics[width=1\linewidth]{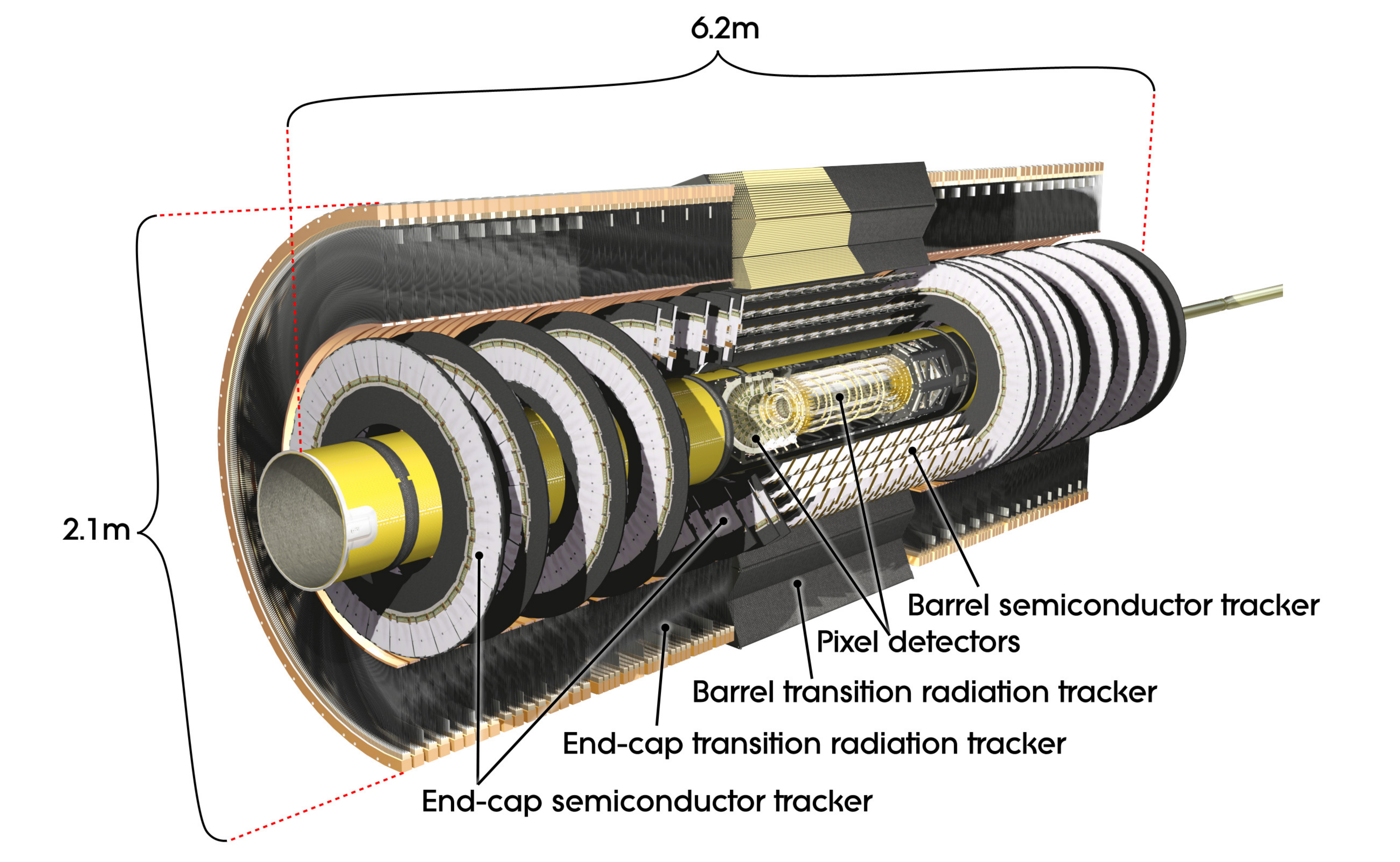}}  
   \end{minipage}
   \hfill
   \begin{minipage}[h]{0.36\linewidth}
   \subfloat[~]{\includegraphics[width=1\linewidth]{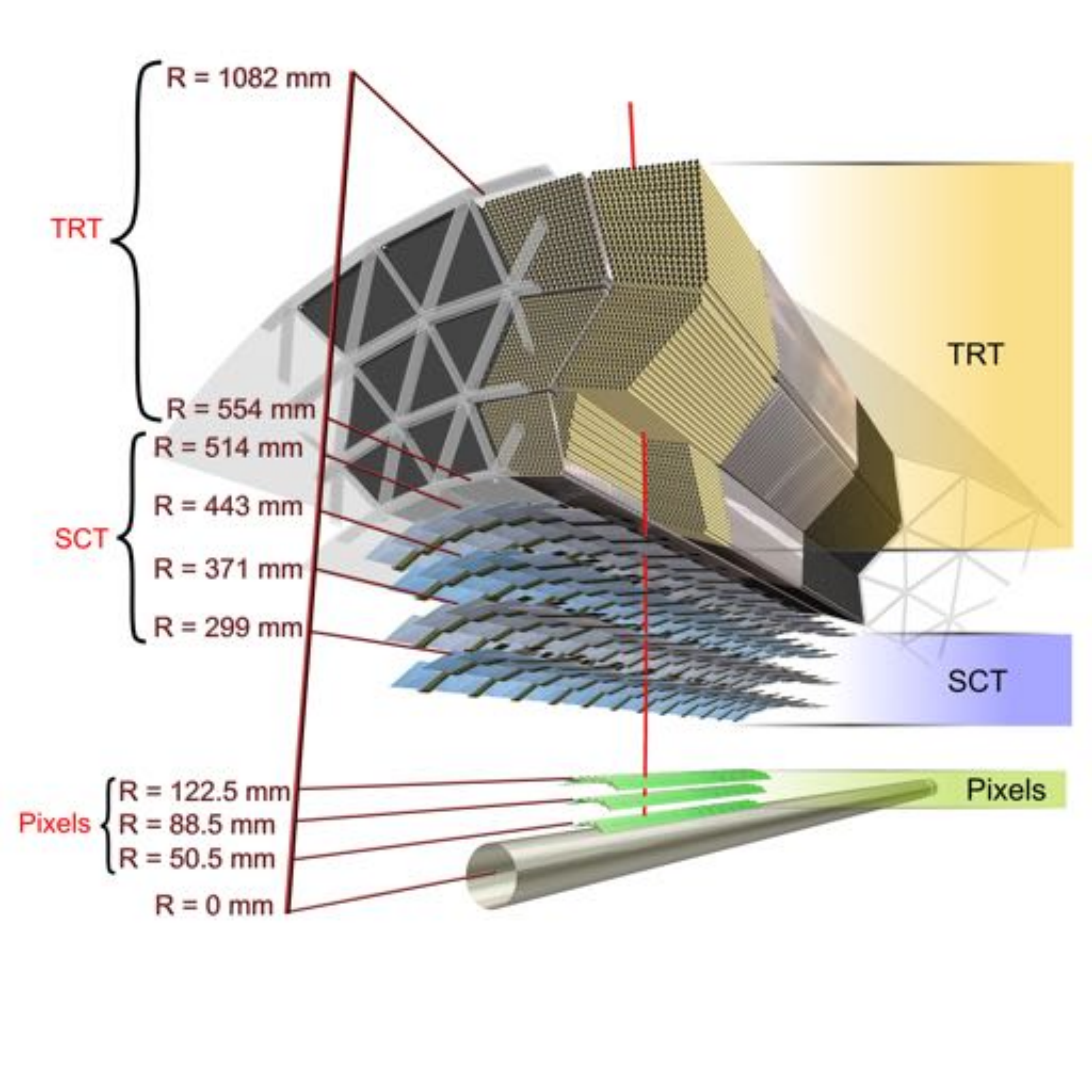}}  
   \end{minipage}
   \caption{ Schematic views of the ATLAS Run 1 inner detector:
  (a) barrel and end-cap sections; (b)  cross section of the barrel section showing the TRT, SCT, and pixel sub-detectors.}
   \label{fig:InnerDetBarrel}
 \end{figure}

The inner detector covers the pseudorapidity\footnote{The ATLAS experiment uses a right-handed
coordinate system with its origin at the nominal interaction point (IP) in
the centre of the detector and the $z$-axis along the beam direction. The
$x$-axis points from the IP to the centre of the LHC ring, and the $y$-axis
points upward. Cylindrical coordinates $(r,\phi)$ are used in the
transverse $(x,y)$ plane, $\phi$ being the azimuthal angle around the $z$-axis. 
The pseudorapidity is defined in terms of the polar angle
$\theta$ as $\eta=-\ln\tan(\theta/2)$.} range $|\eta| <$ 2.5. 
 Schematic views of the Run 1 inner detector are presented in Fig.~\ref{fig:InnerDetBarrel}.
Particle trajectories are identified using  the combined information from the sub-detectors of the ID:
 the innermost silicon pixel detector, the surrounding silicon microstrip
 semiconductor tracker (SCT), and  the transition radiation tracker (TRT), made of straw tubes filled with a Xe-$\textrm C\textrm O_2$ gas mixture \cite{InnerDetectorTDR}.
 All three sub-systems are divided into a barrel section and two end-caps. The barrel sections consist of
 several cylindrical layers, while the end-caps are composed of radial disks and  wheels.
The sensitive regions of the three sub-detectors  cover radial distances in the barrel section
from  $50.5$ to $122.5$, $299$ to $514$, and $554$ to $1082$~mm.
Typical position resolutions are $10$, $17$, and $130 \;\mu{\textrm m}$ for the transverse coordinate
in the pixel detector, the SCT, and the TRT respectively. In the case of the pixel and SCT,
the resolutions in the $z-$coordinate are $115$ and $580 \;\mu {\textrm m}$. 
The superconducting solenoid coil around the tracking system produces a
 $2$~T axial magnetic field.  
A track from a charged particle traversing the barrel detector would 
typically have $11$ measurements in the silicon detector\footnote{Measurements of
charged particle trajectories in the pixel, SCT and TRT are called \em{ID hits}.}
($3$ pixel clusters and $8$ strip clusters) and more than $30$ measurements in the TRT \cite{ChargedPmultiplicities}.
\subsection{\label{sec::trigger}The minimum-bias trigger}
A minimum-bias trigger was used to select the data presented in this paper.
This trigger is designed to record a random selection of bunch crossings, unbiased by any hard physics produced in the bunch crossing, by using a signal from
 the minimum-bias trigger scintillators (MBTS). The MBTS
 are mounted at each end of the detector in front of the liquid-argon 
end-cap calorimeter cryostats at $z = \pm 3.56$~m, covering the range $2.09 < |\eta| < 3.84$.
 The MBTS trigger used for this 
paper requires one hit above threshold from either side of the 
detector, referred to as a single-arm trigger \cite{ChargedPmultiplicities}.
\subsection{\label{sec::PileUp}Determination of pile-up interactions}
Depending on the length of the read-out window of a sub-detector, signals from
neighbouring bunch crossings can be   present simultaneously when the detector is read 
out. The impact of interactions from the neighbouring bunch crossings is referred
 to as out-of-time pile-up, while in-time pile-up results from the
presence of multiple  $pp$ interactions in the same bunch crossing.

During most of Run 1 of the LHC, the separation of proton bunches was $50$~ns.
The timing resolution of the inner detector components is about $25$~ns.
This is  sufficient for the out-of-time pile-up to have a much smaller impact on ID measurements than the in-time pile-up.
As a consequence the number of reconstructed vertices is a direct measure of the amount of in-time pile-up on an event-by-event
basis.

The instantaneous luminosity, $L$, can be expressed in terms of the
visible interaction rate,  $R^{\textrm vis}_{\textrm{inel}}$, and the visible inelastic cross section, $\sigma^{\textrm vis}_{\textrm{inel}}$, as:
\begin{equation}
\label{eq:basic_lumi}
L = \frac{R^{\textrm vis}_{\textrm{inel}}}{\sigma^{\textrm vis}_{\textrm{inel}}}.
\end{equation}
The inelastic cross section, $\sigma _{\textrm inel}$,  and  the visible inelastic cross section are related through: 
$\sigma ^{\textrm vis} _{\textrm inel}~=~\epsilon \sigma_{\textrm inel}$. Here  $\epsilon$ is the efficiency of the
detector to record an inelastic collision.  The inelastic cross section is defined as the 
total cross section minus the elastic cross section.

In practice, the full  rate of inelastic collisions is never directly measured.  Only
a fraction of it is observable in the detector due to the $\eta$ acceptance. 
The luminosity is measured using a set of dedicated detectors which allow  bunch-by-bunch
measurements. The luminosity detectors
are calibrated using dedicated Van der Meer scans \cite{Aad:2013ucp}.
The uncertainty in the luminosity measurement is $1.9\%$ \cite{Lumi8}. 

The number of $pp$ inelastic interactions per bunch crossing follows a Poisson distribution with mean value~$\mu$.
Assuming that the $pp$ collider operates at a revolution
frequency $f_{\textrm r}$  with $n_{\textrm b}$ interacting bunches per beam,
the luminosity can also be expressed as:
\begin{equation}
\label{eq:freq_lumi}
L = \frac{ \mu \; n_{\textrm b} \; f_{\textrm r}}{\sigma_{\textrm{inel}}}.
\end{equation}

The value of $\mu$ changes during data-taking as a function of time: it decreases with decreasing beam intensity 
and increasing emittance. The highest  value  is at the start of the stable beam period of the fill.  
For the studies presented in this paper, $\mu$ is calculated using Eq.~\ref{eq:freq_lumi}.
The  value of the inelastic cross section at 8~TeV centre-of-mass energy is 71.5 mb, taken from
 the PYTHIA8 MC generator \cite{Sjostrand:2007gs}.  
Experimental measurements \cite{1748-0221-3-08-S08007, Alfa} are found to be compatible with the cross section predicted by PYTHIA8.
The overall uncertainty in $\mu$ is $4\%$, which is derived from the quadratic sum of the 
uncertainties in the luminosity and in the inelastic cross section.

\subsection{Parameters affecting the luminous region at the LHC}
The size, position and shape of the luminous region, or beam spot,
 are determined by the operating parameters of the beams and magnets of the LHC \cite{Evans:2008zzb}.
The transverse size is determined by the focusing of the LHC beams near the interaction region and by the spread in position--momentum phase space of 
the protons within the colliding bunches.
The latter is quantified by the geometric emittance \emitjb{} of the beams, or equivalently by the normalised emittance defined as 
$\emitnjb{} = \beta_v \, \gamma \, \emitjb{}$, where $\beta_v$ and $\gamma$ are the relativistic functions $\beta_v = v/c \simeq 1$ and $\gamma = \Ebeam/m_p$, \Ebeam 
is the beam energy and $m_p$ is the  mass of the proton.
The focusing of the beams is characterised by the $\beta$-function, and especially its minimum value \bst. The longitudinal size of the luminous region is determined by the bunch length and by the angle \xang (full crossing angle) at which the two beams are brought into collision. In the following discussion it is assumed that 
the emittances and $\beta$-functions in the horizontal and vertical direction are the same for each of the two beams.
These assumptions lead to a circular transverse beam profile, as has been observed to be approximately the case at the LHC.

The particle densities in proton bunches can be described by three-dimensional Gaussian distributions
 with transverse and longitudinal sizes given by
$\bsigj{x} =\bsigj{y} = \sqrt{\emitjb{} \, \btajb{}}$ and $\bsigz{} = c \, T_z / 4$ respectively, where $T_z$ is the "four $\sigma$ bunch length" (in ns) customarily quoted 
for the LHC. Because the ratio $\bsigz{}/\bst$ was small during Run 1, the quadratic form of the $\beta$-function around the interaction region had a negligible effect over the length of the luminous region and the transverse beam size along the beam axis remained constant.
As a result the luminous region is described well by a three-dimensional Gaussian distribution. With the assumption of pair-wise equal bunch sizes mentioned above, the transverse size \lumsigx (and equivalently \lumsigy) of the luminous region is given by $\lumsigx = \bsigx{} / \sqrt{2}$.
For a crossing angle in the vertical plane as is the case for ATLAS, and assuming equal longitudinal bunch sizes \bsigz{} in both beams, the longitudinal size of the luminous region is given by:
\begin{equation}
\lumsigz = \frac{c \, T_z / 4}{\sqrt{2} } \frac{1}{\sqrt{1 \, + \, (\frac{\bsigz{}}{\bsigy{}}\frac{\xang}{2})^2}} \; .
\end{equation}

A summary of typical LHC parameters for $pp$ collisions at $\sqrt{s} = 7\,\TeV$ in 2011 and at $\sqrt{s} = 8\,\TeV$ in 2012 is shown in Table~\ref{tab:lhcparam} together with the resulting expected sizes of the luminous region.  The measured sizes of the luminous region are discussed in  Section~\ref{BeamSpotStability} and Table~\ref{tab:beamspot-averages}.
\begin{table}[tp]
\begin{center}
\begin{tabular}{lcc}
\hline \hline
Year                                                & 2011       & 2012 \\ \hline
Beam energy [TeV]                                   & 3.5        &   4.0   \\
$\bst$ [m]                                          & 1.5, 1.0  &   0.6   \\
Normalised emittance $\varepsilon_{\textrm N}$ [$\mu$m rad]   & 2.5        &   2.5   \\
Full crossing angle \xang [$\mu$rad]                & 240        &   290   \\
$4\sigma$ bunch length $T_z$ [ns]                   & 1.20        &   1.25  \\
Bunch length \bsigz{} [mm]                          & 90         &   94   \\
\hline
Expected transverse beam-spot size \lumsigx, \lumsigy [$\mu$m]  & 22, 18  &  13  \\
Expected longitudinal beam-spot size \lumsigz [mm]              & 60, 59  &   54  \\
\hline \hline
\end{tabular}
\end{center}
\caption{\label{tab:lhcparam}Summary of LHC parameters for typical $pp$ collision fills and corresponding expected sizes of the luminous region. Emittance and bunch length values (and the corresponding beam-spot sizes) refer to values expected at the start of a fill. The two values given for expected transverse and longitudinal beam-spot size in 2011 correspond to the two $\bst$ settings of 1.5~m and 1.0~m. Measured average beam-spot parameters are presented in Table~\ref{tab:beamspot-averages} (Section~\ref{BeamSpotStability}).}
\end{table}

%% file: Datasets.tex
\begin{figure}[tp]
\centering
\subfloat{\includegraphics[width=0.55\textwidth]{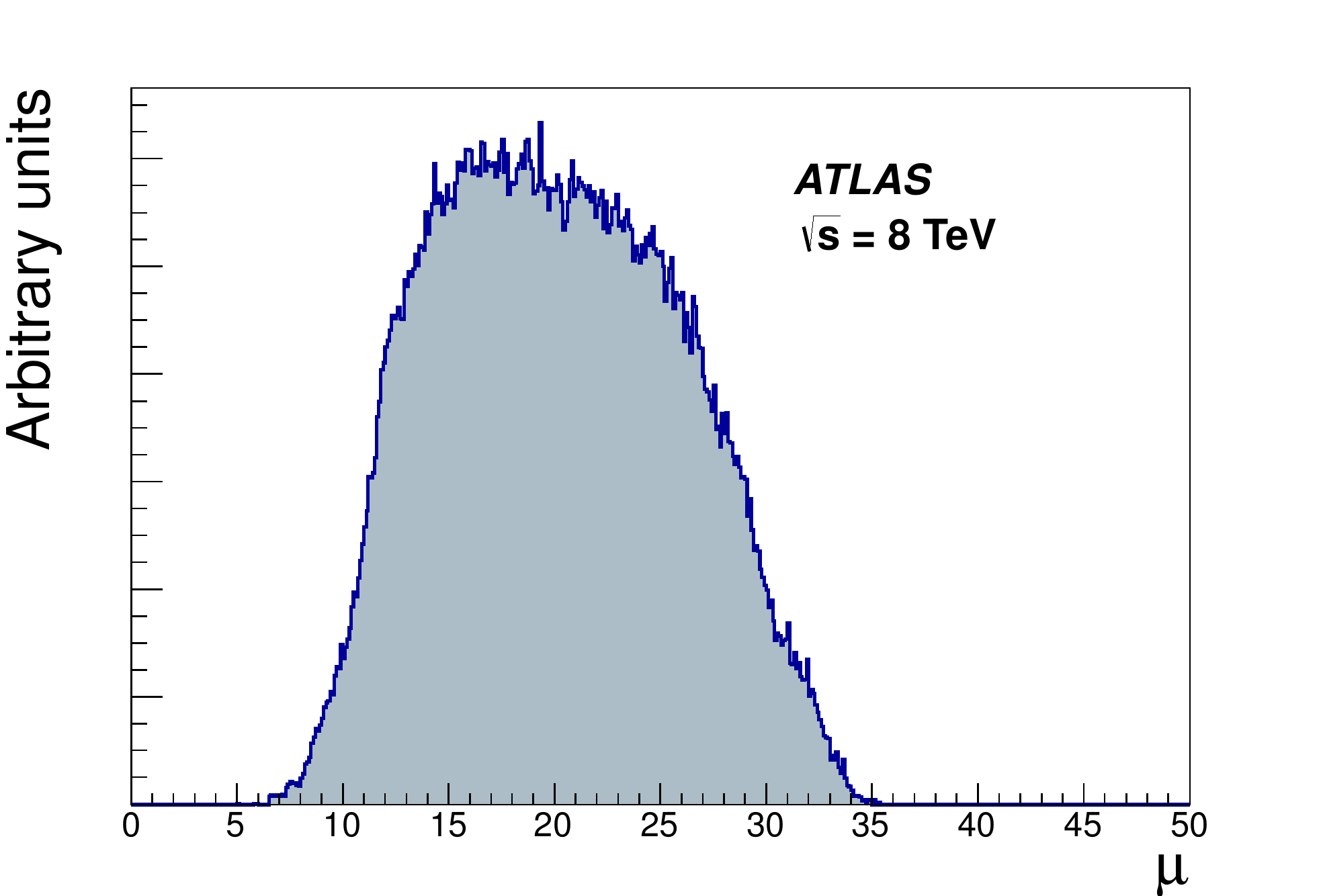} }
\caption{ The average number of interactions per proton bunch crossing, $\mu$, during 8 TeV data-taking in Run 1, weighted by the luminosity.
}
\label{fig:AveInt}
\end{figure}

This paper uses $pp$ collision data with $\sqrt{s} = 8$ TeV recorded during the LHC Run 1 period.
Data were collected  using the minimum-bias triggers described in Section~\ref{sec::Atlas}.
The data-taking conditions 
of the corresponding data samples are summarised in Table~\ref{tb:Datasets}.  
 \begin{table}[tp]
\begin{center}
 \begin{tabular}{lccc}
 \hline \hline 
 Pile-up conditions & $\mu$ range &Date \\ 
\hline
Low $\mu$ &         0--1 & April 2012\\
High $\mu$ &      55--72 & July 2012  \\
Run 1 data range & 7--40 & 2012 \\
\hline \hline
\end{tabular}
\end{center}
\caption{\label{tb:Datasets}The data-taking conditions  of the $pp$ collision data samples used in this paper.}
\end{table}
 The studies presented  here aim to cover the full range of Run 1 $\mu$ values and use both a special high-$\mu$ data sample 
 as well as a range of  lower-$\mu$ data. The distribution of the average number of interactions
per $pp$ bunch crossing in Run 1 is shown in Fig.~\ref{fig:AveInt}. 
This does not include the special
high and low $\mu$ runs listed in Table~\ref{tb:Datasets}. Most data taken in Run 1 had pile-up near $\mu = 20$.
The low pile-up dataset was taken at average $\mu$ around $0.01$, while the special high pile-up run featured
 peak collision multiplicities up to $\mu=72$.

The results presented in this paper use  MC simulation of 
hard-scatter interactions and soft inelastic $pp$ collisions.
The collection of soft inelastic interactions is referred to here as the 
minimum-bias sample.  These are events that would have been
collected with the minimum-bias trigger, described in
Section~\ref{sec::trigger}, and they represent an average beam crossing, without selection 
of a specific hard-scatter interaction.

Minimum-bias samples were
simulated with the PYTHIA8 MC generator, with the A2  set of tuned parameters~\cite{ATL-PHYS-PUB-2012-003}  
and the MSTW2008LO 
parton density function set \cite{Martin:2009iq}. 
The  PYTHIA8  model 
for soft QCD uses a phenomenological adaptation of  $2\to 2$  parton scattering to 
describe low transverse momentum processes.   
Samples were generated for non-diffractive, 
single-diffractive, and double-diffractive interactions. These contributions
 were combined according to the  PYTHIA8 generator cross sections.

 To study the collective effects of multiple primary vertices reconstructed in one beam crossing, 
MC simulation with no hard-scattering process but only pile-up was created for $\mu$ up to $72$.
 These samples mimic randomly triggered events, and were also generated with PYTHIA8 using the A2 tune.
A special configuration was used to match 2012 data-taking conditions,
including the beam spot with  $z$-direction size equal to the average observed in data.

Hard-scatter interactions were simulated with POWHEG  \cite{Alioli:2010xd} interfaced to  PYTHIA8 
 for the  $Z\to \mu\mu$  and $H\to \gamma\gamma$ processes, and  MC@NLO \cite{Frixione:2002ik}, HERWIG \cite{Herwig}
and Jimmy~\cite{Jimmy} for  top-quark pair production  ($t\bar{t}$). 
The CT10 parameterisation \cite{PDF-CT10} of the parton density functions was used.
 The top-quark pairs were generated with a lepton filter, requiring a lepton in the final state.
The hard-scatter interaction samples were generated  for a range of pile-up between $\mu=0$ and $38$.
The overlaid pile-up collisions were simulated with the soft QCD processes of PYTHIA8 in the manner of
the minimum-bias simulation described above.

All generated events are processed with the  ATLAS detector  simulation framework \cite{Aad:2010ah}, 
using the GEANT 4~\cite{Agostinelli:2002hh} toolkit. After full detector simulation, the MC events are 
reconstructed and analysed in the same manner as data.

When comparing data with simulation in the presence of pile-up interactions, 
the average number of collisions per bunch crossing in simulation is re-weighted
to match that measured in data.
In order to obtain the same visible cross section for $pp$ interactions for the simulation and data,
a $\mu$-rescaling is also applied before the re-weighting. 
The rescaling factor is calculated by comparing 
the ratio of the visible cross section to the total inelastic cross section,  $\epsilon_{\xi} = \sigma^{\textrm vis}_{\textrm inel} / \sigma_{\textrm inel}$,
for data with that for simulation.
The value of $\epsilon_{\xi}^{\textrm data}$ is computed from independent measurements of these
cross sections in data \cite{Aad:2011eu, Antchev:2011vs}. The value of $\epsilon_{\xi}^{\textrm MC}$ is
 computed from events simulated with the PYTHIA8 MC generator with the A2 tune.
The final scale factor is corrected to match the visible cross section within 
the ATLAS inner detector acceptance, resulting in $\epsilon^{\textrm MC}_{\xi}/ \epsilon^{\textrm data}_{\xi}=1.11$.
The uncertainty in this scale factor is $5\%$. It is calculated from the quadrature sum of the uncertainties in the cross-section measurements,
 $3.5\%$ and $2.6\%$ from Ref.~\cite{Aad:2011eu} and Ref.~\cite{Antchev:2011vs} respectively, 
and a $2\%$ uncertainty in the extrapolation from $7~$TeV to $8~$TeV and to the inner detector acceptance.

%% file: Reconstruction.tex
This section describes the method for reconstructing primary vertices. 
The input to the vertex reconstruction is a collection of reconstructed tracks.
A brief summary of the main steps of track reconstruction is presented 
 in Section \ref{sec::TrackReco}.
The vertex reconstruction is presented in Section \ref{sec::FitFind}.
This is followed by a description of how primary vertices are used to reconstruct
the shape of the luminous region, or beam spot, in Section \ref{BeamSpot}, and
a description of the stability of the beam spot in Section \ref{BeamSpotStability}.

\subsection{\label{sec::TrackReco}Track reconstruction}
The reconstruction of charged-particle trajectories in the inner detector is based on 
fitting a trajectory model to a set of  measurements. 
The reconstructed charged-particle trajectories are hereafter referred to 
 as tracks. The general structure and performance of ATLAS track reconstruction 
is described in detail in Refs.~\cite{atlasConf:pileup, NEWT} and a brief overview is 
given below.

Track seeds consist of three measurements in different layers of 
the pixel detector and SCT. Tracks are propagated out from the seed 
towards the TRT (``inside-out'')  using a combinatorial
Kalman filter \cite{NEWT}, and additional silicon hits are added to the seed. 
An ambiguity solving procedure  is applied to remove track candidates 
with incorrectly assigned hits.
 The  candidate tracks are scored in a reward--penalty schema with respect to one another.
To favour fully reconstructed tracks over short track segments, each additional measurement
associated with a track leads to a better score value. The measurements  from different sub-detectors 
are weighted differently, preferring the precision measurements (e.g. pixel clusters) 
and downgrading measurements from less precise detector parts. To provide a realistic description of
detector  acceptance and efficiency, the concept of  a  hole  on a track is introduced. A hole represents 
a measurement on a detector surface 
 that is expected, given the trajectory predictions, but not observed
 (holes are not considered on the first and last surfaces
in the measurement). 
The presence of holes reduces the overall track score.   The $\chi^{2}$ of the track fit is also
used to penalise poor-quality candidates.  Finally, the logarithm
of the track transverse momentum  $\ln({p_{\textrm T}})$ is considered as a criterion to promote energetic tracks and 
to suppress the larger number of tracks formed from incorrect combinations of clusters, which tend to have  low  
measured $p_{\textrm T}$. After the reconstruction of tracks in the pixel and the SCT detectors, the successful candidates are 
extrapolated into the TRT volume and combined with measurements there. 

During data-taking at $\sqrt{s}=8$~TeV, the input to the vertex reconstruction 
algorithms consisted of charged-particle tracks selected according to the following criteria:
\begin{itemize}
\item $\pT>400$~MeV; $|d_{0}|<4$~mm; $\sigma(d_{0})<5$~mm; $\sigma(z_{0})<10$~mm;
\item  At least four hits in the SCT detector;         
\item  At least nine silicon (SCT or pixel) hits;
\item  No pixel holes.
\end{itemize}
Here the symbols $d_{0}$ and $z_{0}$ denote the transverse and longitudinal
impact parameters of tracks with respect to the centre of the luminous region,
and $\sigma(d_{0})$ and $\sigma(z_{0})$ denote the corresponding uncertainties \cite{atlasConf:pileup}.
The impact parameter requirements are  applied to reduce contamination from  tracks originating from secondary interactions.
The above requirements are tighter than the standard ATLAS track selection criteria
in order to maintain a low rate of fake tracks  
(tracks mistakenly reconstructed from a random combination of hits)
at Run 1 pile-up levels (up to $\mu = 40$).   The track reconstruction efficiency under this selection 
is between $75\%$ and $85\%$ for central rapidities ($|\eta| < 1.5$) and track $\pT$ above 500 MeV; the
efficiency falls to about $60\%$ at higher rapidities or about $65\%$ for  tracks with $\pT$ between 400 and 500 MeV.
\subsection{\label{sec::FitFind}Primary vertex finding and fitting}
The procedure of primary vertex reconstruction is divided into two 
stages: vertex finding and  vertex fitting \cite{Piacquadio:2008zzb}. The former stage
generally denotes the pattern recognition process: the association of reconstructed tracks to 
vertex candidates. The vertex fitting stage deals with  reconstruction of the actual vertex position 
and its covariance matrix. 
The strategy is explained in detail in this section, and can be 
briefly outlined in these steps:
\begin{itemize}
\item  A set of tracks satisfying the track selection criteria is defined.
\item  A seed position for the first vertex is selected.
\item  The tracks and the seed are used to estimate the best vertex position with a fit. The fit is an iterative procedure, and in each
iteration less compatible tracks are down-weighted and the vertex position is recomputed.
\item  After the vertex position is determined, tracks that are incompatible with the vertex are removed from it and allowed to be used in the determination of another vertex.
\item  The procedure is repeated with the remaining tracks in the event. 
\end{itemize}
Each of these steps (except the track selection described in the previous section) is expanded on below.
\begin{enumerate}
\item{
The seed position of the vertex fit is based on the beam spot in the transverse plane.
  The $x$- and $y$- coordinates of the starting point are
 taken from the centre of the  beam spot, reconstructed as discussed in Section \ref{BeamSpot}.
 The $z$-coordinate of the starting point
 is calculated as the mode of the $z$-coordinates of tracks at their respective points of closest
 approach to the reconstructed  centre of the beam spot. The mode is calculated using 
 the Half-Sample Mode algorithm \cite{HSM}. 
}

 \item{After the seed has been determined, the iterative primary vertex finding procedure  begins. 
The vertex position is determined using 
  an adaptive vertex fitting algorithm with an annealing  procedure~\cite{Fruhwirth:2007hz}.
Using the seed position as the starting point and parameters   of reconstructed 
tracks as input measurements, the algorithm performs an iterative
$\chi^2$ minimisation, finding the optimal vertex position. Each input track
is assigned a weight, reflecting its compatibility with the vertex estimate.  The vertex position is 
recalculated using the weighted tracks, and then the procedure is repeated, recalculating track
weights with respect to the new vertex position.
  The individual track weights are calculated according to the following equation:
\begin{equation}
  \omega ( \hat{\chi}^2 ) = \frac{1}{1 + \exp \left( \frac{\hat{\chi}^{2} - \chi^{2}_{\textrm cutoff}}{2T}\right)}.
  \label{eq:weight}
\end{equation}
Here  $\hat{\chi}^{2}$ is the  $\chi^2$ value calculated in three dimensions between the last 
estimated vertex position and the respective point of the closest approach of the track.
Tracks with lower weights are less compatible with the vertex and will have less influence on the position calculation.
The  constant $\chi^{2}_{\textrm cutoff}$ defines the threshold where the weight 
of an individual track becomes equal to $0.5$. Tracks with low weights are not removed, but will have less
impact on the calculated vertex position.
The value of  $\chi^{2}_{\textrm cutoff}$   is set to nine, which 
corresponds to about three standard deviations.   The temperature $T$ controls the smoothness of the weighting 
procedure. For low values of $T$, $\omega ( \hat{\chi}^2 )$  approaches a step function, and for large values of $T$ the
function flattens, progressively losing its $\chi^2$ dependence.
To avoid convergence in local minima, the weighting procedure 
is applied  progressively by decreasing the temperature $T$ during the fit iterations. 
The temperature is lowered from some high starting value in a  pre-defined sequence of steps 
that converges at $T=1$.   A typical distribution of track weights is shown in  Fig.~\ref{fig:trackweights}.
It widens as $T$ decreases, reaching an optimal separation of track outliers for $T = 1$.
\begin{figure}[tp]
\centering
\subfloat{\includegraphics[width=0.55\textwidth]{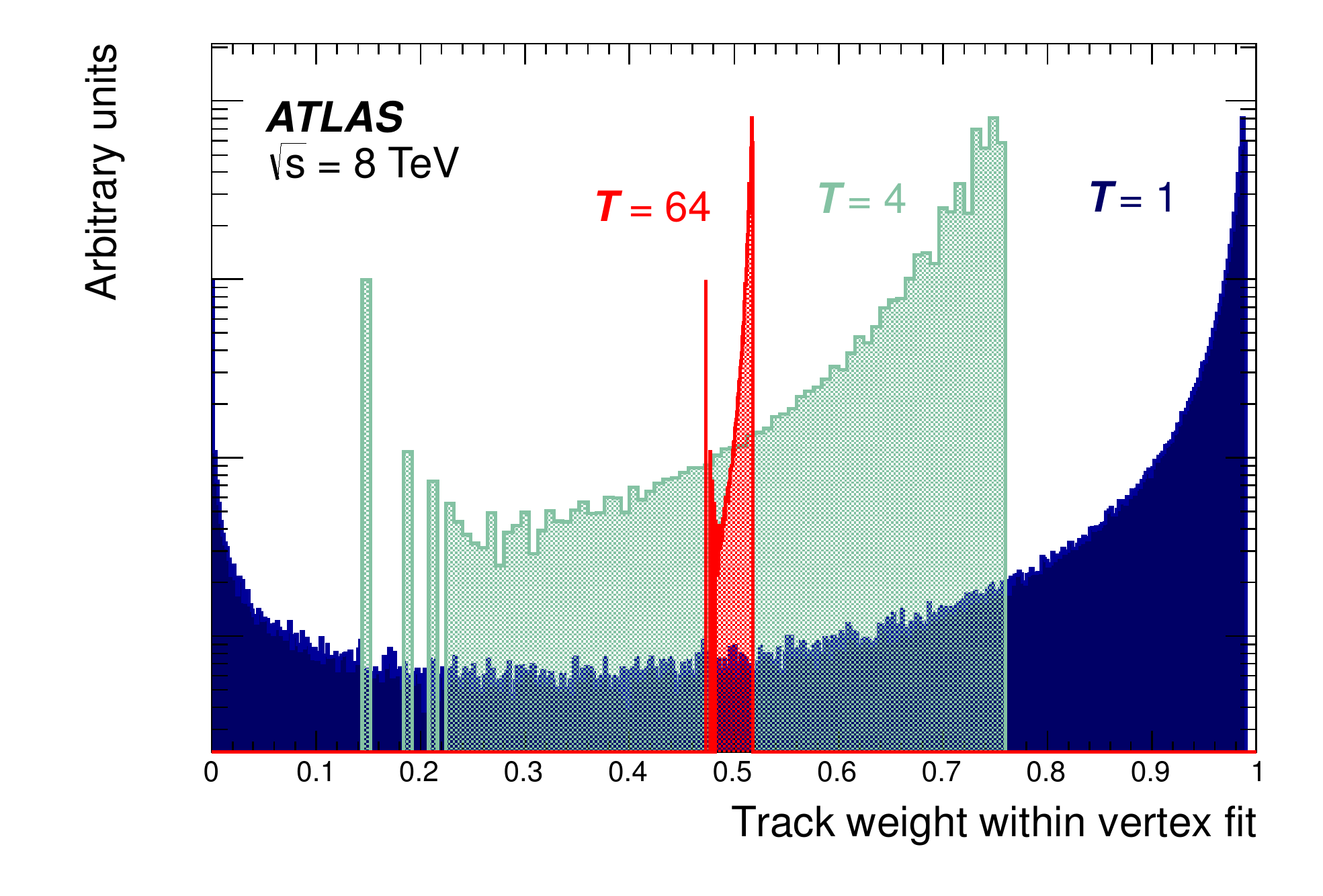} }
\caption{ Histogram showing the weights applied to tracks in the vertex reconstruction fit.  The fitting algorithm iterates through
progressively smaller values of the temperature $T$, effectively down-weighting outlying tracks in the vertex fit.
The vertical axis is on a logarithmic scale.
}
\label{fig:trackweights}
\end{figure}}
 \item{After the last iteration, the final weight of each track used in the vertex fit is evaluated.
  Tracks found incompatible with the vertex by more than seven standard deviations are 
 removed from the vertex candidate and returned to the pool of unused tracks.
This loose  requirement  is intended to reduce the number of single $pp$ interactions which are 
reconstructed as two distinct primary vertices due to the presence of track outliers,
while maintaining a high efficiency. }   
\item{After the vertex candidate is created, the rejected tracks are considered 
as input for a new vertex finding iteration. 

The procedure described above is then repeated starting from step $1$, calculating the new starting position from remaining tracks,
 until no unassociated tracks are left in the event or no additional vertex  can be found in the remaining set of tracks.}
\end{enumerate}

\begin{figure}[tp]
\centering
\subfloat{\includegraphics[width=0.55\textwidth]{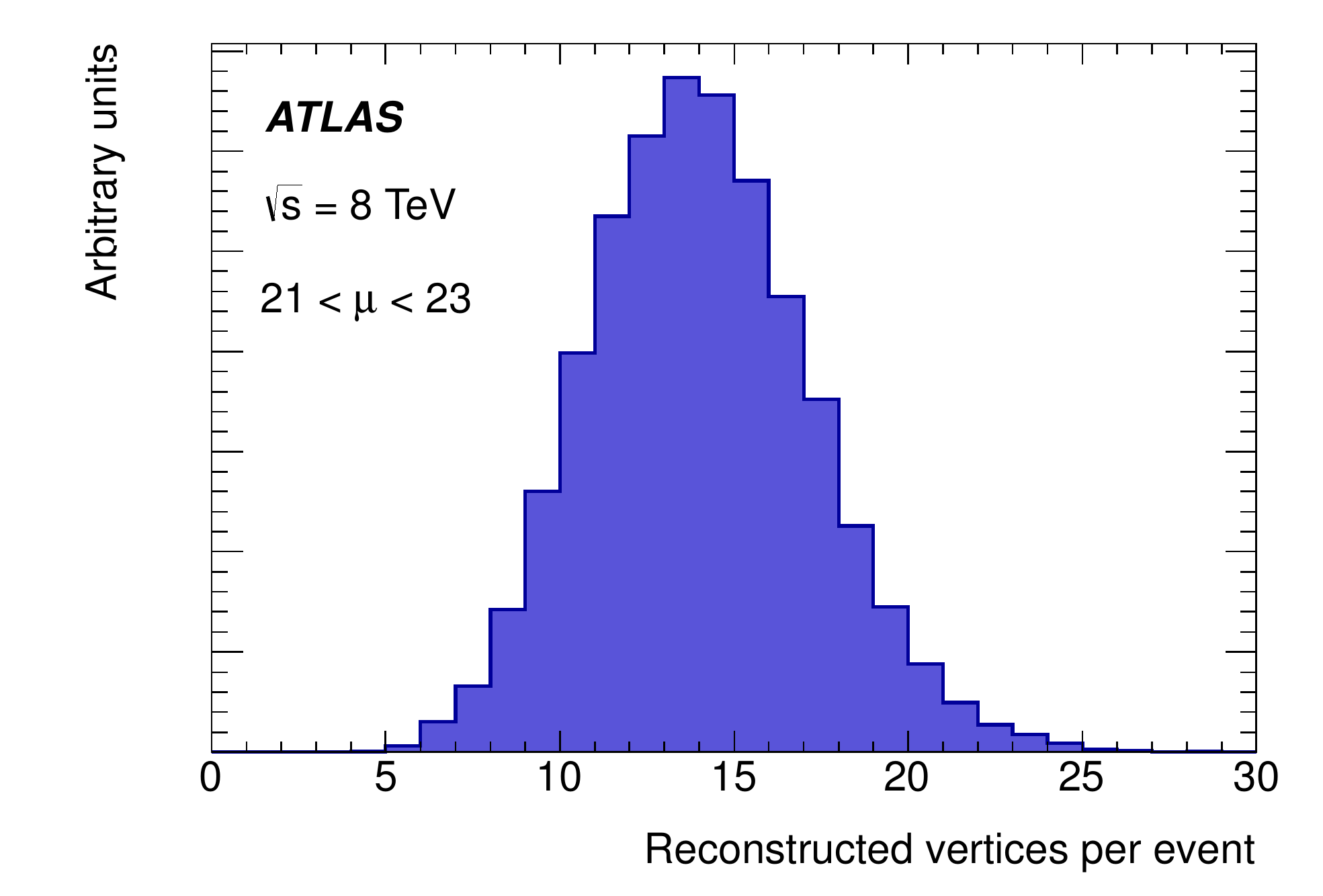} }
\caption{ Distribution of the number of reconstructed vertices per event in a sample of $\sqrt{s}= 8$ TeV minimum-bias data for the pile-up range $ 21< \mu <23 $.}
\label{fig:nVtx}
\end{figure}

All vertices with at least two  associated tracks are retained as valid primary vertex candidates.
The output of the vertex reconstruction algorithm is a set of three dimensional vertex positions and their covariance matrices.
Figure~\ref{fig:nVtx} shows a typical distribution for the number of reconstructed vertices per event in Run 1
for  minimum-bias data  collected in the pile-up range $21< \mu <23$.

\begin{figure}[tp]
\centering
\subfloat[]{\includegraphics[width=0.6\textwidth]{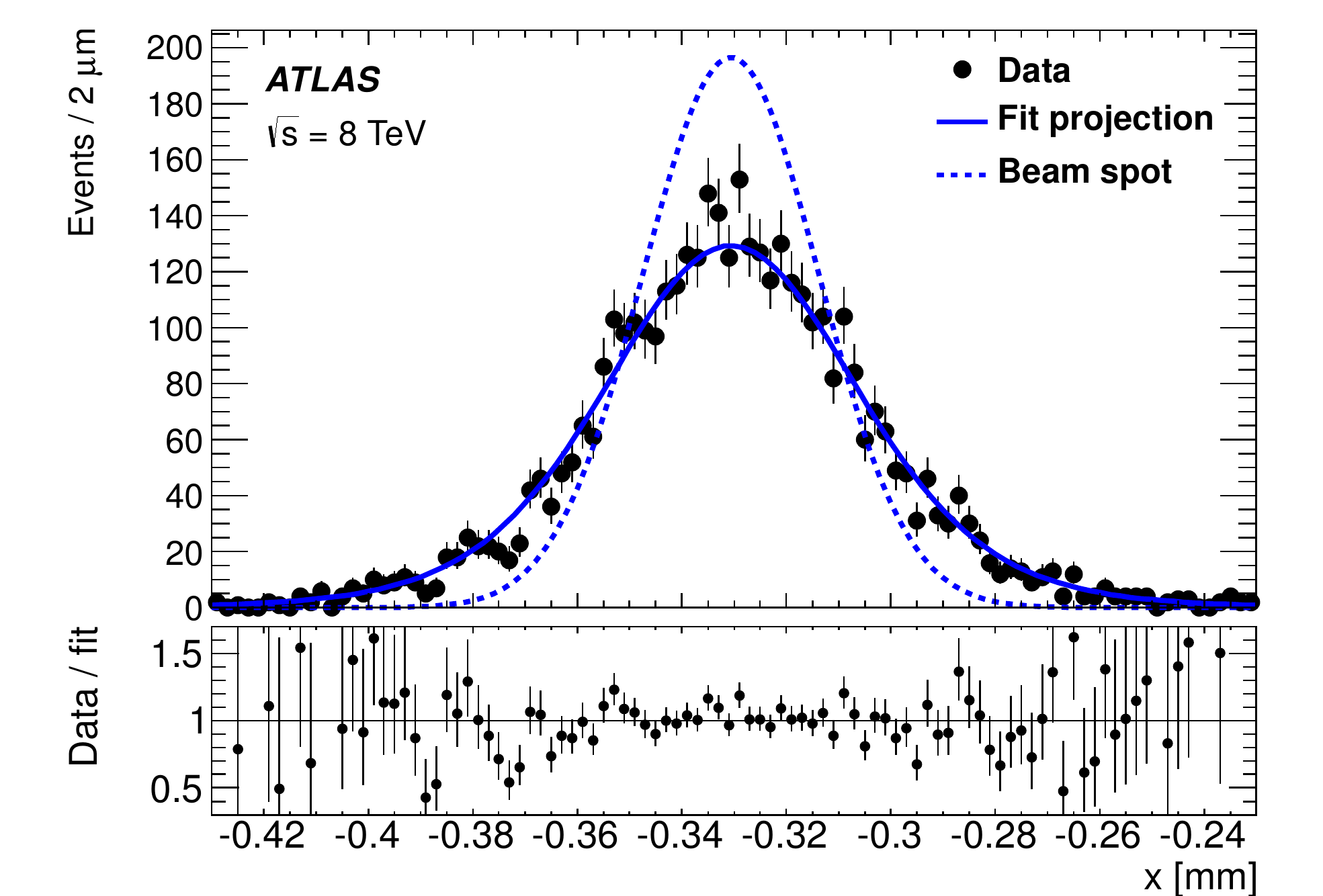}}\\
\subfloat[]{\includegraphics[width=0.6\textwidth]{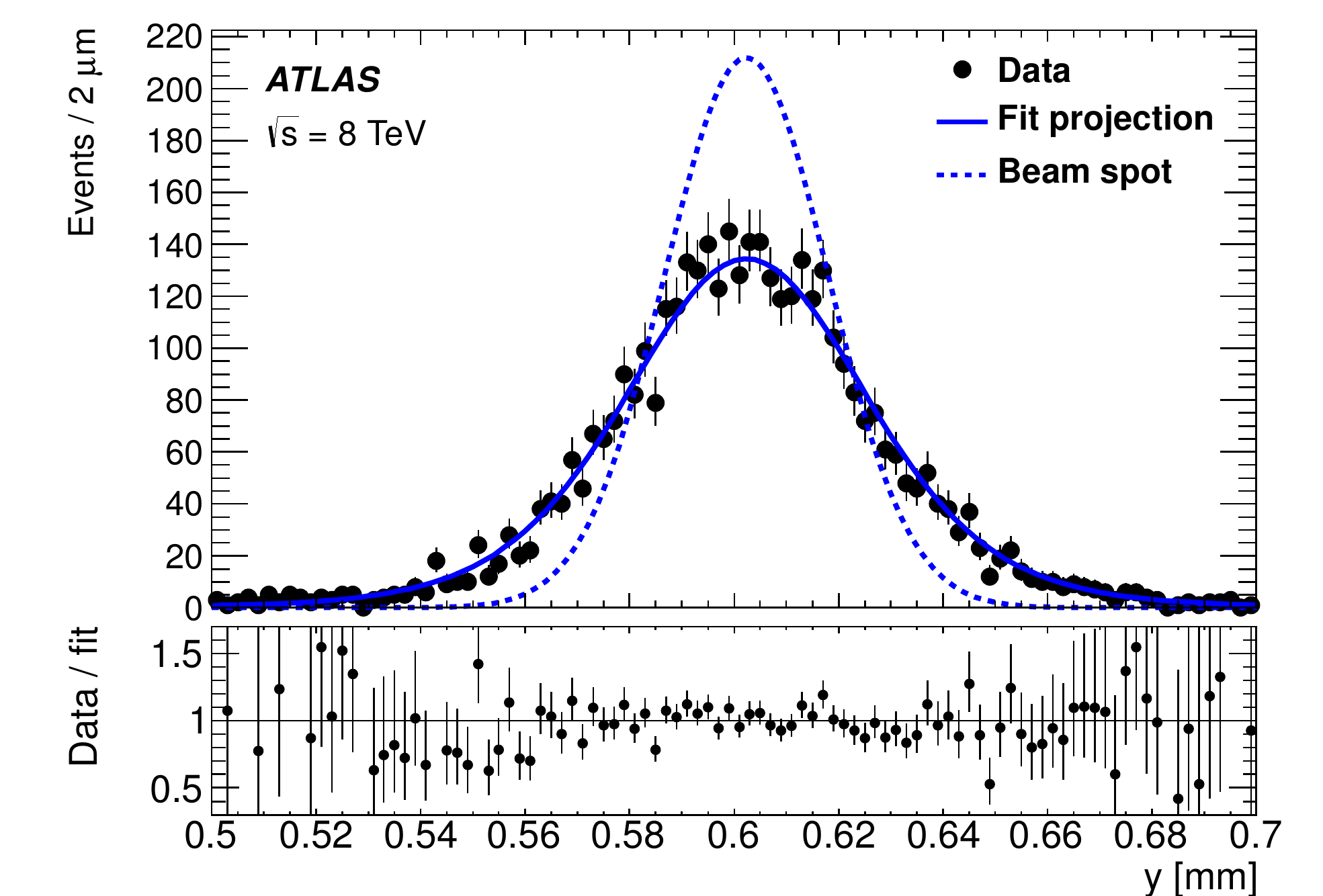}}\\
\subfloat[]{\includegraphics[width=0.6\textwidth]{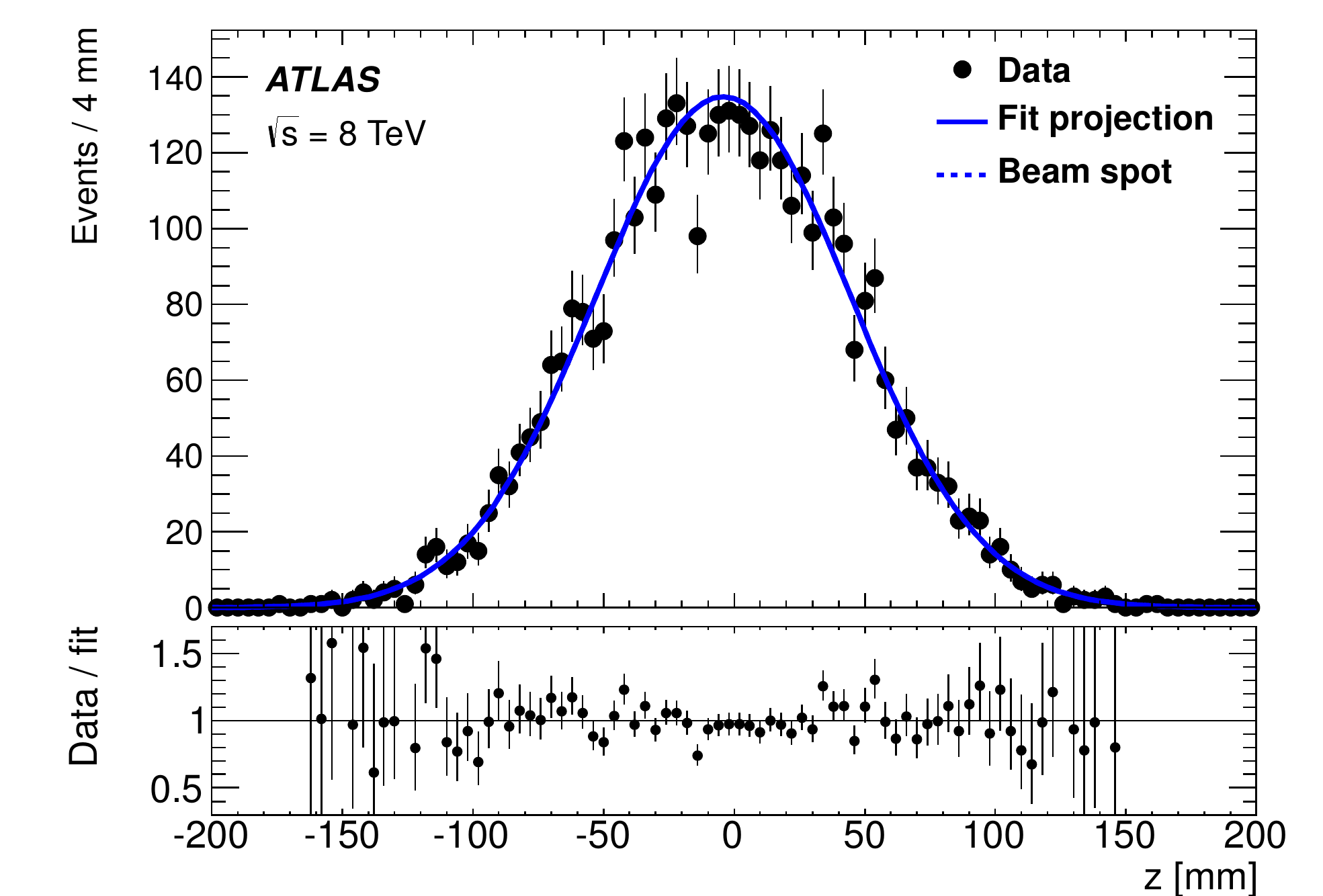}}
\caption{\label{fig:beamspot-run215456-fitman-y}Distribution in (a) $x$, (b) $y$ and (c) $z$ of the reconstructed primary vertices used for a typical single beam-spot fit, projection of the 3D Gaussian beam-spot fit result, and fitted beam spot. The fit projection and beam spot curves are identical in (c).}
\end{figure}

The reconstructed position and width of the beam spot can be used as an additional measurement during the
primary vertex fit.
It is taken as a three-dimensional Gaussian measurement centred around the 
beam-spot centre and with the beam-spot size as the width. 
Tracks outside the beam spot have low compatibility 
with the vertex fit and are thus removed in the iterative fitting procedure.
   This procedure is hereafter referred to as the beam-spot constraint.
Figure~\ref{fig:beamspot-run215456-fitman-y} shows typical distributions of the $x$, $y$, and $z$ coordinates of  
primary vertices without the beam-spot constraint.
The transverse position resolution of vertices reconstructed from a small number of tracks                                         
may exceed $100\;\mu{\textrm m}$. For these vertices the application of the beam-spot 
constraint significantly improves their transverse position resolution.
In the $z$-direction, the length of the luminous region has no significant impact on the   
 resolution of primary vertices. The longitudinal resolution of primary vertices is determined by the intrinsic resolution 
of the primary tracks. However, knowledge of the  longitudinal beam-spot size still helps to remove far outlying tracks.               
\subsection{\label{BeamSpot}Beam-spot reconstruction}
The beam-spot reconstruction is based on an unbinned maximum-likelihood fit to the spatial distribution of primary vertices collected from many events. These 
primary vertices are reconstructed without beam-spot constraint from a representative subset of the data called the express stream during the 
detector calibration performed approximately every ten minutes.
In each event only the primary vertex with the highest sum of squares of transverse momenta  of contributing tracks, denoted hereafter as
$\sum{\pt^2}$, is considered. In order to be used in the beam-spot fit, this vertex must
 include at least five tracks and must have a probability of the $\chi^2$ of the vertex fit greater than 0.1\%. The requirement of at least five tracks ensures that most vertices have 
 a transverse vertex resolution better than $50\,\um$ with a most probable value of about $15~\um$ that is comparable to the transverse beam-spot size.
At least $100$ selected vertices are required to perform a beam-spot fit, and in a typical fit several thousand vertices collected over a time period
 of about ten minutes are available.
The fit extracts the centroid position (\lumposx, \lumposy, \lumposz) of the beam spot (luminous centroid), the tilt angles
$\lumposx'$ and $\lumposy'$ in the $x$--$z$ and $y$--$z$ planes respectively, and the luminous sizes 
(\lumsigx, \lumsigy, \lumsigz), which are the measured sizes of the luminous region with the vertex resolution deconvoluted from the measurements.

In the transverse plane the width of the distribution of primary vertices
is the convolution of the vertex resolution with the width of the
luminous region. This is modelled by the transverse covariance matrix
\begin{eqnarray}
  V_{i}   = V^{\textrm B} +  k^2 \;  V^\mathrm{V}_{i} \; ,
\end{eqnarray}
where $V^\mathrm{B}$ describes the transverse beam-spot size and allows for a rotation of the
luminous-region ellipsoid in the transverse plane in case of non-circular beams. The transverse vertex 
resolution $V^\mathrm{V}_{i}$ estimated by the vertex fit for each primary vertex $i$ is scaled by a parameter $k$ determined 
by the beam-spot fit in order to account for any differences between fitted and
expected vertex resolutions. The parameter $k$ is expected to be close to unity as long as 
the vertex fitter provides good estimates of the vertex position uncertainty, the contamination from secondary vertices among the 
primary vertex candidates used in the beam-spot fit is small, and the
Gaussian fit model provides an adequate description of the beam-spot shape.
During 2012, the average value of $k$ was $1.16$. No vertex resolution correction and no error scaling is applied in the longitudinal 
direction because the longitudinal beam-spot size of about $50$~mm is much larger than 
the typical $z$ resolution of $35$~\um for the vertices selected for the beam-spot fit.

The beam-spot fit assumes a Gaussian shape in $x$, $y$ and $z$ and the corresponding
probability density function (PDF)
is maximised using the {\normalfont \scshape Minuit}  \cite{Minuit} minimisation package after an 
iterative procedure removes a small number of outliers incompatible with the fit. The effect of this outlier removal on the fitted beam-spot parameters is 
negligible  but brings the error scaling factor $k$ closer to $1$.

As an example of the beam-spot fit, Fig.~\ref{fig:beamspot-run215456-fitman-y} shows the distribution of primary vertices selected as input to the beam-spot fit (before outlier removal), together with the projection of the fit result. The fitted beam spot, i.e.~the distribution of primary vertices after unfolding of the vertex position resolution, is also shown. The impact of the vertex position resolution is clearly seen in the transverse direction, whereas in the longitudinal ($z$) direction the vertex resolution is negligible compared to the beam spot and therefore fitted beam spot and fit projection are identical.
\subsection{\label{BeamSpotStability}Beam-spot stability}
\begin{figure}[tp]
\centering
\subfloat[]{\includegraphics[width=0.5\textwidth]{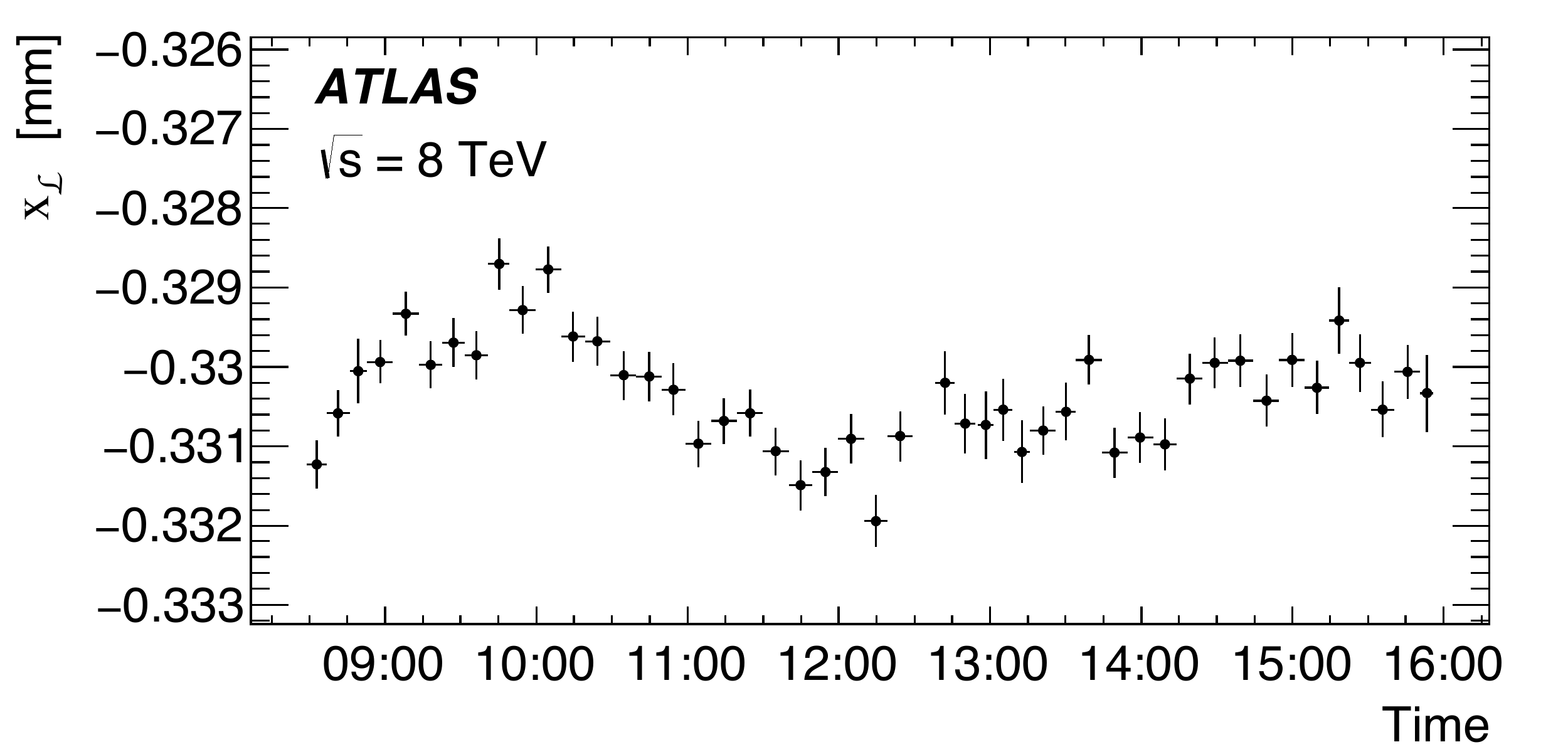}}
\subfloat[]{\includegraphics[width=0.5\textwidth]{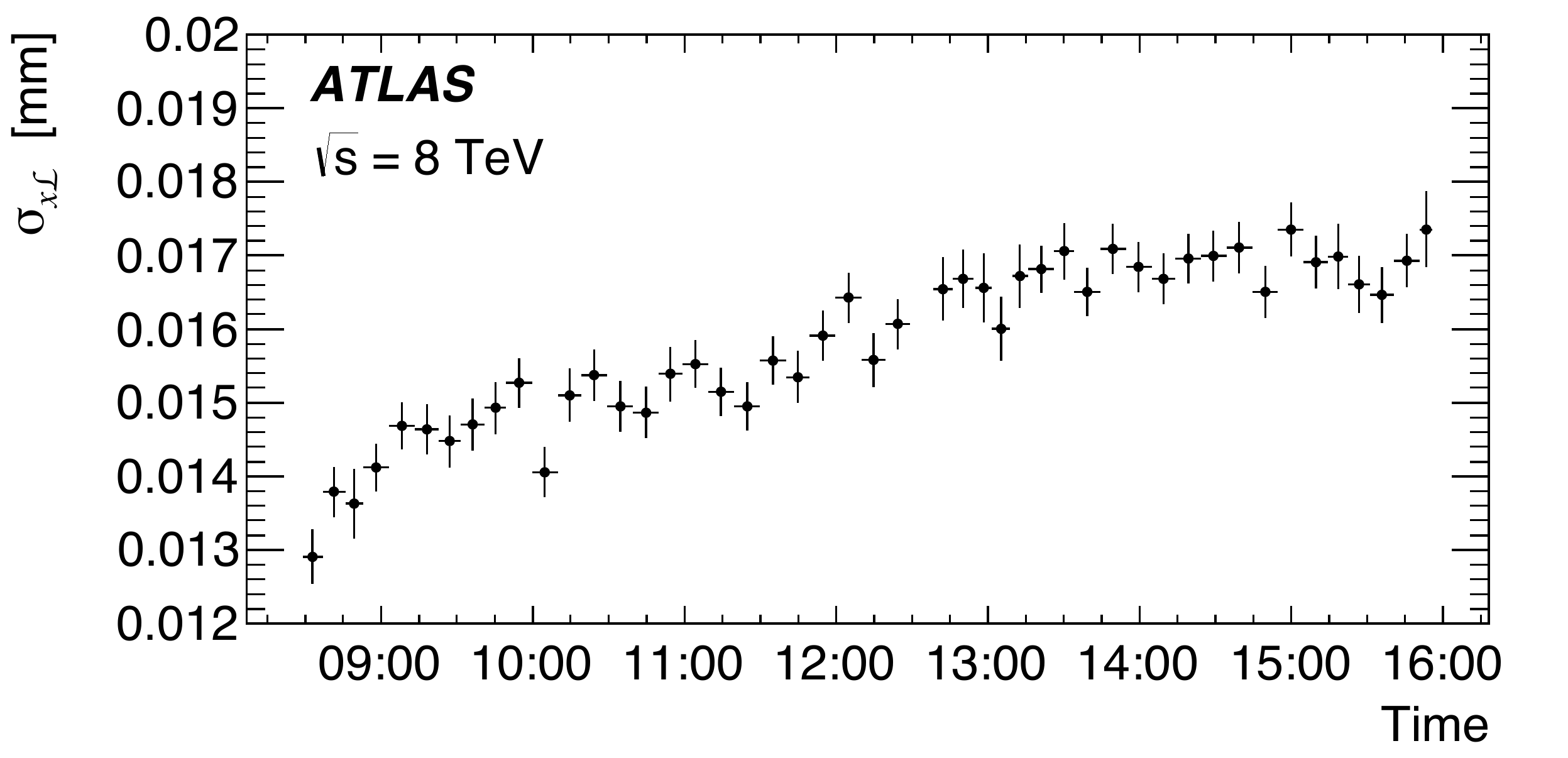}}\\
\subfloat[]{\includegraphics[width=0.5\textwidth]{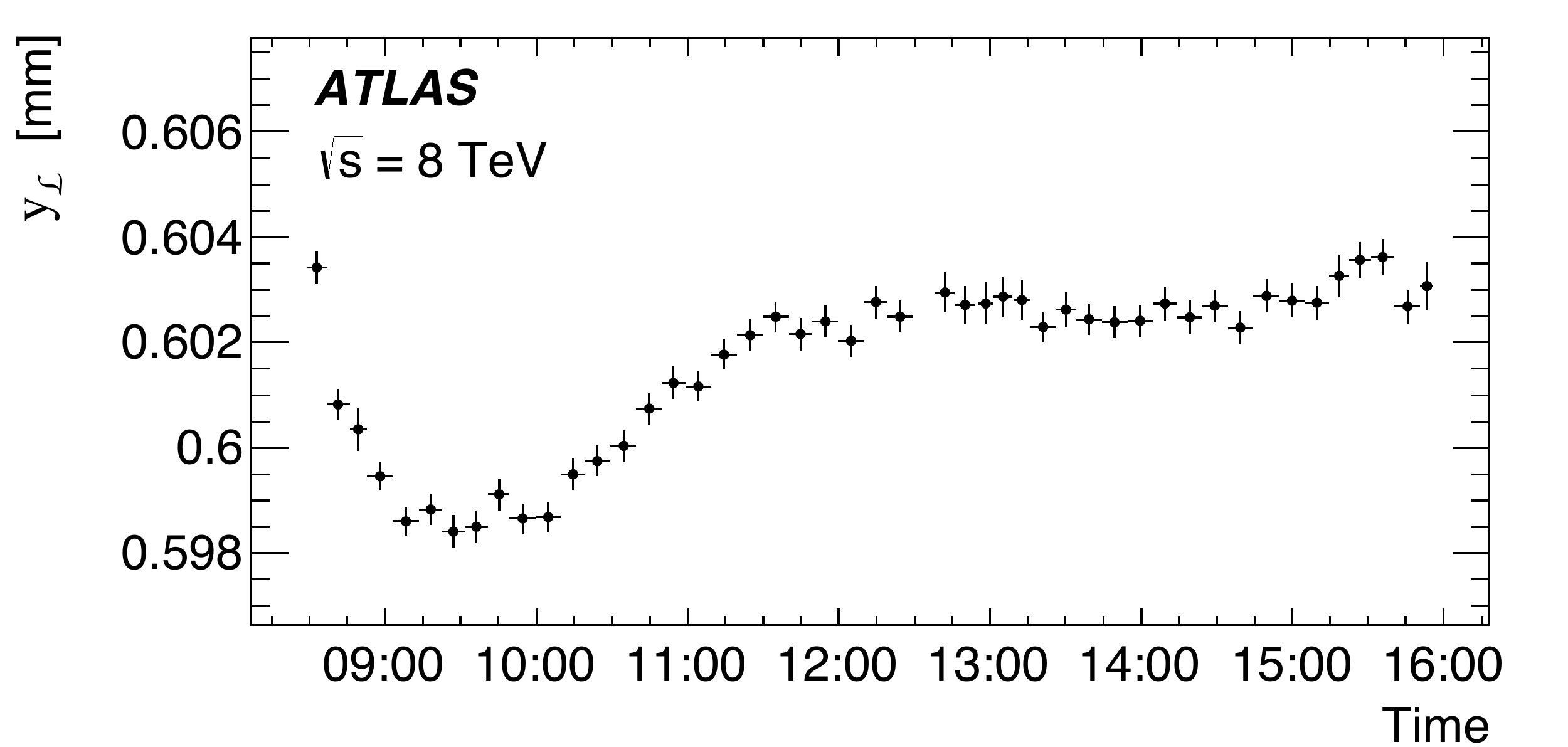}}
\subfloat[]{\includegraphics[width=0.5\textwidth]{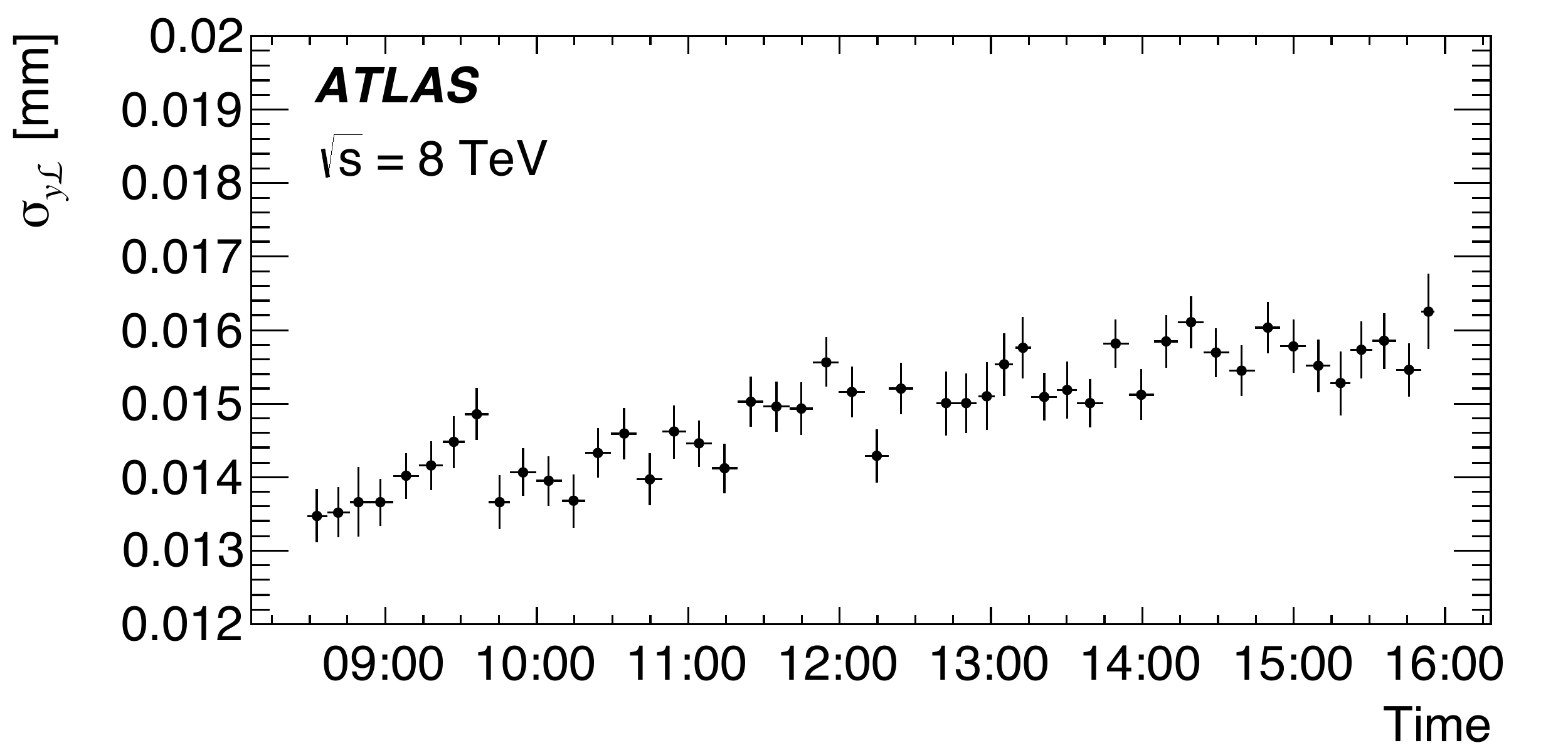}}\\
\subfloat[]{\includegraphics[width=0.5\textwidth]{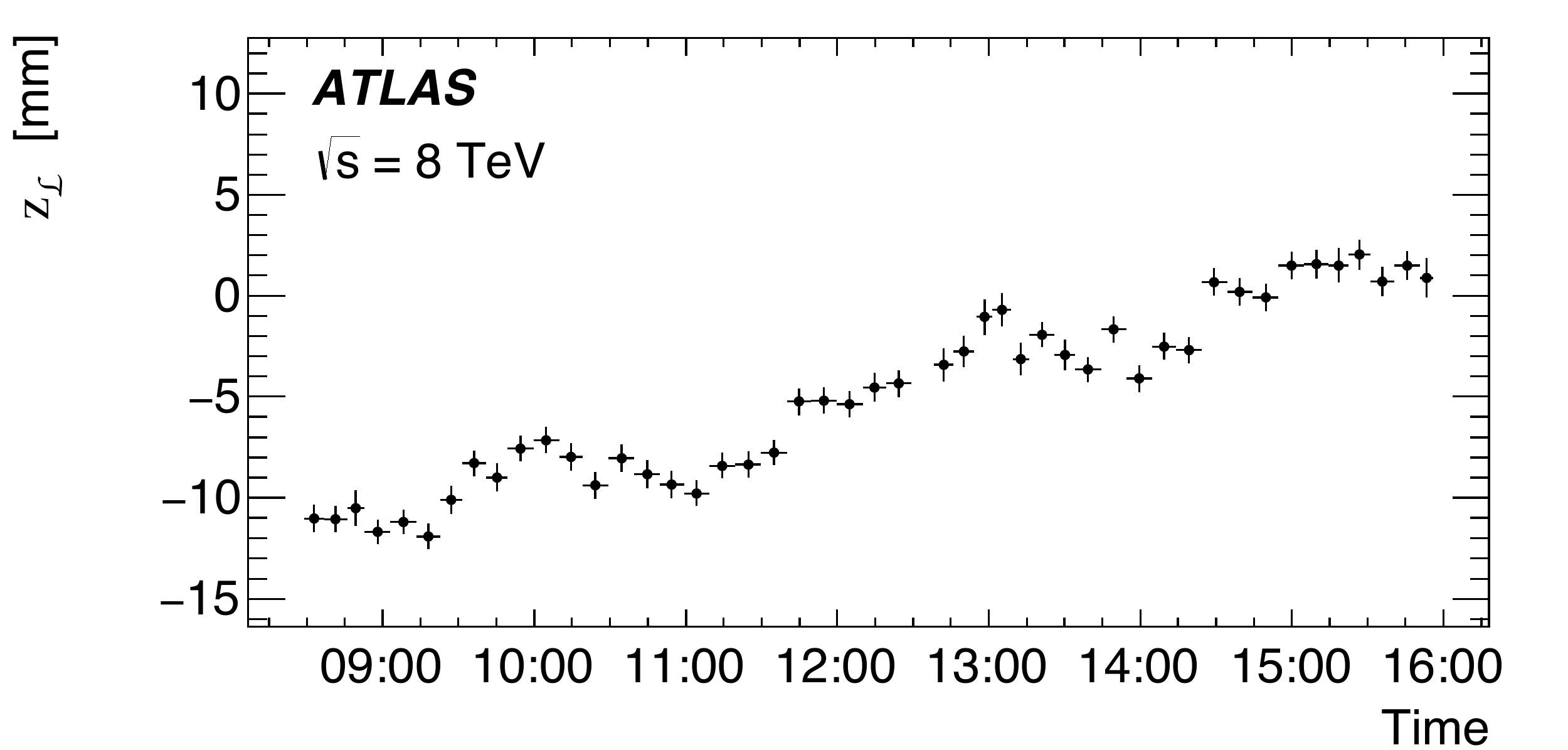}}
\subfloat[]{\includegraphics[width=0.5\textwidth]{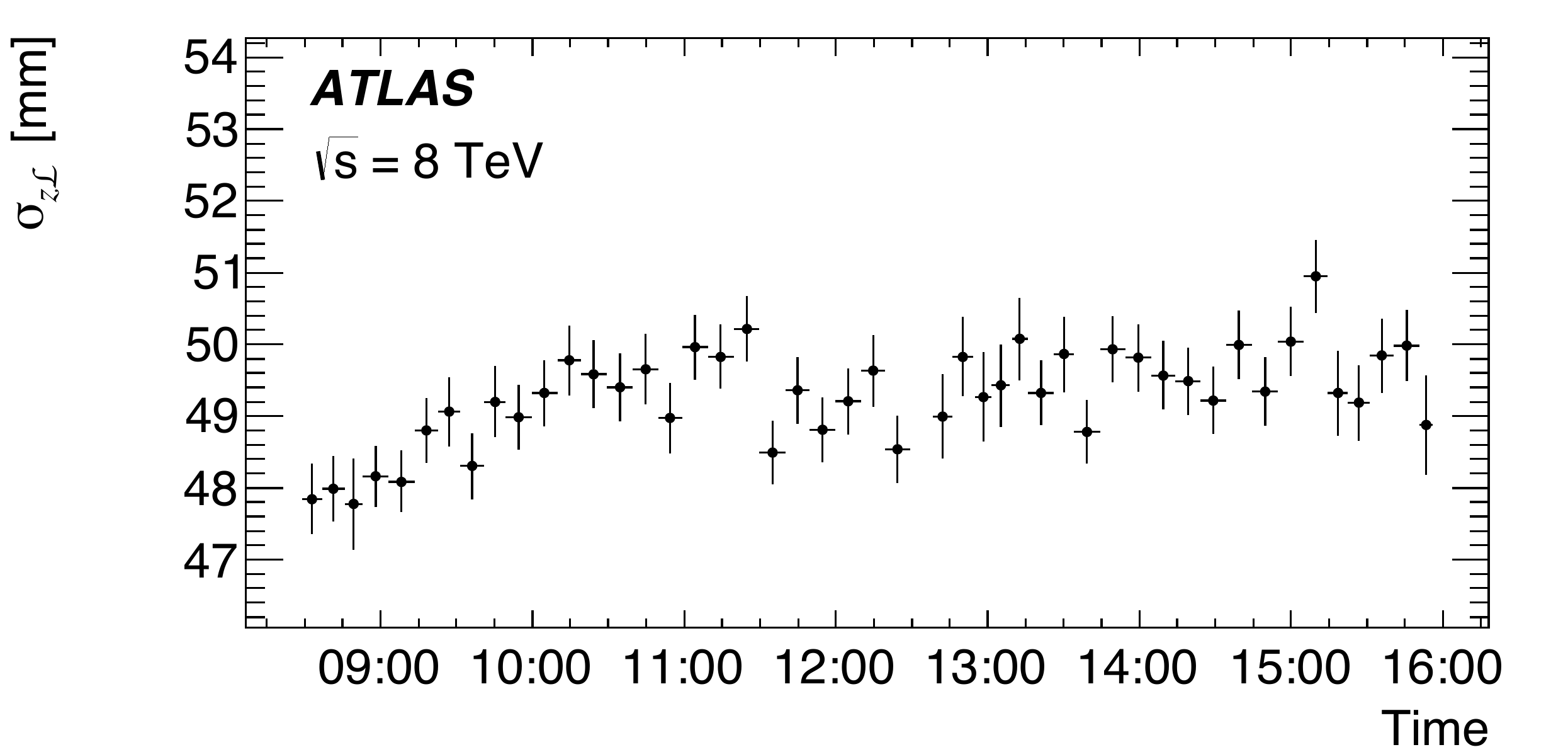}}
\caption{\label{fig:beamspot-run215456}Position ((a) ${\ensuremath{{x}_{\mathcal{L}}}\xspace}$, (c) ${\ensuremath{{y}_{\mathcal{L}}}\xspace}$, (e) ${\ensuremath{{z}_{\mathcal{L}}}\xspace}$) and size ((b) $\lumsigx$, (d) $\lumsigy$, (f) $\lumsigz$)
 of the luminous region in ATLAS  during a typical fill at $\sqrt{s} = 8$~TeV. The transverse sizes are corrected for the transverse vertex resolution.}
\end{figure}

The evolution of the beam-spot position and size as a function of time during a typical LHC fill is shown in Fig.~\ref{fig:beamspot-run215456}. The coordinates of the beam-spot position are given with respect to the ATLAS coordinate system. The precise origin location and the orientation of the ATLAS coordinate system is defined through the detector alignment procedure. The origin was chosen to be at the nominal interaction point with a $z$-axis along the beam direction, ensuring that the coordinates of the beam-spot centroid position are close to zero. In the early Run 1 data, a tilt angle of $\lumtiltx \approx 500\, \mu$rad was observed. In 2011 the ATLAS coordinate system was rotated in order to align the coordinate system more precisely with the beam line.

The downward movement of the beam-spot position during the first 40 minutes of the run followed by a gradual rise as seen in Fig.~\ref{fig:beamspot-run215456}(c) is typical and is attributed to movement of the pixel detector after powering up from standby. The increase in transverse size during the fill (Figs.~\ref{fig:beamspot-run215456}(b) and \ref{fig:beamspot-run215456}(d)) is expected from the transverse-emittance growth of the beams.
The magnitude of the changes in longitudinal beam-spot position (Fig.~\ref{fig:beamspot-run215456}(e)) is typical and is understood to be due to relative RF phase drift.
The increase in longitudinal size (Fig.~\ref{fig:beamspot-run215456}(f)) reflects bunch lengthening in the beams during the fill.
The tilt angles \lumtiltx and \lumtilty (not shown in Fig.~\ref{fig:beamspot-run215456}) were stable at the level of about $10\,\mu$rad.

\begin{figure}[tp]
\centering
\subfloat[]{\includegraphics[width=0.8\textwidth]{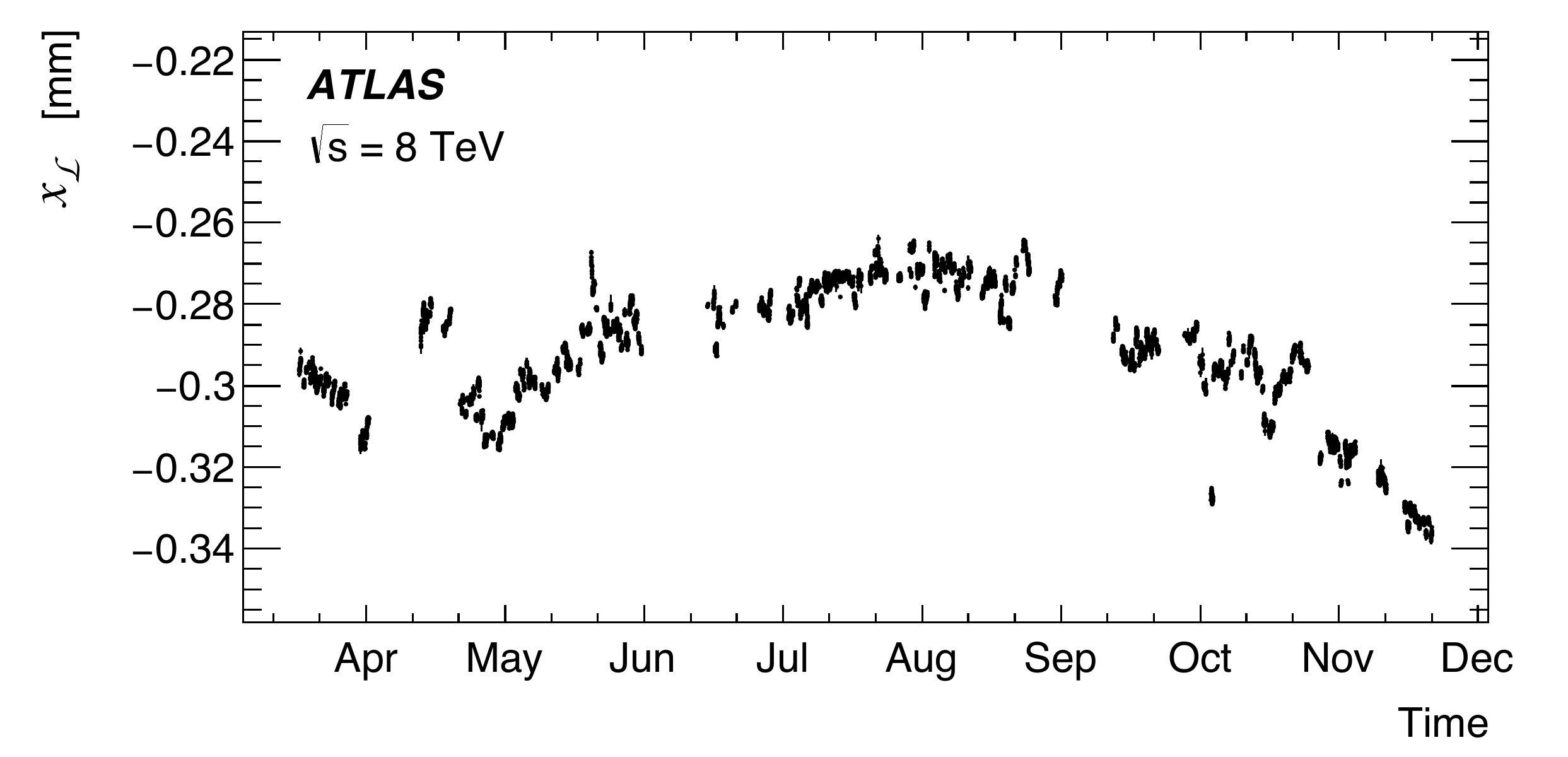}}\\
\subfloat[]{\includegraphics[width=0.8\textwidth]{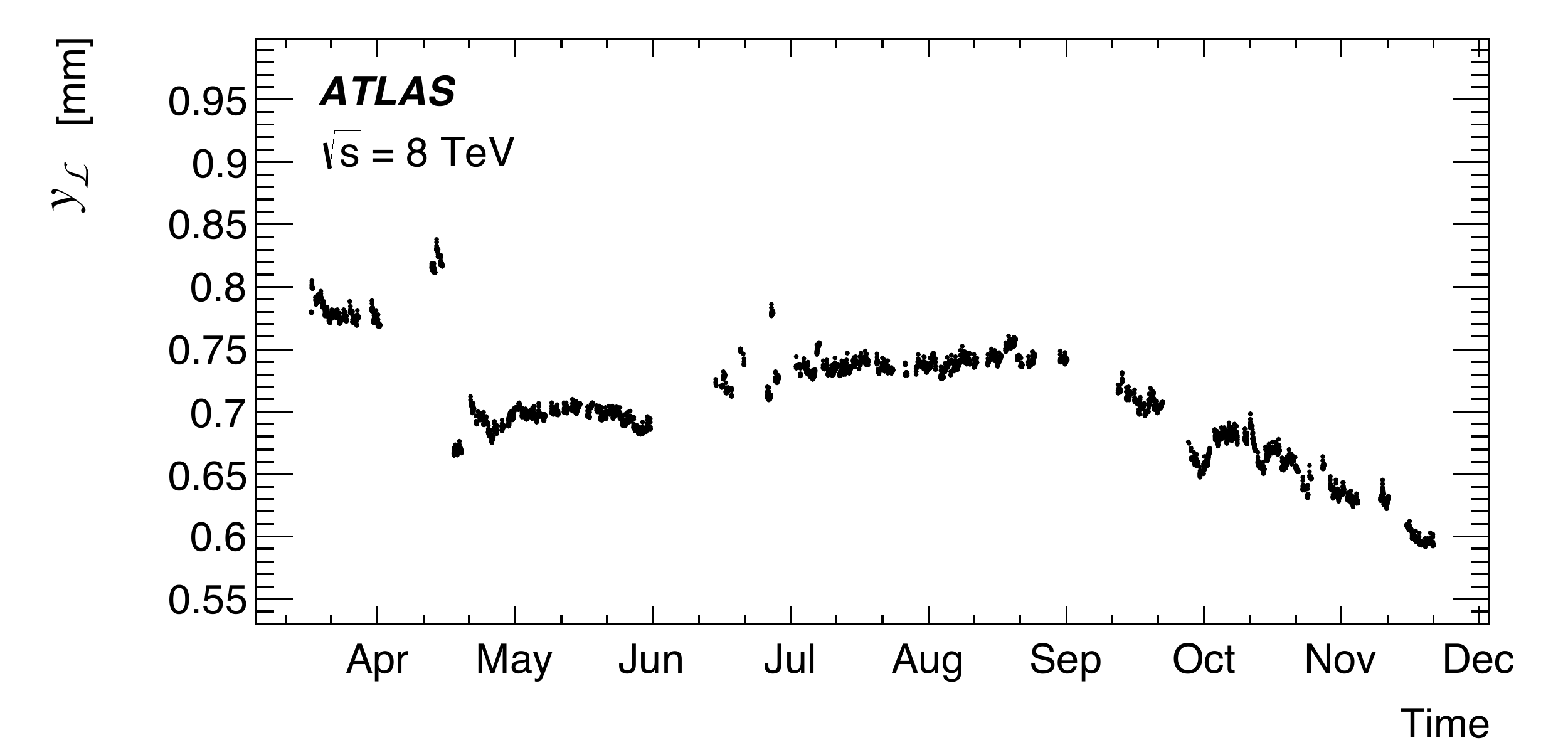}}\\
\subfloat[]{\includegraphics[width=0.8\textwidth]{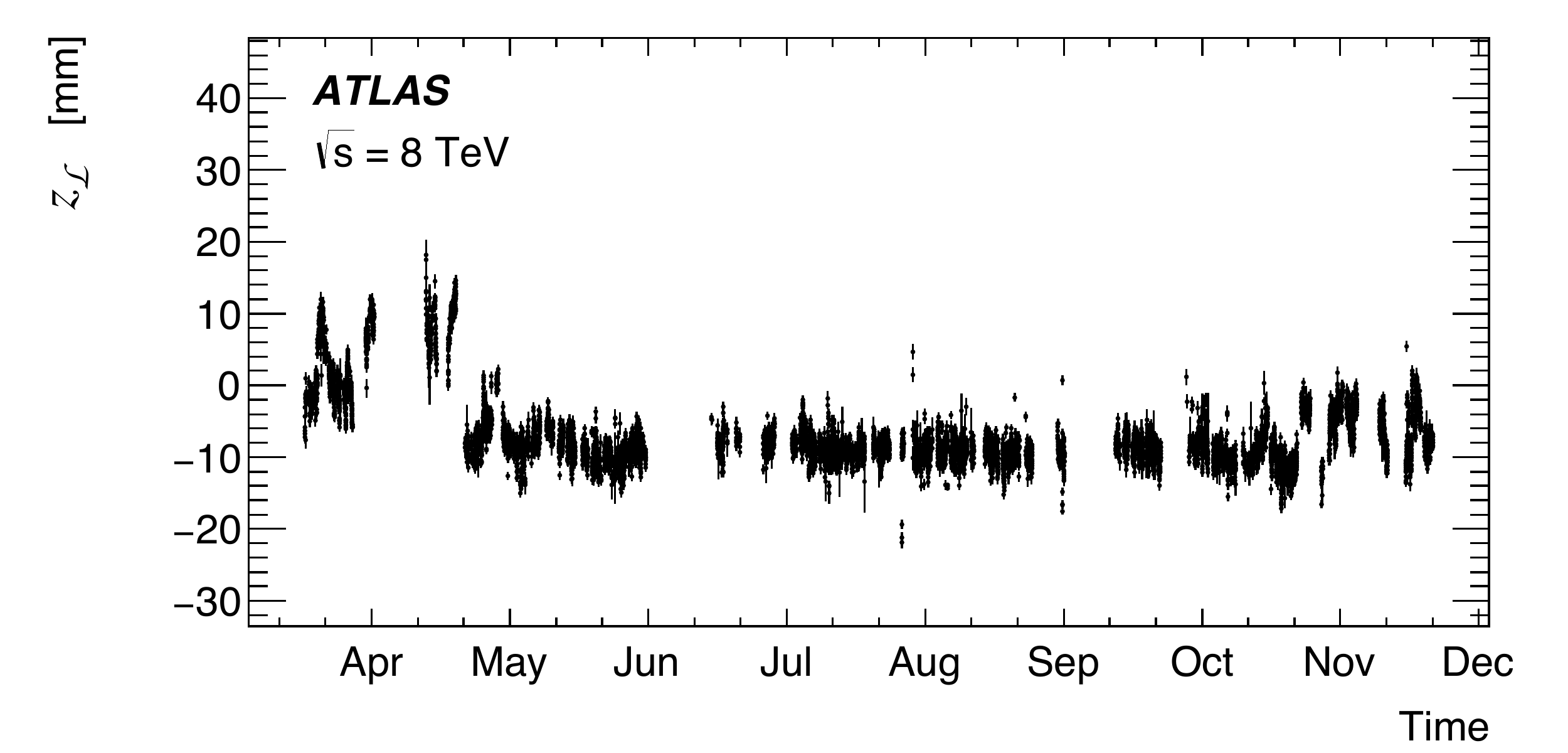}}\\
\caption{\label{fig:beamspot-2012}Position of the luminous region in ATLAS over the course of $pp$ running in 2012 at $\sqrt{s}=8$~TeV. The data points are the result of a maximum likelihood fit to the spatial distribution of primary vertices collected over ten minutes. Errors are statistical only.}
\end{figure}

The long-term evolution of the beam-spot position during 2012 is shown in Fig.~\ref{fig:beamspot-2012}. The large vertical movement at the beginning of May visible in 
Fig.~\ref{fig:beamspot-2012}(b) was associated with movement of the ID.
 Apart from variations in each fill due to transverse-emittance growth and bunch lengthening, both the transverse and longitudinal beam-spot sizes remained unchanged during 2012.

Table~\ref{tab:beamspot-averages} summarises the beam-spot position and size in 2010, 2011 and 2012 for $pp$ collision data. 
\begin{table}[tp]
\begin{center}
\begin{tabular}{lcccccccc}
\hline \hline
Year  & $\beta^*$ [m] & \lumposx [mm] & \lumposy [mm] & \lumposz [mm] & \lumsigx [$\mu$m] & \lumsigy [$\mu$m] & \lumsigz [mm]  \\ \hline
2010 &$11$&$-0.347\pm     0.015$&$0.611 \pm     0.018$&$\phantom{-}0.9 \pm       3.5$&$49 \pm          8$&$60 \pm        12$&$29 \pm         3$\\
2010 &$2$& $-0.364\pm      0.031$&$0.647 \pm     0.009$&$-1.2 \pm        2.2$&$30 \pm          5$&$39 \pm    12$&$36  \pm          3$\\
2010 &$3.5$&$\phantom{-}0.081\pm  0.033$&$1.099 \pm      0.029$&$-3.0\pm        4.6$&$41 \pm          4$&$        44 \pm         6$&$       63 \pm          3$\\
2011 &$1.5$&$-0.050\pm      0.018$&$  1.059 \pm     0.051$&$-6.2 \pm        3.8$&$         26 \pm          2$&$   24 \pm          2$&$      57 \pm         3$\\
2011 &$1.0$&$-0.052 \pm      0.009$&$      1.067 \pm     0.013$&$-6.7 \pm       1.5$&$        21 \pm          2$&$         20 \pm          1$&$         56 \pm          3$\\
2012 &$0.6$&$-0.291 \pm      0.016$&$      0.705 \pm     0.046$&$ -7.3\pm        4.7$&$         15 \pm          2$&$         15 \pm         1 $&$         48 \pm          2$\\
\hline \hline
\end{tabular}
\end{center}
\caption{\label{tab:beamspot-averages}Average beam-spot position and size for $pp$ collision data in 2010, 2011 and 2012 for different \bst settings. The errors given in the table are the RMS spread of the parameters during the corresponding time period.}
\end{table}
Data from special runs is excluded. As expected, the average transverse beam-spot size scales approximately with $\sqrt{\bst/\Ebeam}$, but is also influenced by changes in the normalised emittance and by the amount of emittance growth during the fills. In 2010 and 2011 
the centre-of-mass energy was $7$~TeV. In 2012 it increased to 8 TeV. During this time the crossing angle \xang was increased from zero at the start of 2010 to $290\,\mu$rad in 2012.

The measured transverse size of the beam spot at the start of a run is in good agreement with the values expected 
from the LHC machine parameters at the start of a fill (Table~\ref{tab:lhcparam}). This can be seen in Fig.~\ref{fig:beamspot-run215456}. The average  transverse size in 2012 shown in Table~\ref{tab:beamspot-averages} 
($15\; \mu{\textrm m}$) is larger than the expected size of $13$~\um from Table~\ref{tab:lhcparam} due to emittance growth during the run. Within the relatively large uncertainty expected for the $4\sigma$ bunch length $T_z$ due to instrumental and non-Gaussian effects, the longitudinal beam-spot size is in reasonable agreement with expectations from the LHC parameters shown in Table~\ref{tab:lhcparam}.

%% file: HardScatter.tex
This section describes how both the reconstruction and identification efficiencies 
of hard-scatter primary vertices are evaluated using simulation. The impact of pile-up 
tracks and vertices on the performance is also estimated.
A classification scheme based on MC generator-level information, denoted hereafter as truth-level information, is used to 
describe the level of pile-up contamination in reconstructed vertices from hard-scatter processes.

\subsection{\label{sec::Truth}Monte Carlo truth matching and classification of vertices}
To study the performance of primary vertex reconstruction using MC
 simulation, a truth-matching algorithm has been developed, based on the generator-level particles 
 associated to tracks contributing to reconstructed vertices.    The procedure first classifies each reconstructed 
track used in a vertex fit.   The compatibility criteria for track truth-matching are 
based on the fraction of hits used to reconstruct the track in each sub-detector that 
were produced by the generated primary particle as discussed in Ref.~\cite{atlasConf:pileup}.  Each reconstructed track is classified as one of the following: 
\begin{itemize}
\item{A track matched to a hard-scatter interaction.}
\item{A track matched to a pile-up interaction.}
\item{An unmatched track.  Such a tracks are considered random combinations of detector hits falsely identified as 
charged particle trajectories.  These are referred to as fake tracks.}
\end{itemize}

Tracks are matched to their primary generating interaction, i.e.\ tracks
from secondary interactions are traced back to a hard-scatter or pile-up interaction.
Based on the above classification, reconstructed  vertices can be categorised.
For each vertex, the sum of the weights assigned to all contributing tracks is
normalised to unity. The fractional weights of individual tracks in each vertex are calculated.
Vertices can then be put into one of the following exclusive categories:
\begin{itemize}
\item \label{itm:matched}
Matched vertex: Tracks identified as coming from the same generated interaction contribute at least 70\% of the total  weight
of tracks fitted to the reconstructed vertex.
\item \label{itm:merged} Merged vertex: No single generated interaction contributes more than 70\%
of track weight to the reconstructed vertex. Two or more generated interactions contribute to the reconstructed vertex.
\item \label{itm:split} Split vertex: The generated interaction with the largest contribution to the reconstructed 
vertex is also the largest contributor to one or more other reconstructed vertices.  In this case, the reconstructed vertex with the 
highest fraction of track $\Sigma p_{\textrm T}^{2}$  is categorised as matched or merged and the vertex or vertices with lower $\Sigma p_{\textrm T}^{2}$   are categorised as split.
\item \label{itm:fake} Fake vertex:  Fake tracks contribute more weight to the reconstructed vertex than any generated interaction.
\end{itemize}

This classification schema allows detailed studies of vertex reconstruction in a pile-up environment.
The effects of splitting and merging of primary vertices as well as the influence of these effects
on the vertex reconstruction efficiency and primary vertex resolution can be studied. This schema
also allows the reconstructed vertices to be associated either with the primary hard-scatter $pp$ collision  or
with pile-up interactions.

When studying the hard-scatter $pp$ collisions, the reconstructed events are classified based on the following
mutually exclusive definitions:
\begin{itemize}
\item {Clean}: The event contains one matched vertex corresponding to the hard-scatter interaction. 
The hard-scatter interaction does not contribute more than $50\%$  of the accumulated track weight to 
any other vertex.
\item {Low pile-up contamination}: The event contains one and only one merged vertex where the 
hard-scatter interaction contributes more than $50\%$ of the accumulated track weight.
\item  {High pile-up contamination}: The event does not contain any vertex where the hard-scatter 
interaction contributes more than $50\%$ of the accumulated track weight. It does however contain at 
least one merged vertex in which the hard-scatter interaction contributes between $1\%$ and $50\%$ of the accumulated track weight.
\item  {Split}: The event contains at least two merged vertices in which the hard-scatter 
interaction contributes more than 50\% of the accumulated track weight.
\item  {Inefficient}: The event does not contain any vertex where the hard-scatter interaction contributes more than 1\% of the accumulated track weight. 
\end{itemize}
In the current analysis, all categories except ``Inefficient" are considered as successful in reconstructing the  hard-scatter
primary vertex. All of these categories thus contribute to the calculation of total vertex reconstruction efficiency.   

\subsection{\label{sec::HSefficiency}Vertex reconstruction and selection efficiency for hard-scatter interactions}
\begin{figure}[tp]
\centering
{\includegraphics[width=0.45\textwidth]{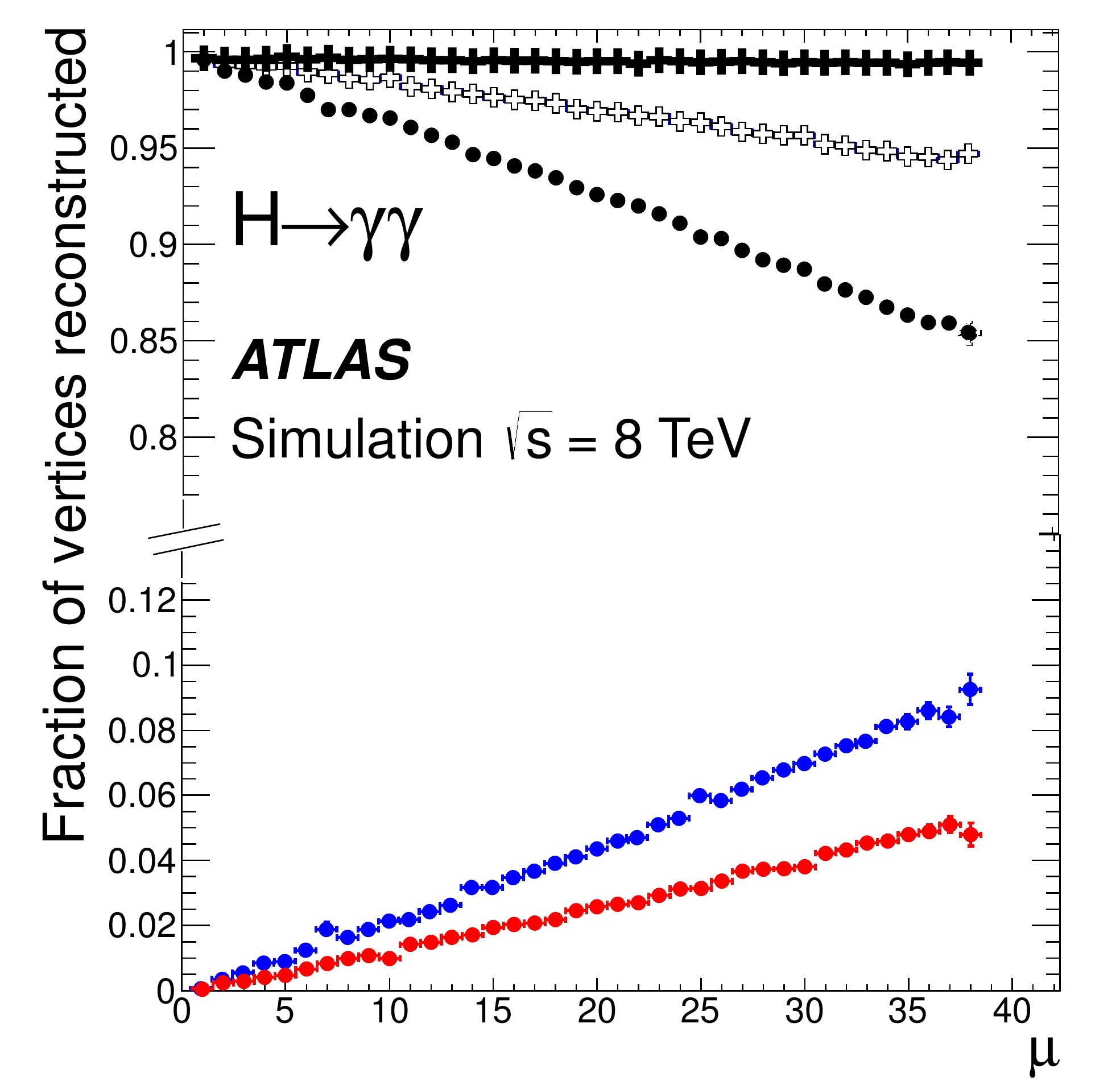}}
{\includegraphics[width=0.45\textwidth]{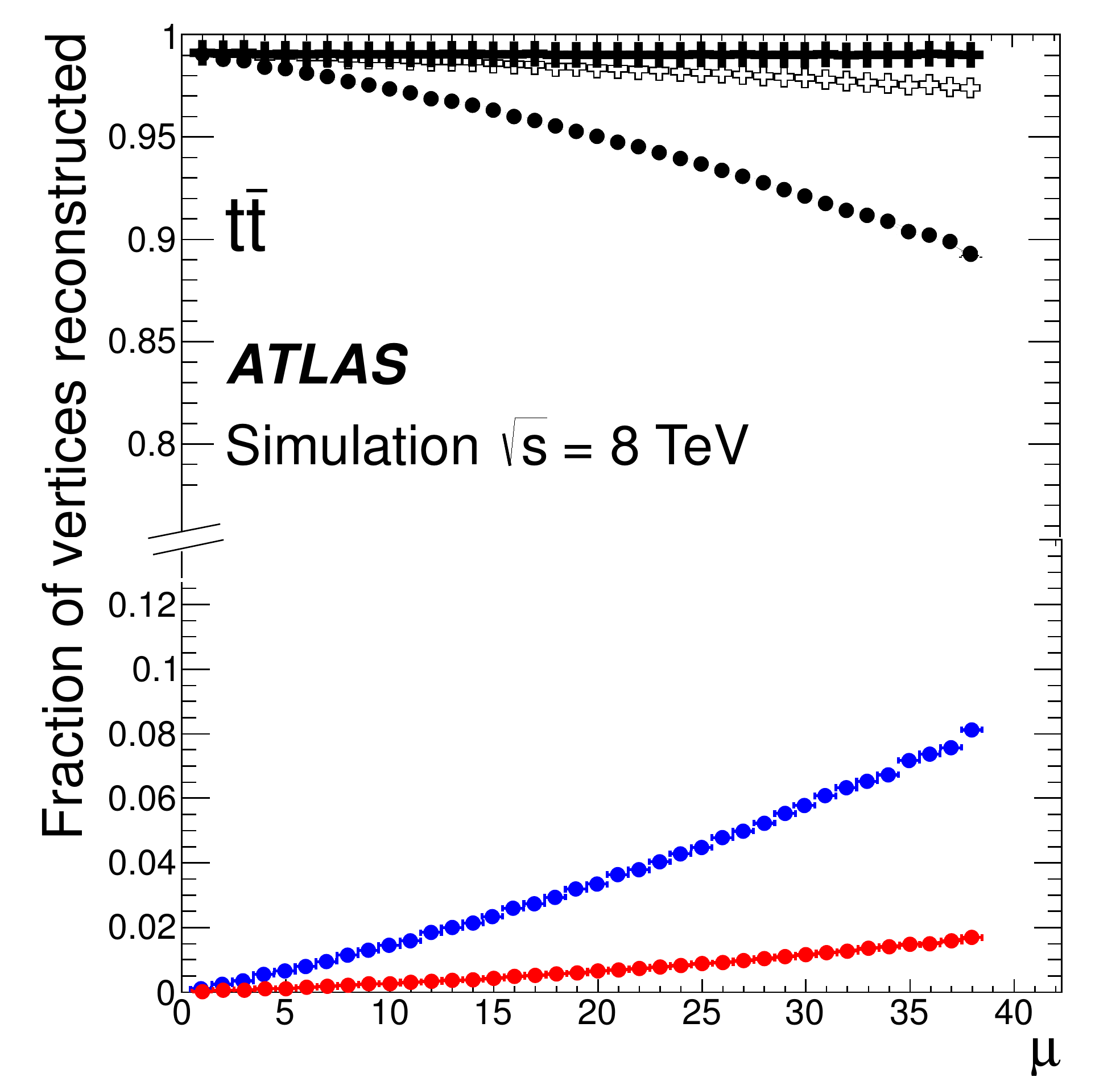}}\\
{\includegraphics[width=0.45\textwidth]{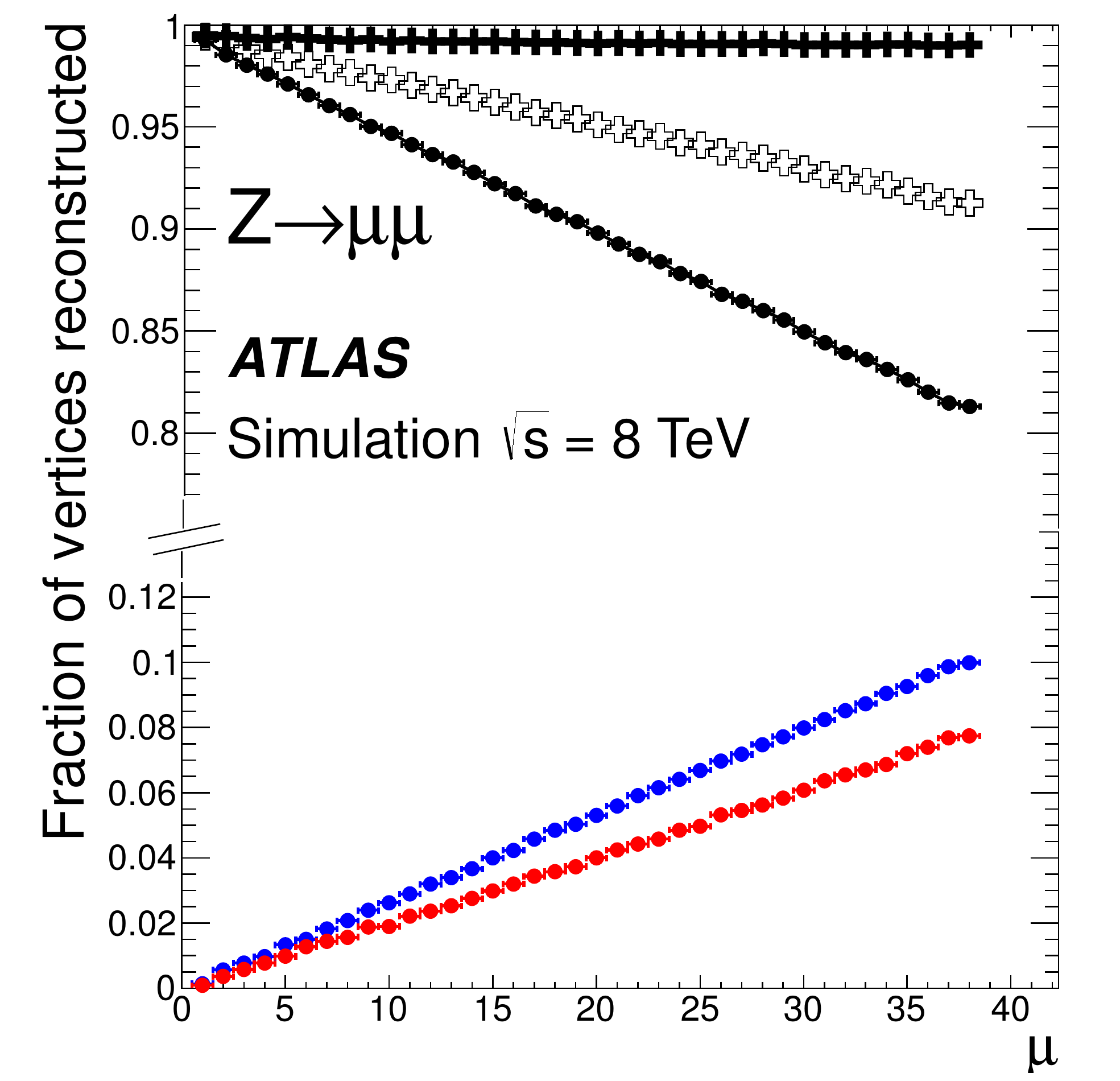}}
{\includegraphics[width=0.45\textwidth]{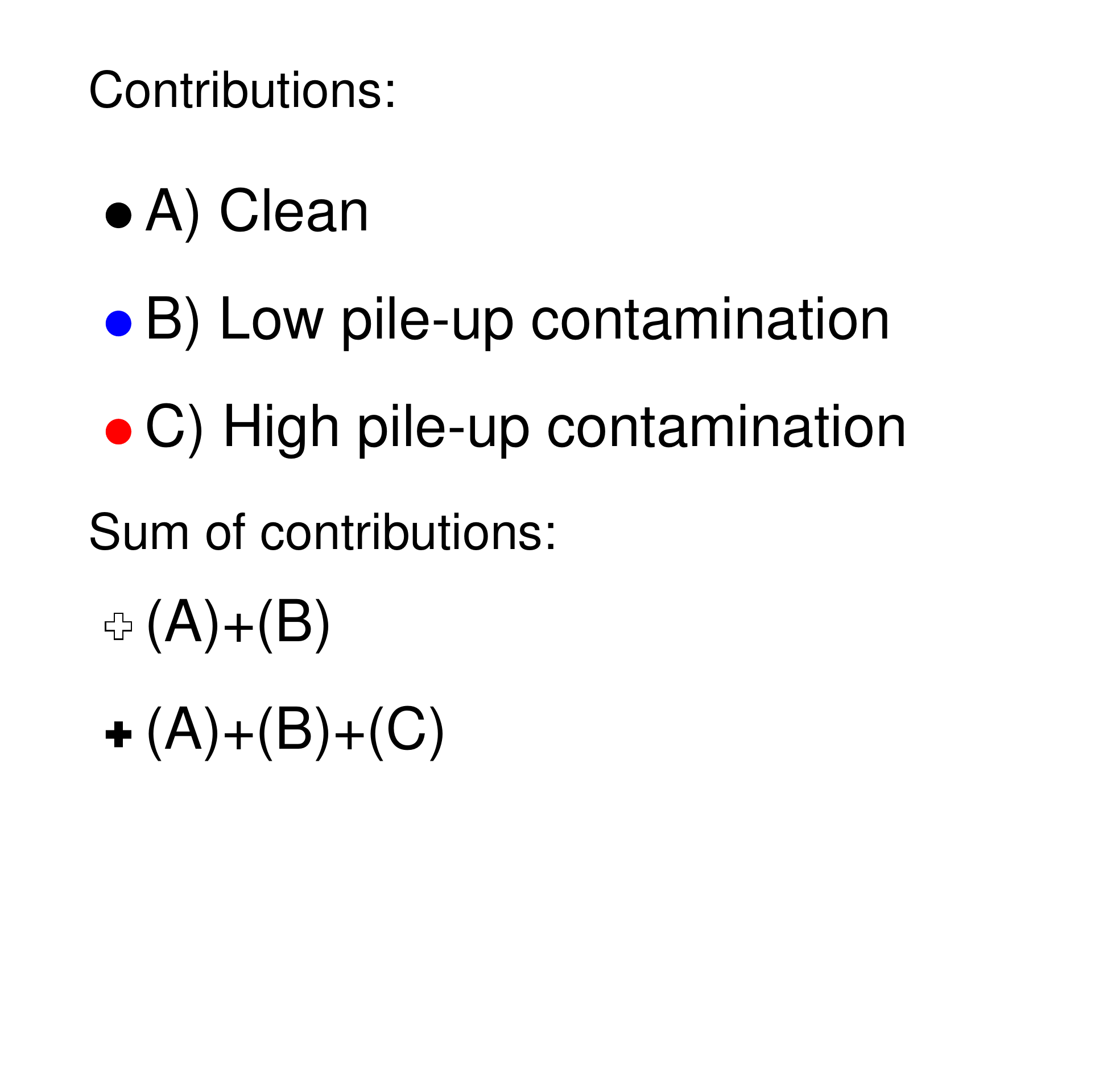}} \\
\caption{\label{fig:hs_cat_z} 
Contributions to the predicted primary vertex reconstruction efficiency 
as a function of the average number of interactions per bunch crossing, $\mu$.
The mutually exclusive categories of events are defined in Section~\ref{sec::Truth}. 
The black circles show the contribution to the efficiency from events categorised as clean, and
the blue and red circles show the contributions from events with low and high pile-up contamination respectively.
The open crosses show the sum of the contributions from events that are clean and those with low pile-up contamination;
the filled crosses show the sum of the contributions from all categories and represent the overall efficiency.
The hard-scatter processes considered are Higgs-boson decay into $\gamma \gamma$, $t\bar{t}$ production with a lepton in the decay, and $Z$-boson decay into $\mu \mu$. 
}
\end{figure}

The efficiency  to reconstruct and also to correctly identify the hard-scatter primary vertex 
is used to quantify the impact of pile-up contamination.
Assuming that the  hard-scatter primary vertex produces reconstructed tracks, 
the efficiency of hard-scatter primary vertex reconstruction is predicted to be larger than $99\%$.  
This includes interactions with low or high pile-up contamination, and split event categories as defined in Section~\ref{sec::Truth}.
The corresponding contributions to the reconstruction efficiencies as a function of simulated $\mu$ are shown in
Fig.~\ref{fig:hs_cat_z} for the processes  $Z\to\mu\mu$, $H\to \gamma \gamma$ and 
$t\bar{t}\rightarrow l+X$ ($t\bar{t}$ decays that include a lepton).

The fraction of events with low and high
 pile-up contamination increases with growing  $\mu$, while the fraction of clean  events 
decreases with $\mu$.  The fraction of  events containing split vertices remains negligible for all $\mu$.
For $\mu = 38$ the fraction of high pile-up contamination 
vertices is $8\%$ for $Z\to\mu\mu$ events, $5\%$ for $H\to \gamma \gamma$ events,
and $2\%$ for $t\bar{t}$ events.

The effect of pile-up contamination on the reconstruction efficiency for the 
hard-scatter primary vertex clearly depends on the nature of the physics process under study.
The hard-scatter interactions corresponding to  $Z$-boson production  leave on average fewer 
charged particles within the detector acceptance than those corresponding to  $t\bar{t}$ production.
Hard-scatter vertices from  $Z$-boson production
can therefore be expected to be more affected by pile-up contamination than those from  $t\bar{t}$
events. Indeed, Fig.~\ref{fig:hs_cat_z} 
shows that the low and high pile-up contamination fractions are always higher for $Z\to \mu\mu$ than for $t\bar{t}$ events.

\begin{figure}[tp]                                                                  
\centering
\subfloat[]{\includegraphics[width=0.4\textwidth]{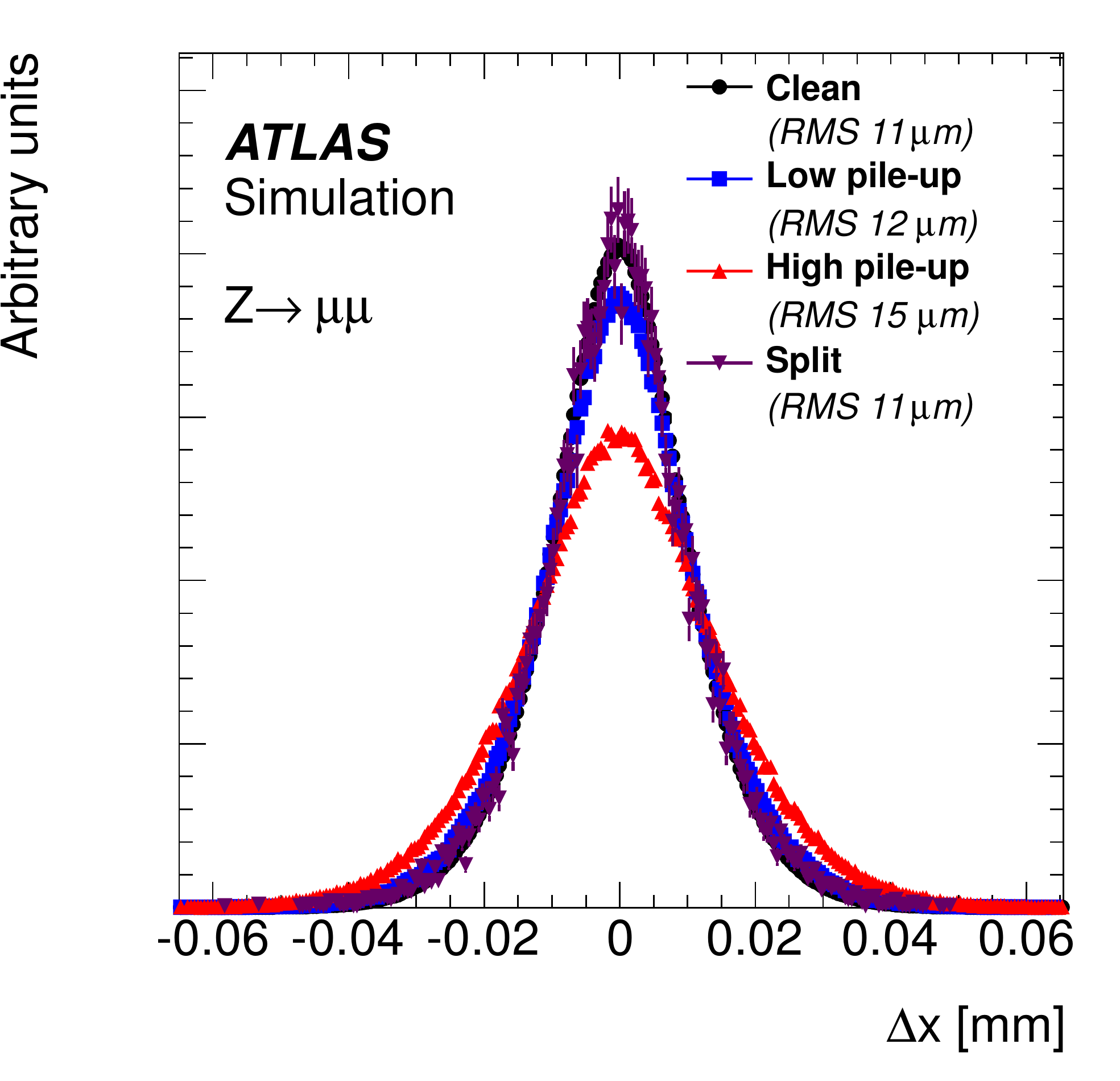}}
\subfloat[]{\includegraphics[width=0.4\textwidth]{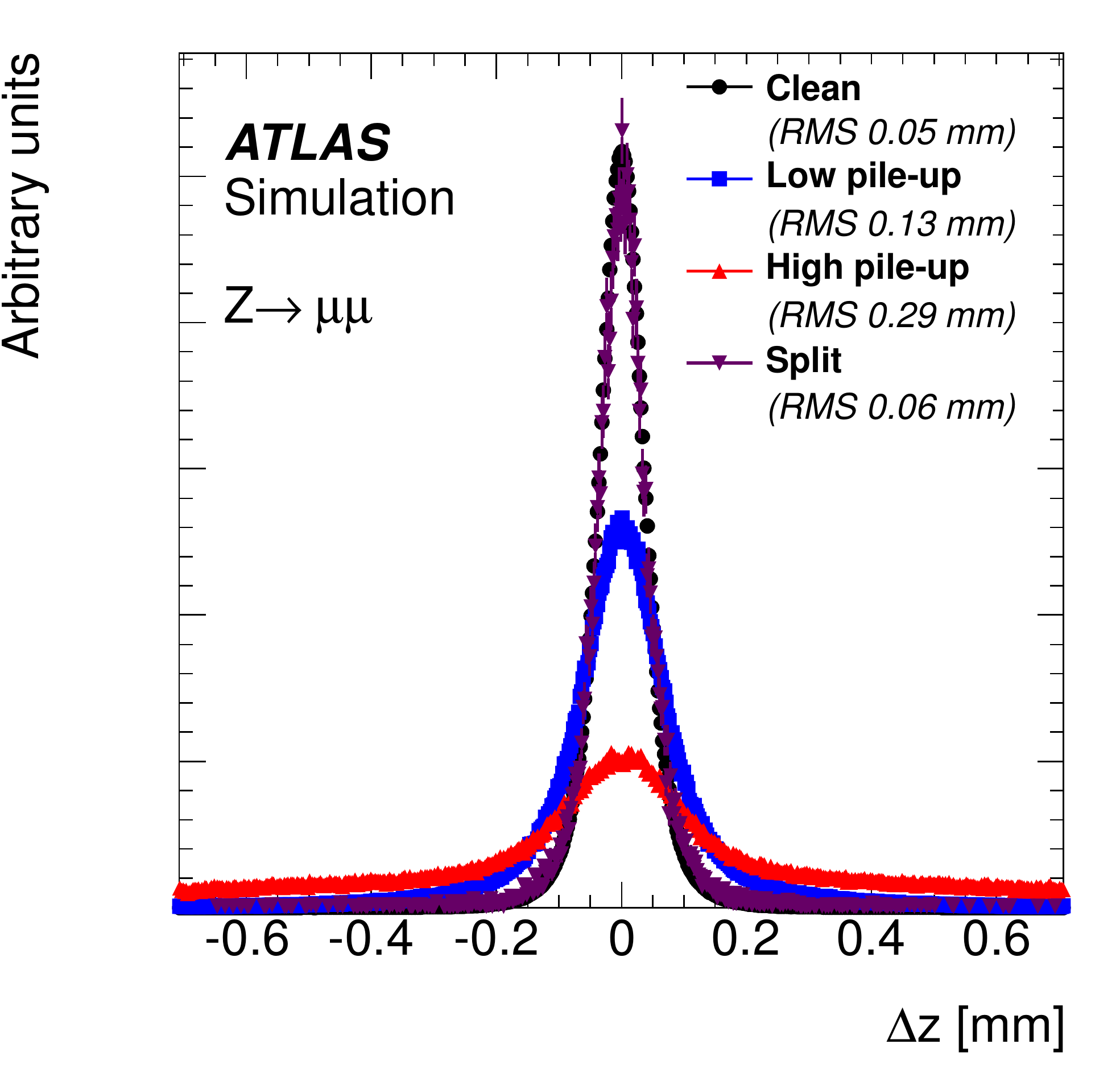}}
\caption{ \label{fig:Res_Zmumu} The residual distributions in (a) $x$ and (b) $z$   coordinates 
for reconstructed primary vertices in a sample of simulated $Z\rightarrow\mu\mu$ events for the four classes of events 
defined in Section~\ref{sec::Truth}. The distributions are normalised to the same area. The RMS values of these residuals are provided for each class.}
\end{figure}

Pile-up tracks contaminating reconstructed hard-scatter vertices lead to a degradation of 
 position resolution. Figure~\ref{fig:Res_Zmumu} shows 
the distribution of residuals of the primary vertex position in a $Z\to \mu\mu$ sample
for different classes.
The residuals are  calculated as the distance between the position of the
  hard-scatter primary vertex at generator level and its reconstructed
 position obtained from the primary vertex reconstruction as described 
in Section~\ref{sec::FitFind}. Only the  vertices matched according to the definition 
presented in Section~\ref{sec::Truth} are taken into account. The results are obtained using the MC simulation 
including detector acceptance without further selection criteria. The categories of clean reconstruction, low and 
high pile-up contamination show progressively degrading resolution.
This effect is visibly largest for the $z$-coordinate, because the transverse coordinates are constrained by the beam-spot
width.  The events categorised as containing split vertices do not suffer from a degraded resolution compared to the clean event 
category. 
 
\begin{figure}[tp]
\centering
\subfloat{\includegraphics[width=0.55\textwidth]{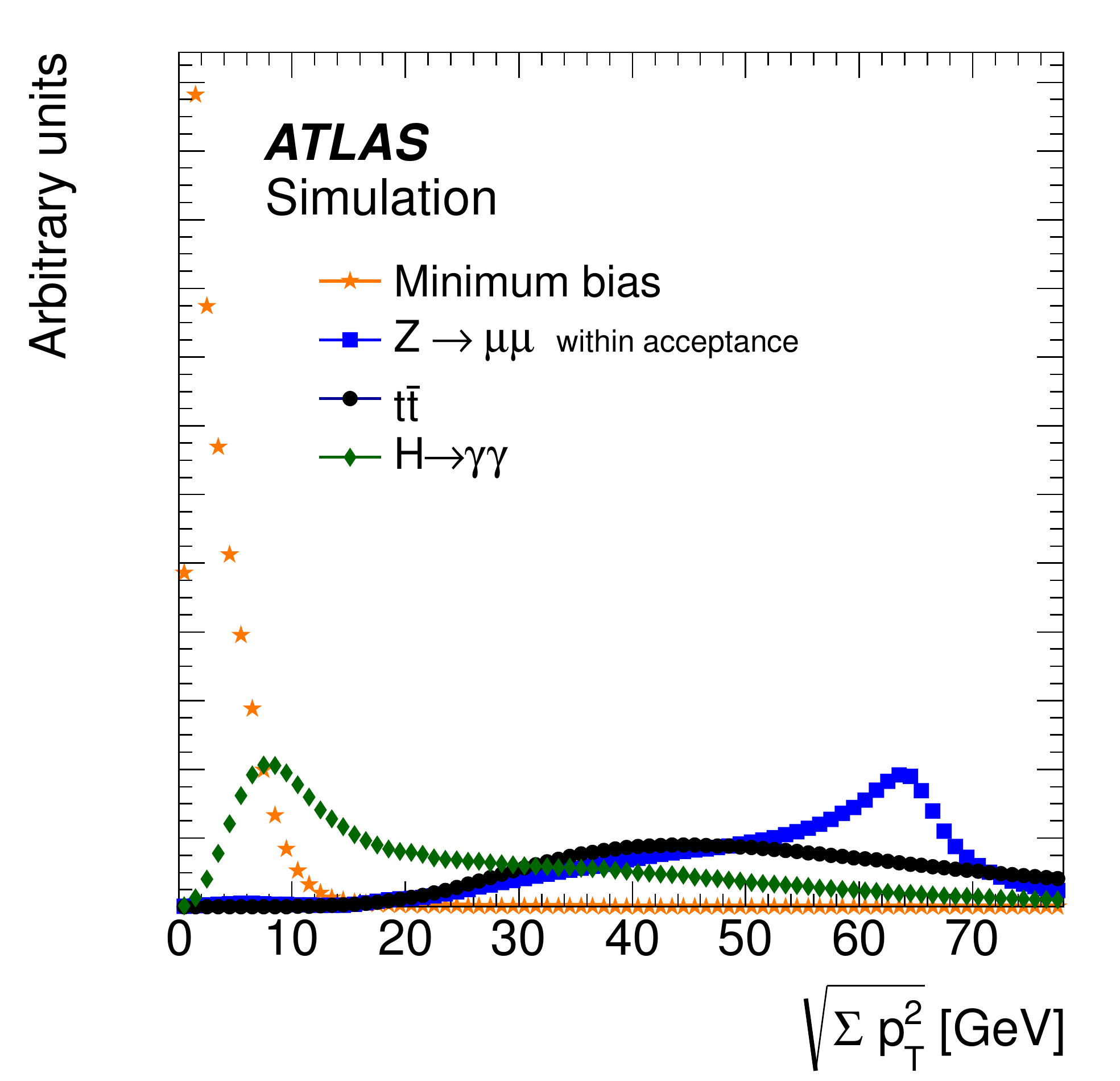} }
\caption{\label{fig:hs_mb_sumpt2} The distributions of the sum of the squared transverse momentum for
 tracks from primary vertices, shown for simulated hard-scatter processes and a minimum-bias sample. In the case of the $Z\to \mu\mu$
 process, only events  with at least two muons with $p_{\textrm T}>15$~GeV reconstructed within the ATLAS inner detector acceptance
 are shown. The $t\bar{t}$ process is filtered to select decays with leptons. The distributions are normalised to the same area.}
\end{figure}

In addition to the degradation of the spatial resolution, the presence of significant pile-up 
makes it more difficult to correctly identify the hard-scatter primary vertex among the
many pile-up vertices reconstructed in most bunch crossings.
   For most hard-scatter physics processes, it is effective to  identify the 
hard-scatter primary vertex  as the primary vertex with the highest sum of the squared transverse
momenta of contributing tracks:  $\sum \pt^2$. This criterion 
is based on the assumption that the charged particles produced in  hard-scatter 
interactions have on average a  harder transverse momentum spectrum than
 those produced in  pile-up collisions. The efficiency of the hard-scatter identification using this criterion depends
on the kinematics of the hard-scatter process. Distributions of 
 $\sqrt{\sum \pt^2}$ of the tracks in various hard-scatter processes 
 are shown in Fig.~\ref{fig:hs_mb_sumpt2}, including $H \to\gamma\gamma$, $Z\to \mu\mu$, and $t\bar{t}$ decays in which a filter has been applied to select decays with leptons. These are compared to a minimum-bias sample, which can be 
taken to have the same $\sqrt{\sum \pt^2}$ distribution as pile-up.

In the case of $Z\to \mu\mu$  and $t\bar{t}$, there is significant transverse momentum carried by charged particles 
even in the case of inclusive samples.  In contrast, in the case of $H \to\gamma\gamma$ events, most of the transverse 
momentum is carried by the photons from the Higgs boson decay. The remaining charged particles in the acceptance of the detector are produced 
in the underlying event and have a much softer $p_{\textrm T}$ spectrum. The efficiency to correctly select 
 the hard-scatter vertex  among many pile-up vertices by choosing  the vertex with the highest  $\sum \pt^2$ 
is thus inferior for $H\to \gamma\gamma$ decays compared to most other hard-scatter processes.  
A more
 efficient  method for choosing the primary vertex in the case of $H\to \gamma\gamma$ decay is described in Ref.~\cite{Aad:2014eha}.

\begin{figure}[tp]
\centering
\subfloat[]{\label{fig:hs_select_a}\includegraphics[width=0.45\textwidth]{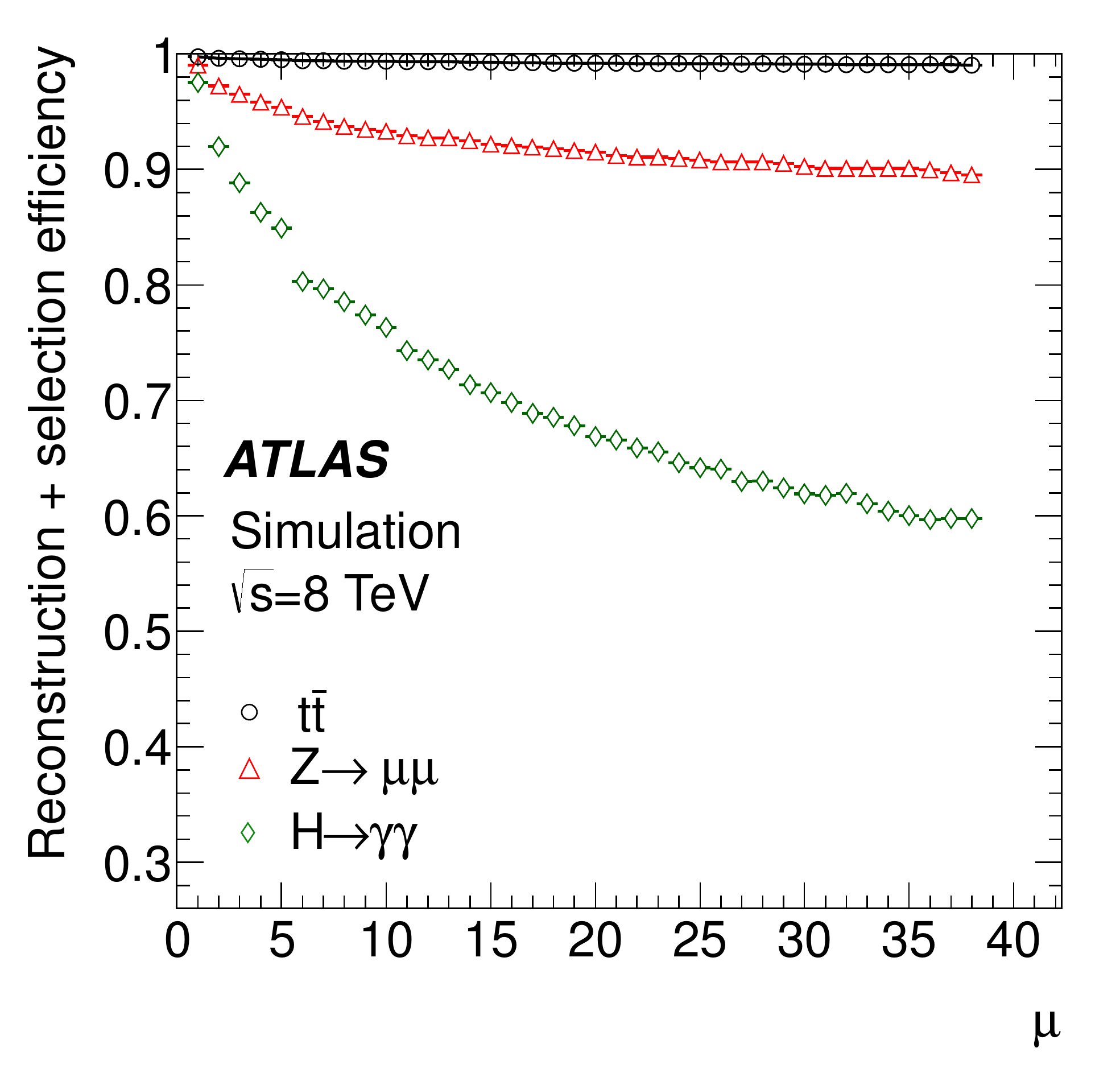} \label{fig:hs_select_nosel}}
\subfloat[]{\includegraphics[width=0.45\textwidth]{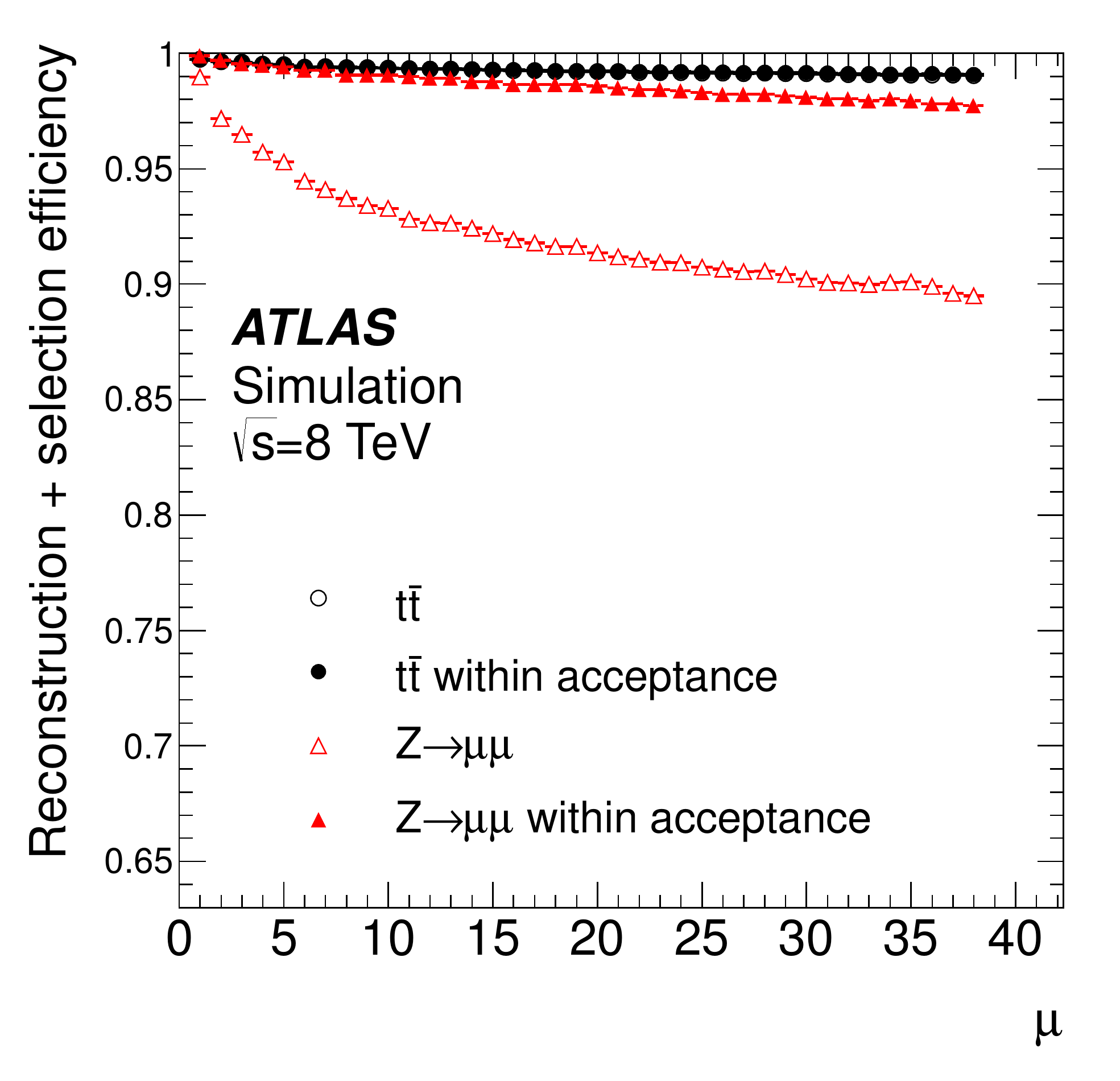} \label{fig:hs_select_dilep}}
\caption{\label{fig:hs_select} Efficiency to reconstruct and then select the hard-scatter primary vertex  as a function of the
average number of $pp$ interactions per bunch crossing, $\mu$, for different physics processes: (a) all reconstructed events;
 (b) events with at least two muons with $p_{\textrm T}>15$~GeV reconstructed within the ATLAS inner detector acceptance. The 
points showing the $t\bar{t}$ efficiency with and without acceptance criteria overlap.}

\end{figure}

For hard-scatter processes, the primary vertex selection efficiency is defined as the fraction of
events in which the highest $\sum \pt^2$  vertex is the vertex associated with the
 MC simulation hard scatter. The MC hard scatter is taken as the vertex with the highest weight of hard-scatter tracks,
as described in Section~\ref{sec::Truth}.  The efficiency to reconstruct and then select the hard-scatter primary vertex 
is shown as a function of $\mu$  in Fig.~\ref{fig:hs_select}(a) for different physics processes.  
The highest efficiency  is achieved for  $t\bar{t}$ events  for all values of $\mu$.  This observation is attributed to the 
high multiplicity of high transverse momentum  tracks produced in top-quark decays.   The selection efficiency for $Z\to \mu\mu$ events
 is greatly improved when additional criteria reflecting the kinematics of the physics process are imposed.
Figure~\ref{fig:hs_select}(b) shows the selection efficiencies after requiring at least two muons with $p_{\textrm T}>15$~GeV 
to be reconstructed within the ATLAS inner detector acceptance.  The $t\bar{t}$ sample shows a selection efficiency
above 99\% with or without the muon acceptance requirement (the points are overlapping in the figure).
   A clear selection efficiency improvement for the 
$Z\to \mu\mu$ process is visible when muons are reconstructed in the acceptance,
 resulting in at most 2\% of events with a wrongly selected hard-scattering primary vertex 
for $\mu$ of 38. 
These  losses are primarily due to the small but non-zero probability that the $\sum \pt^2$ of tracks
from one of the inelastic interactions in the minimum-bias sample
is larger than in the $Z\to \mu\mu$ interaction, as illustrated in Fig.~\ref{fig:hs_mb_sumpt2}.
A more quantitative prediction of this loss is given in Section~\ref{sec::Projection}.

%% file: SingleVertexPerformance.tex
This section presents a study of single primary vertex reconstruction in soft interactions which are characteristic 
of the pile-up events superimposed on the hard-scatter event of interest. This study is based on a 
minimum-bias data sample with a single primary vertex reconstructed in each event and corresponding to an average
 number of interactions per bunch crossing $\mu = 0.01$. 
 These data are compared to a simulation of inelastic interactions using the PYTHIA8 event generator.

The reconstruction efficiency for primary vertices produced in soft 
$pp$ interactions  varies depending on the nature of the soft interaction process.  
If the majority of final-state  charged particles are produced outside  the detector acceptance,
the reconstruction of the corresponding primary vertex may be 
unsuccessful. The vertex reconstruction efficiency may be further 
reduced by the inefficient reconstruction of very low $p_{\textrm T}$ trajectories, characteristic of
these soft interactions. Table~\ref{tb:Efficiencies} shows the efficiencies for reconstructing the primary vertex
in events from a minimum-bias sample with only single interactions.
\begin{table}[tp]
\begin{center}
 \begin{tabular}{lccc}
\hline\hline
&Non- & Single- & Double- \\
&diffractive &diffractive & diffractive \\
 \hline
Efficiency without any selection cuts &$92.9\%$ &$45.7\%$ &$49.0\%$ \\
&&&\\
Efficiency requiring at least two charged  &$96.1\%$ &$92.6\%$ &$90.2\%$ \\
particles with $p_{\textrm T}> 400$~MeV and  $|\eta|< 2.5$ & & & \\
&&&\\
Efficiency requiring at least two charged &$99.6\%$ &$99.5\%$ &$99.3\%$ \\
particles reconstructed in the inner detector & & & \\
\hline\hline
\end{tabular}
\end{center}
\caption{Vertex reconstruction efficiencies, at various selection levels, for 
 non-diffractive, single-diffractive, and double-diffractive interactions in PYTHIA8
 minimum-bias simulation.}
\label{tb:Efficiencies}
\end{table}
These efficiencies are obtained from PYTHIA8 MC simulation  separately for 
 the three processes which produce minimum-bias triggers in the experiment, namely
 non-diffractive, single-diffractive, and double-diffractive interactions.
Without selection cuts the reconstruction efficiency depends strongly on the  process: 
increasing from $46\%$ for single-diffractive to $93\%$ for non-diffractive  interactions.  
Taking into account the relative contributions of each process to inelastic
interactions, the  average efficiency is estimated to be about 
$80\%$.   The difference in the efficiencies   estimated for the different processes is primarily due to the different distributions 
of  transverse momenta and  pseudorapidities of charged particles produced in each process.  In  diffractive processes, the charged particles  are 
 mostly  produced at large pseudorapidities, often outside  the acceptance of  the ATLAS tracking system.
The very soft transverse momentum spectrum of these charged particles is an additional complication in their reconstruction. 
As shown in the second row of Table~\ref{tb:Efficiencies}, basic geometrical and kinematic requirements on the generated particles remove most of 
the differences in efficiency among the non-diffractive, single- and double-diffractive processes. The overall vertex reconstruction efficiency increases to
$95\%$ in this case.  The remaining differences in efficiencies are mostly due to the dependence of the track reconstruction efficiency on $\eta$ and $p_{\textrm T}$.
The third row of Table~\ref{tb:Efficiencies} shows that the primary vertex reconstruction efficiency further increases to about $99\%$ for all processes after 
requiring that at least  two tracks are reconstructed within the inner detector, in addition to the requirements listed in the second row.
The intrinsic efficiency of the ATLAS vertex reconstruction algorithm
is thus expected to be very high if at least two charged particles are produced within the inner detector acceptance.

\begin{figure}[tp]
\centering
\subfloat[]{\includegraphics[width=0.45\textwidth]{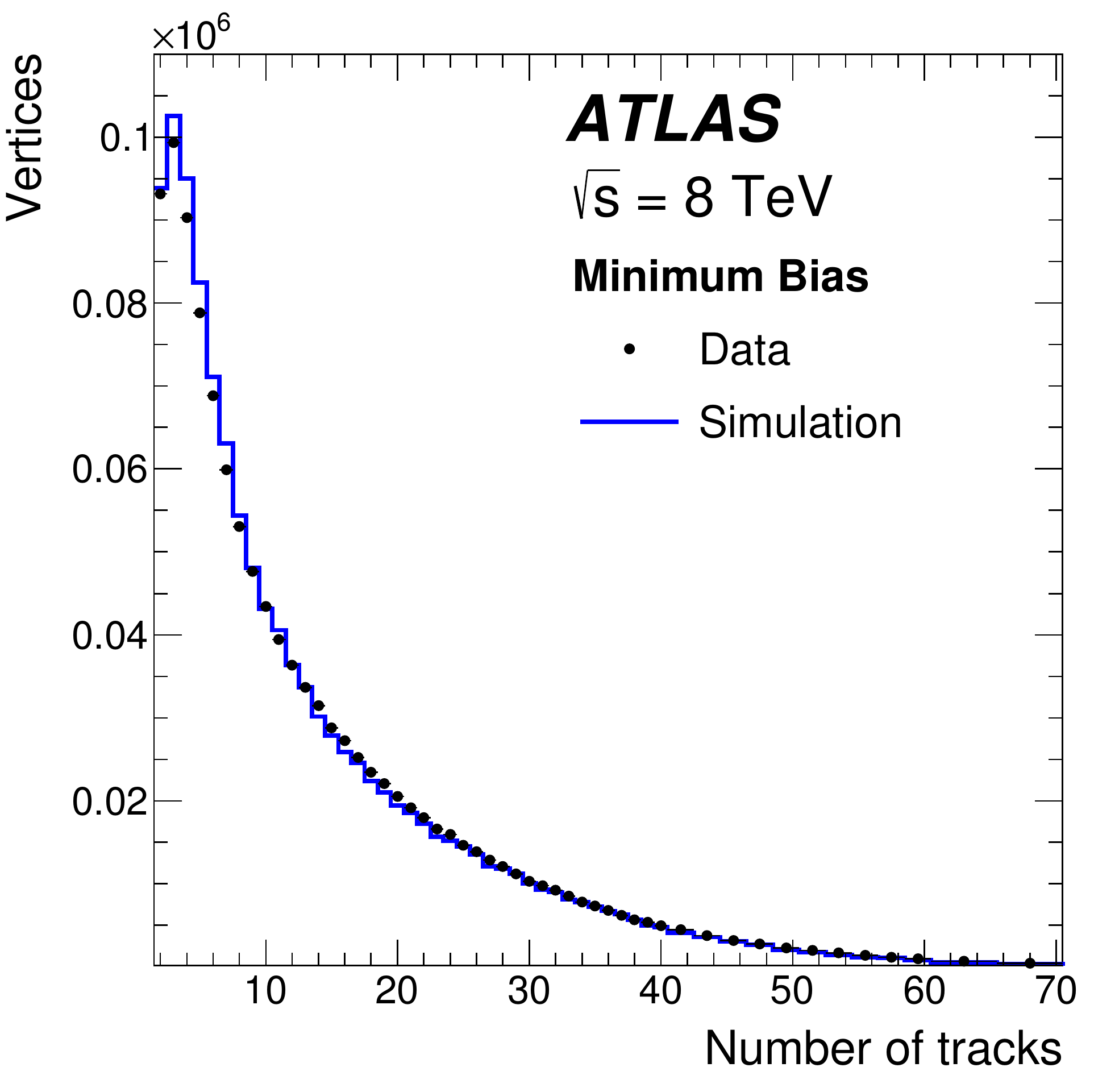}}
\subfloat[]{\includegraphics[width=0.45\textwidth]{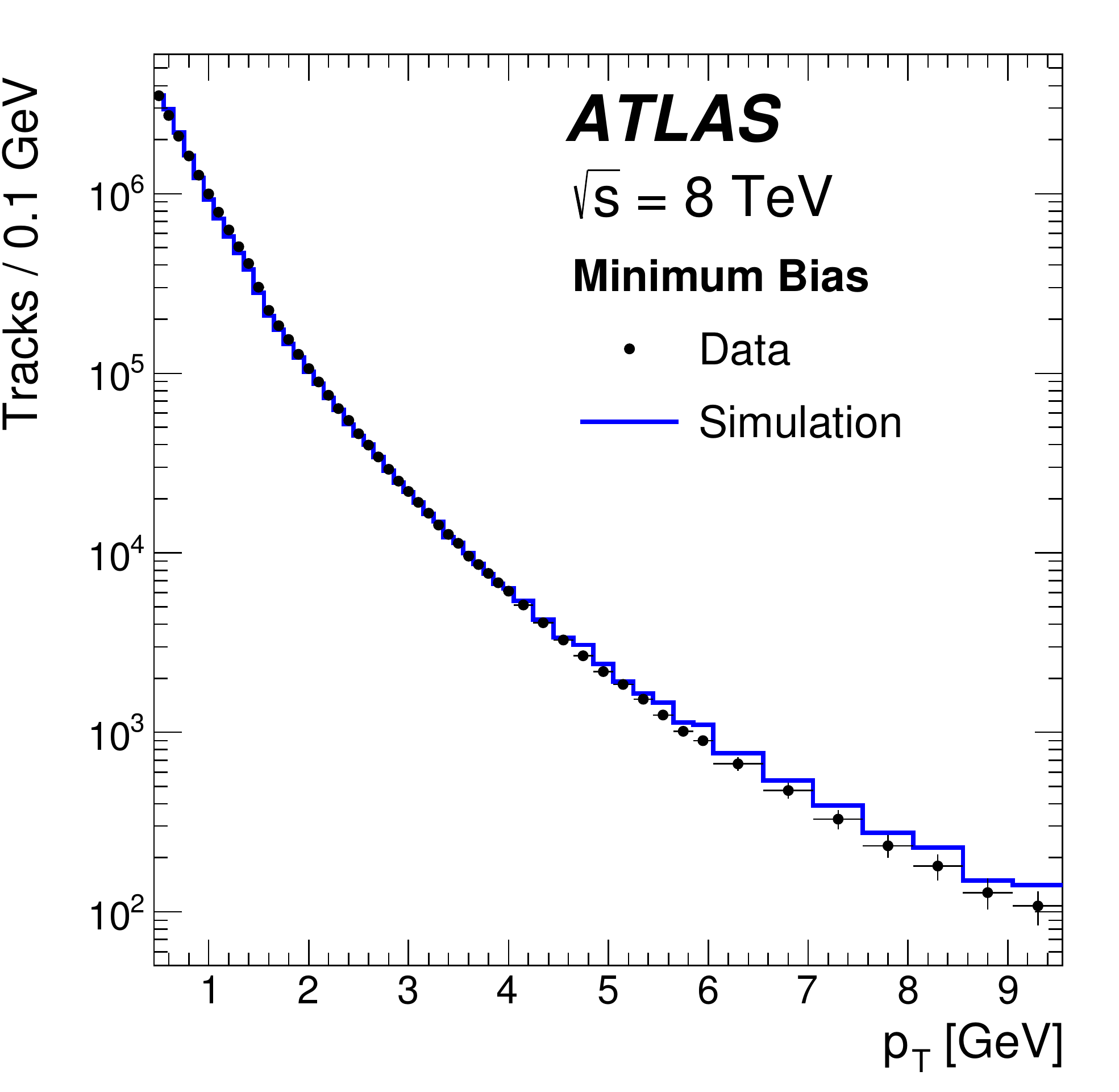}}\
\subfloat[]{\includegraphics[width=0.45\textwidth]{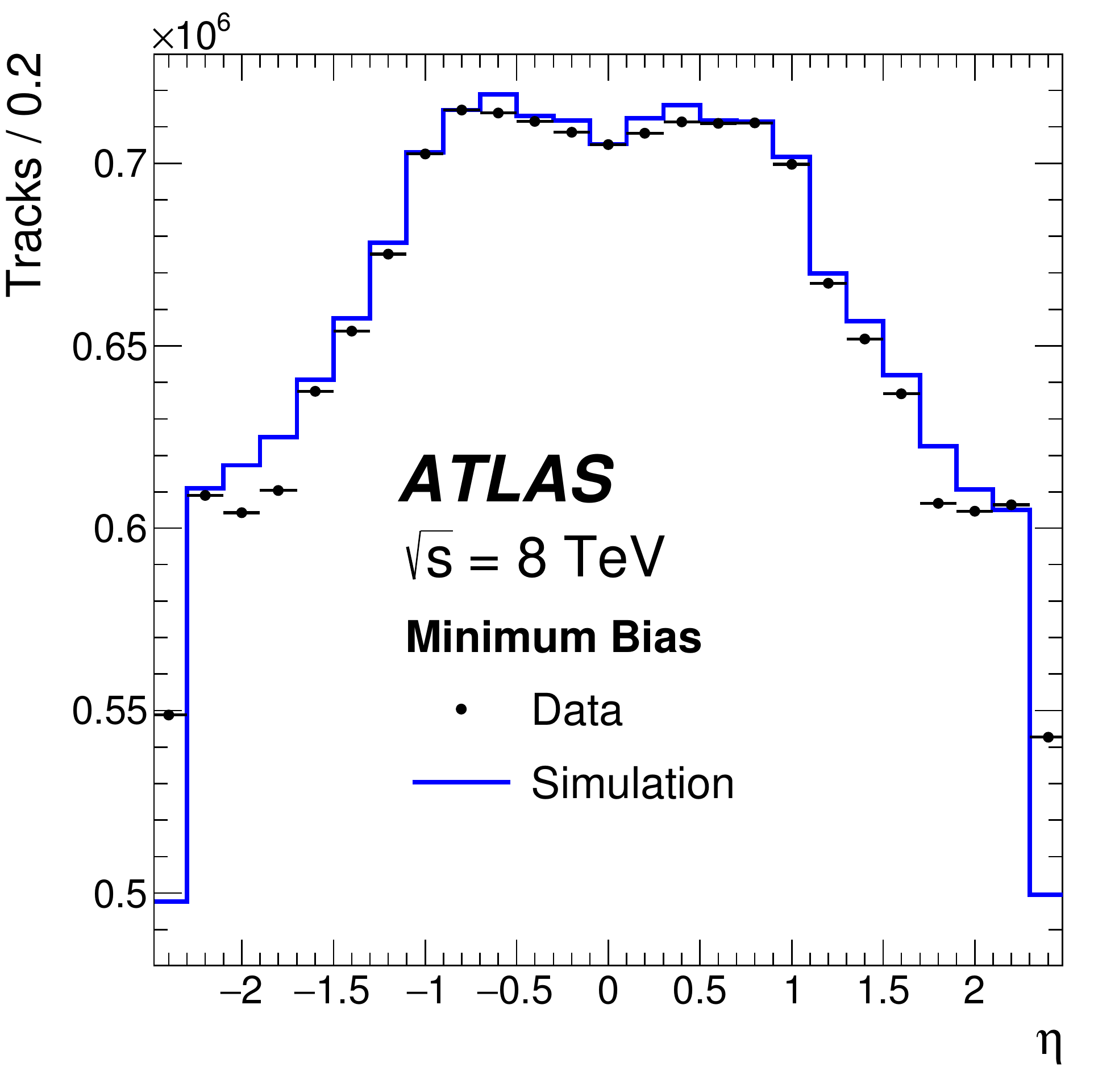}}
\subfloat[]{\includegraphics[width=0.45\textwidth]{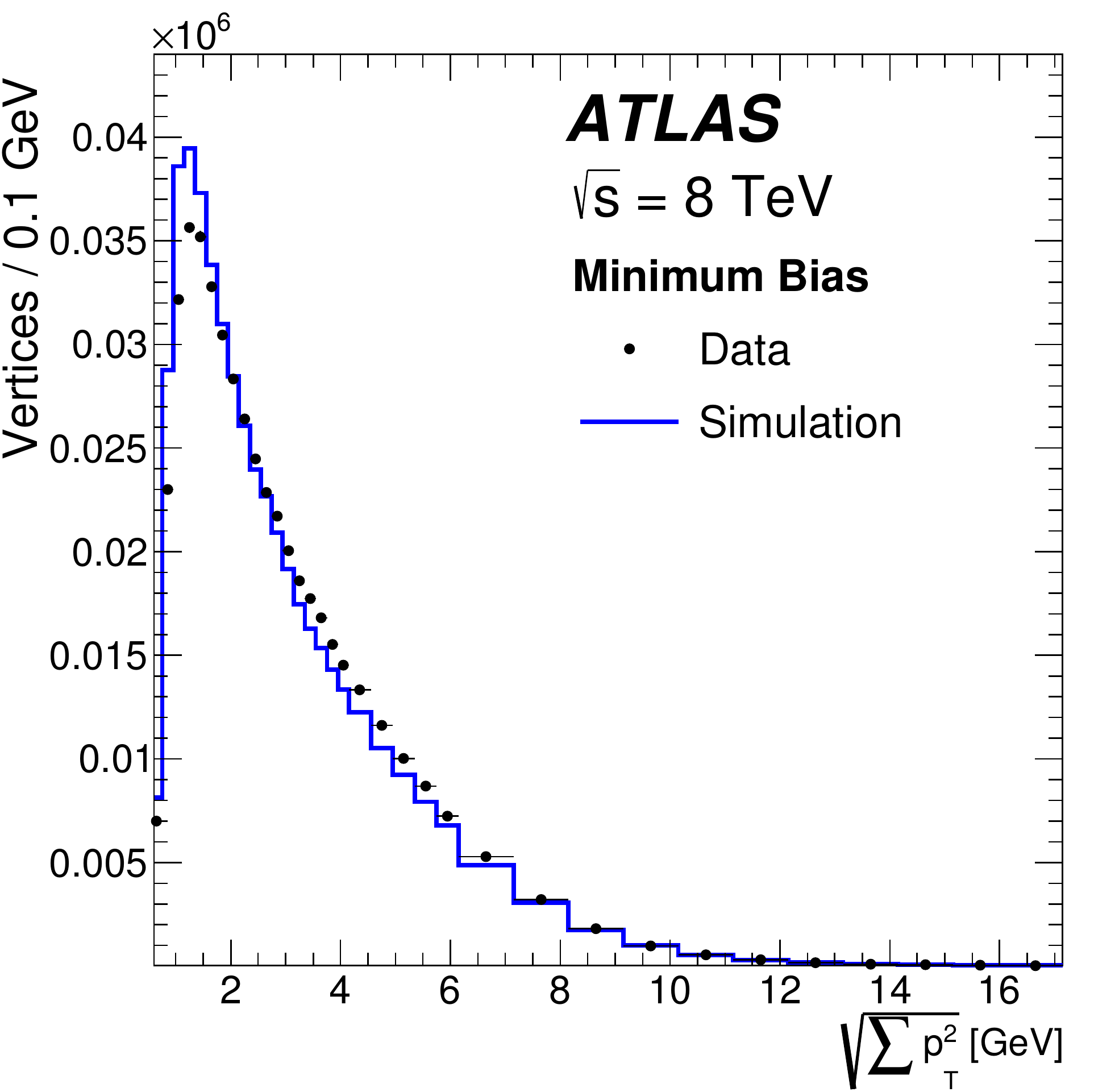}}
\caption{ 
Distributions of (a) number of tracks per vertex, (b) track transverse momentum $p_{\textrm T}$, (c)
 track pseudorapidity $\eta$ and (d) $\sqrt{ \Sigma p_{\textrm T}^{2}}$ of the tracks associated with each vertex.
Distributions are shown for tracks associated with primary vertices in low $\mu$ minimum-bias data and in simulation samples.
\label{fig:DataMC_LowMu_TrackpT_TrackEta}}
\end{figure}

Figure~\ref{fig:DataMC_LowMu_TrackpT_TrackEta} compares the simulation to data for the distributions
of the number of  fitted tracks, the track $p_{\textrm T}$, track $\eta$, and $\sqrt{ \Sigma p_{\textrm T}^{2}}$ of tracks in primary vertices.
The figure illustrates how soft the pile-up interactions are: 
only $0.4\%$ of the tracks belonging to a reconstructed primary vertex have $p_{\textrm T} > 4 $~GeV and only $1.2\%$ of the 
reconstructed vertices have a total $\sqrt{ \Sigma p_{\textrm T}^{2}}$  above $10$~GeV.
There are small discrepancies  between simulation and data  at very high values in the track $p_{\textrm T}$ spectrum and at high $\eta$.
 As described in Refs.~\cite{ChargedPmultiplicities, ATL-PHYS-PUB-2012-003}, these are due to deficiencies in the physics modelling 
 of these distributions and not related to the primary vertex reconstruction algorithm.
The dominant sources of systematic uncertainties relevant to the comparisons in 
Fig.~\ref{fig:DataMC_LowMu_TrackpT_TrackEta}  are the knowledge of the beam-spot size, the modelling of fake tracks, 
and the dependence of the track reconstruction efficiency on $p_{\textrm T}$, $\eta$ and $\mu$. These sources are not included in the 
error bars of the corresponding plots, but contribute to the observed discrepancies between data and simulation.

The position resolution of single vertices is
 estimated either from MC simulation or from data using the split-vertex method (SVM).
In this method the $n$ tracks associated to a
primary vertex are ordered in descending order of their transverse momenta.
The tracks are then split into two groups, one with even-ranking tracks and one with odd-ranking tracks, such that both 
groups have, on average,  the same number of tracks, $n/2$. 
The vertex fit is applied independently to each group. 
The spatial separation between two
resulting vertices gives a measurement of the intrinsic
resolution for a vertex with $n/2$ tracks.  The two split vertices
must be reconstructed independently and therefore no beam-spot
constraint is used during the fit.

\begin{figure}[tp]
\centering 
\subfloat[]{\includegraphics[width=0.475\textwidth]{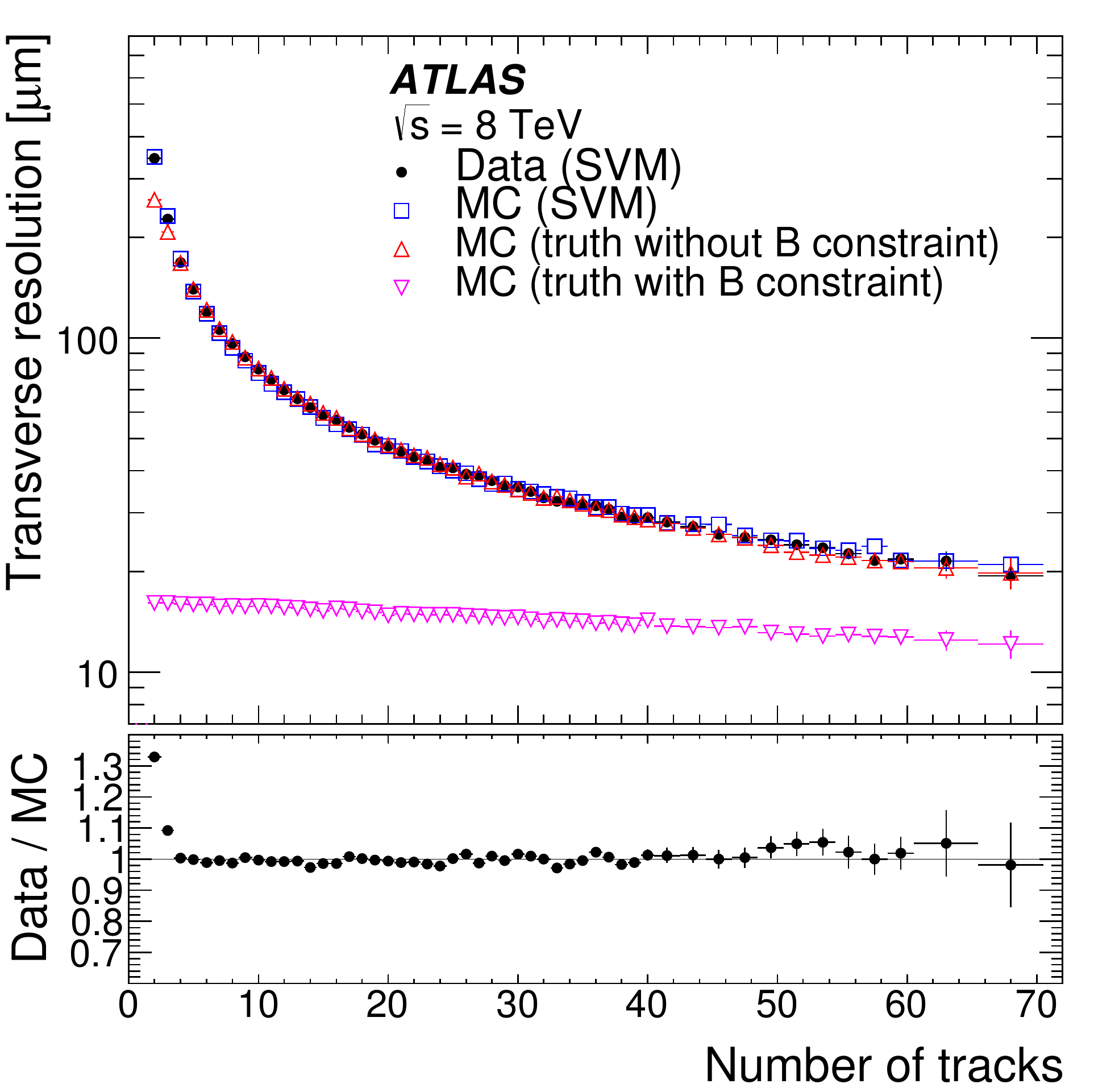}}
\subfloat[]{\includegraphics[width=0.475\textwidth]{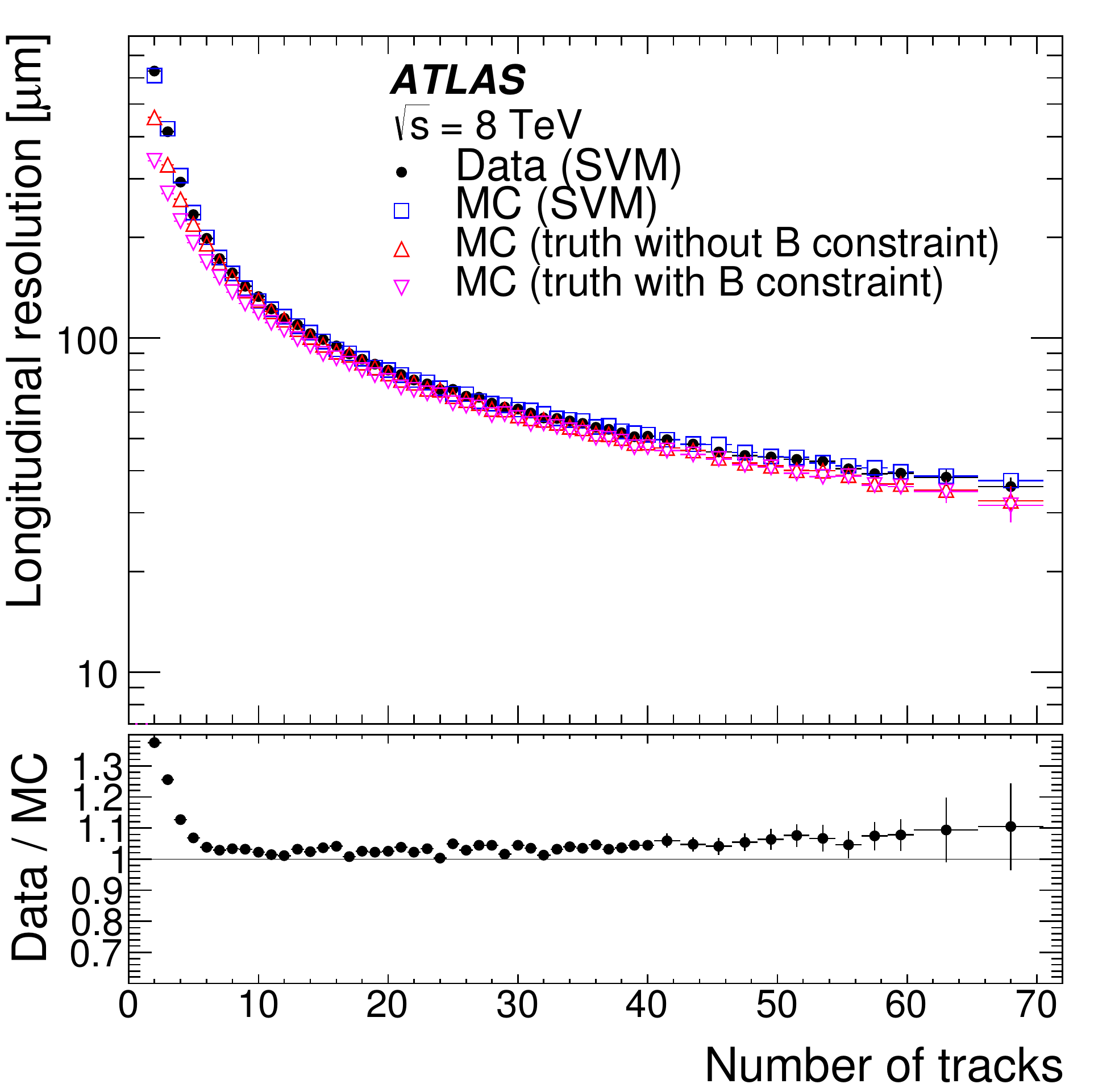}}
\caption{\label{fig:dmtbc_resZ_andX_vs_ntrks}
Resolution of the primary vertex position in (a) $x$ and (b) $z$  as function of the number of 
fitted tracks, estimated using the split-vertex method (SVM)
 for minimum-bias data (black circles) and MC simulation (blue squares).
Also shown is the resolution obtained from the  difference between the generator-level information and 
reconstructed primary vertex position in MC simulation (labeled ``truth''),
 with and without the beam-spot constraint (pink and red triangles respectively).  The 
bottom panel  in each plot shows the ratio of the resolution found using the split-vertex method in 
data to that obtained using the MC generator-level information without the beam-spot constraint. }
\end{figure}

Figure~\ref{fig:dmtbc_resZ_andX_vs_ntrks} shows the resolution in data
calculated with the split-vertex method as a function of the number
of tracks per vertex. 
The split-vertex method is also used to calculate 
the resolution for the minimum-bias simulation sample.
There is    good agreement between the data and simulation distributions, showing that the 
reconstructed track parameters used  in the vertex reconstruction are well modelled in the simulation.  
Figure~\ref{fig:dmtbc_resZ_andX_vs_ntrks} also shows the primary vertex resolution calculated as the
 difference between the true and reconstructed vertex  position in the MC simulation.  The good 
agreement between the split-vertex method and the resolution calculated with the MC generator-level
 information gives confidence that the split-vertex method provides  a reliable measurement of the primary vertex resolution. 
 At very low track multiplicity the result of the  split-vertex method deviates slightly from the resolution obtained using the generator-level 
 information.  Here the resolution obtained from the generator-level information benefits from the perfect knowledge of 
 vertex position decreasing the resolution spread, compared to the resolution obtained from the two reconstructed vertices in the
split-vertex method.  When the beam-spot constraint is included the resolution improves considerably in the transverse 
 direction, staying below 20 $\mu{\textrm m}$ for the full range of $\mu$ studied. The longitudinal resolution 
reaches 30 $\mu$m at high track multiplicity. 
 Figure~\ref{fig:dmtbc_resZ_andX_vs_ntrks} also shows the resolution calculated using MC generator-level information  
 with and without beam-spot constraint.

%% file: PileUp.tex
In this section, the study of the primary vertex reconstruction performance at low $\mu$ 
is extended to the high pile-up regime. A dedicated data sample of minimum-bias events collected 
with values of $\mu$ between $55$ and $72$  was used to study the performance of the primary 
vertex reconstruction in the presence of multiple vertices. 
The simulation samples spanned values of $\mu$ from $0$ to $22$, typical of the standard 2012 data-taking conditions, 
and from  $38$ to $72$ to emulate the high $\mu$ data sample.

\begin{figure}[tp]               
\centering
\subfloat[]{\includegraphics[width=0.46\textwidth]{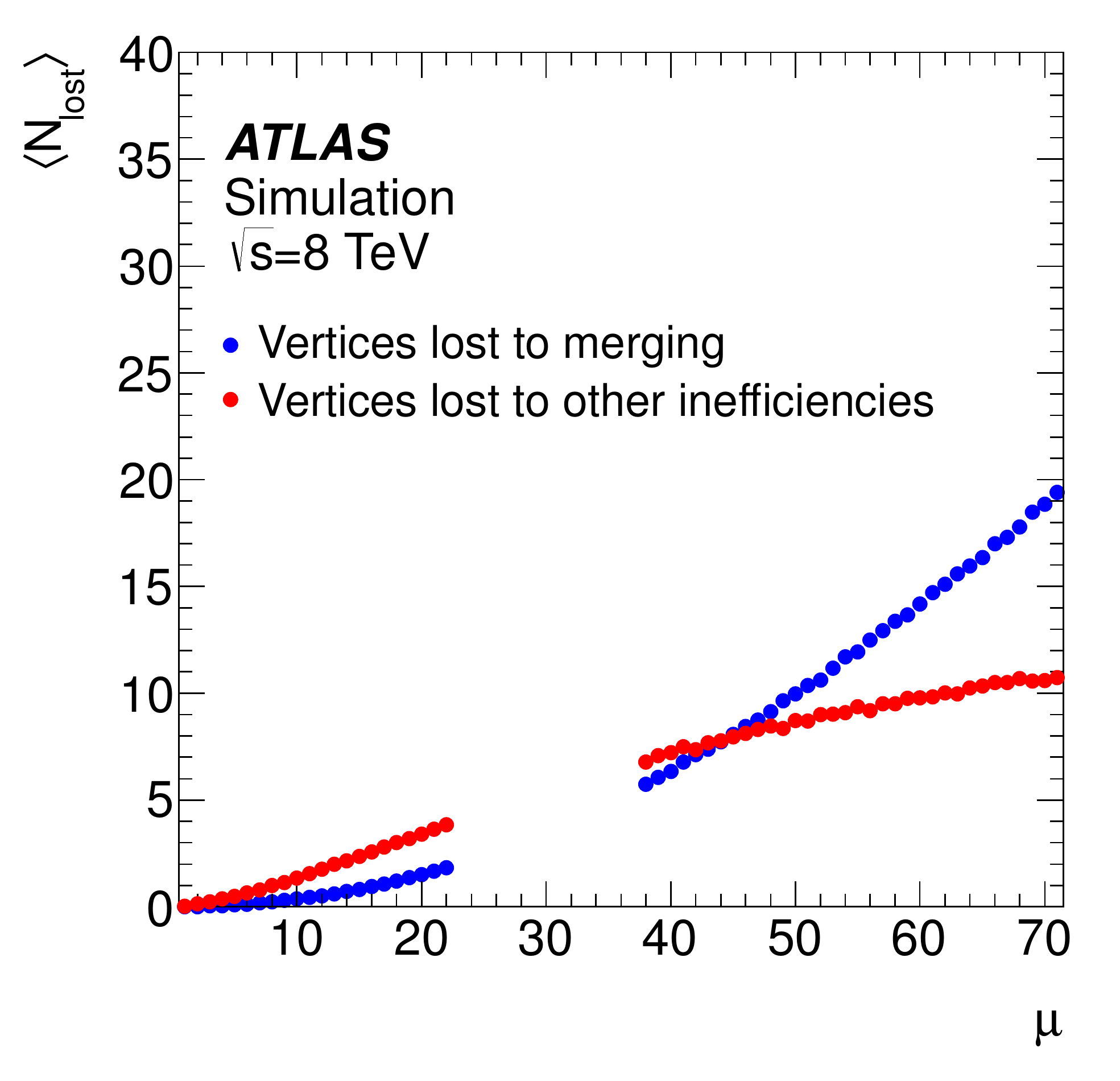}}
\subfloat[]{\includegraphics[width=0.46\textwidth]{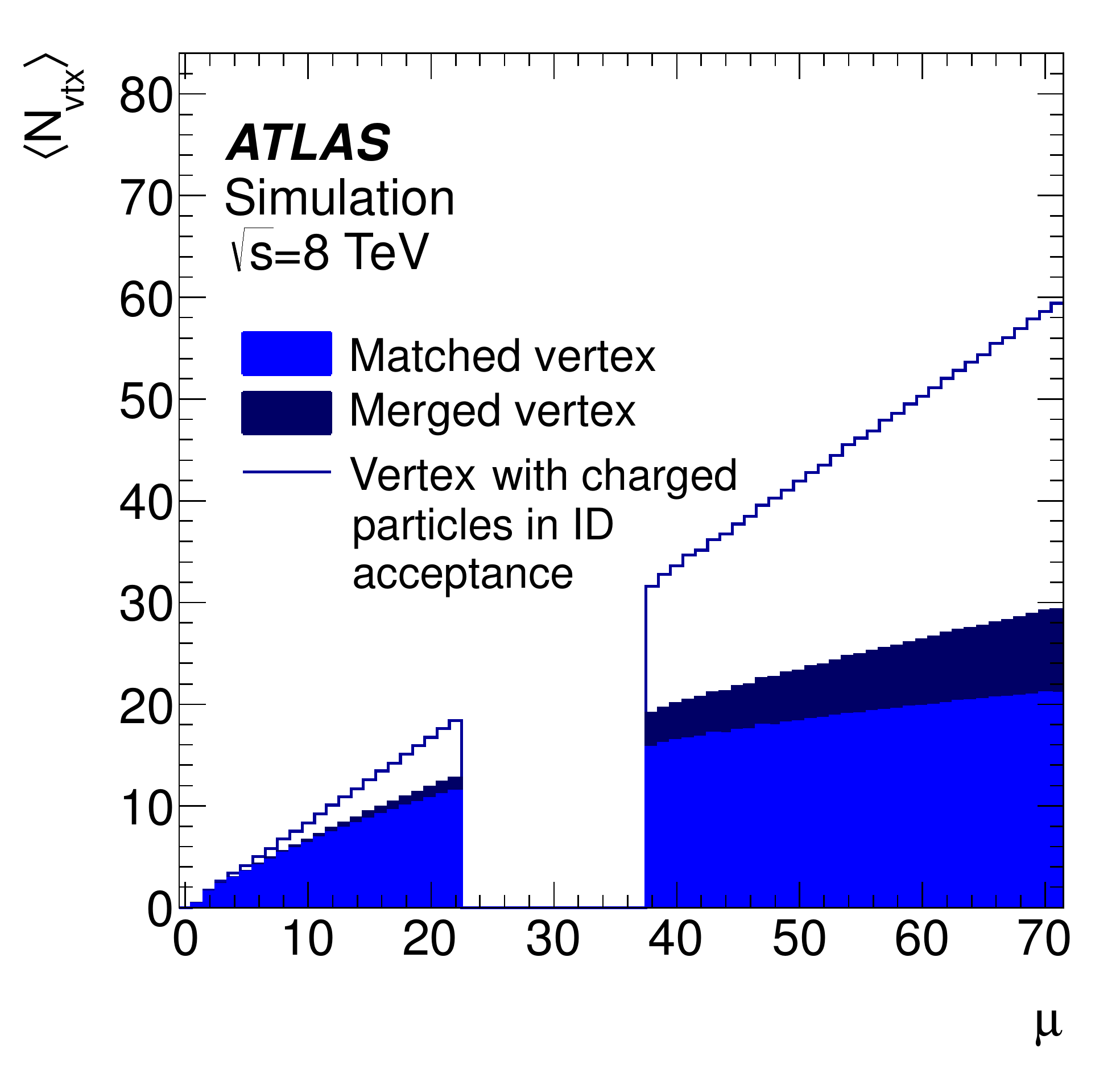}}
\caption{
(a) Average number of generated primary vertices with at least two charged  particles within the detector acceptance,
 that are not reconstructed due to merging (blue) and due to detector inefficiencies (red), as a function of the average number of interactions per bunch crossing, $\mu$.
 (b) Average number of reconstructed primary vertices of each truth-matching category compared to the total number of generated vertices with two particles within 
the detector acceptance, as a function of the average number of interactions per bunch crossing.
The available MC simulation samples were generated with values of $\mu$ below $22$ and above $38$.
\label{fig:nvtxmu}}
\end{figure}

The efficiency of primary vertex reconstruction decreases with increasing pile-up.
In addition to the inefficiencies affecting single vertex reconstruction described in Section~\ref{sec::Perf},
 effects related to the merging of adjacent primary vertices start to play a significant role as pile-up increases.
Figure~\ref{fig:nvtxmu}(a) shows the average number of vertices
lost due to merging and to other effects, such as track reconstruction
and detector acceptance. 
Merging has a small effect on overall vertex reconstruction efficiency for $\mu$ values below  $20$, but it is a dominant effect
 for $\mu$ values above $40$.  Figure~\ref{fig:nvtxmu}(b) shows the average number of expected reconstructed primary vertices as a function 
 of $\mu$, for the two main classes of vertices defined in Section~\ref{sec::HardScatter}, matched vertices,
 consisting of tracks mostly coming from a single interaction, and merged vertices. For the highest values of $\mu$ around $70$, 
 where one expects about $60$ primary vertices with at least two charged particles with $p_{\textrm T} > 400$~MeV within the detector acceptance, 
 a total of $30$ primary vertices are expected to be reconstructed on average, out of which about $10$ are merged vertices. 
 About $20$ additional primary vertices are lost due to merging and about $10$ due to other inefficiencies as shown in Fig.~\ref{fig:nvtxmu}(a).
 Vertices classified as ``Fake" or ``Split", according to the definitions presented in Section~\ref{sec::Truth},  are not shown in Fig.~\ref{fig:nvtxmu}(b), 
 since they represent a very small contribution of at most $2\%$  of the total number of reconstructed vertices at $\mu = 70$.

\begin{figure}[tp] 
\centering
\subfloat[]{\includegraphics[width=0.75\textwidth]{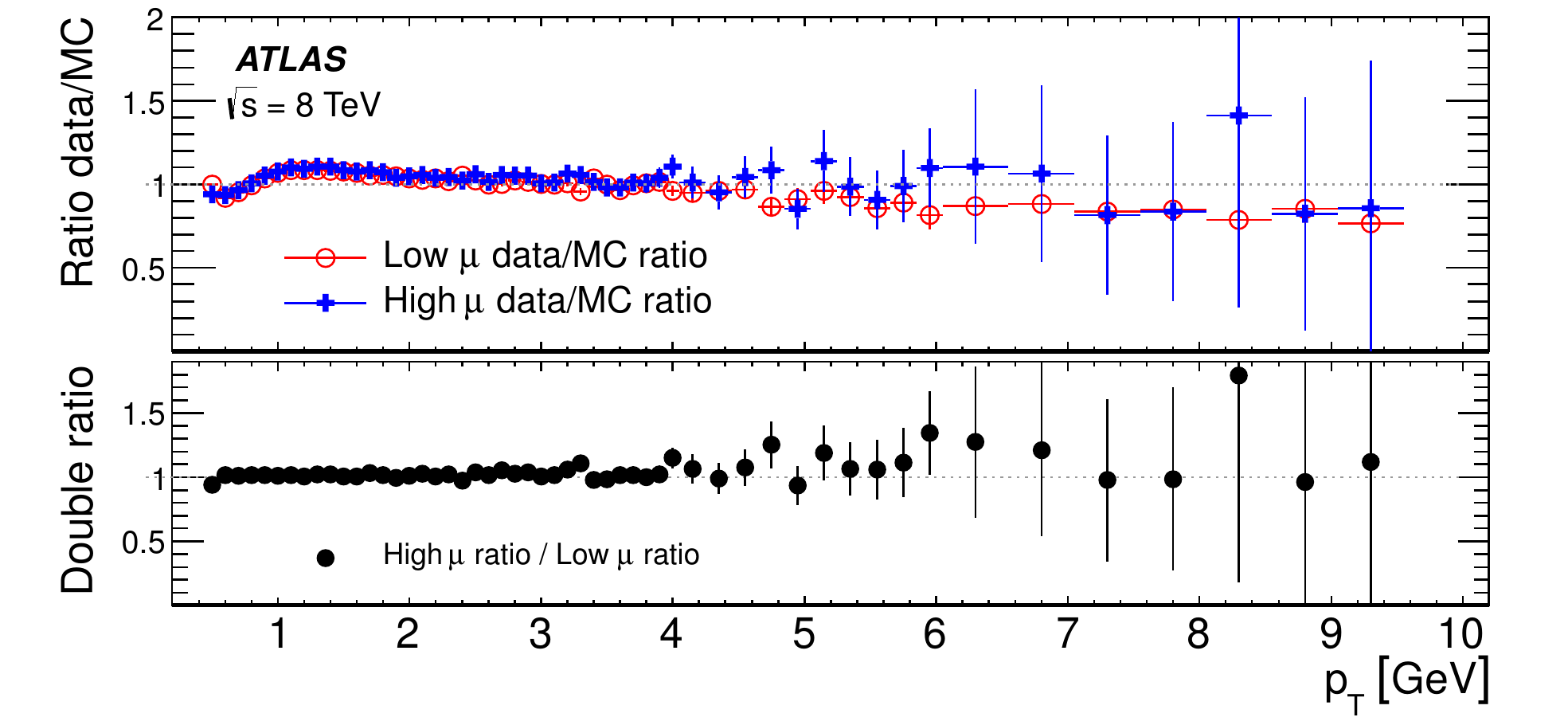}}\\
\subfloat[]{\includegraphics[width=0.75\textwidth]{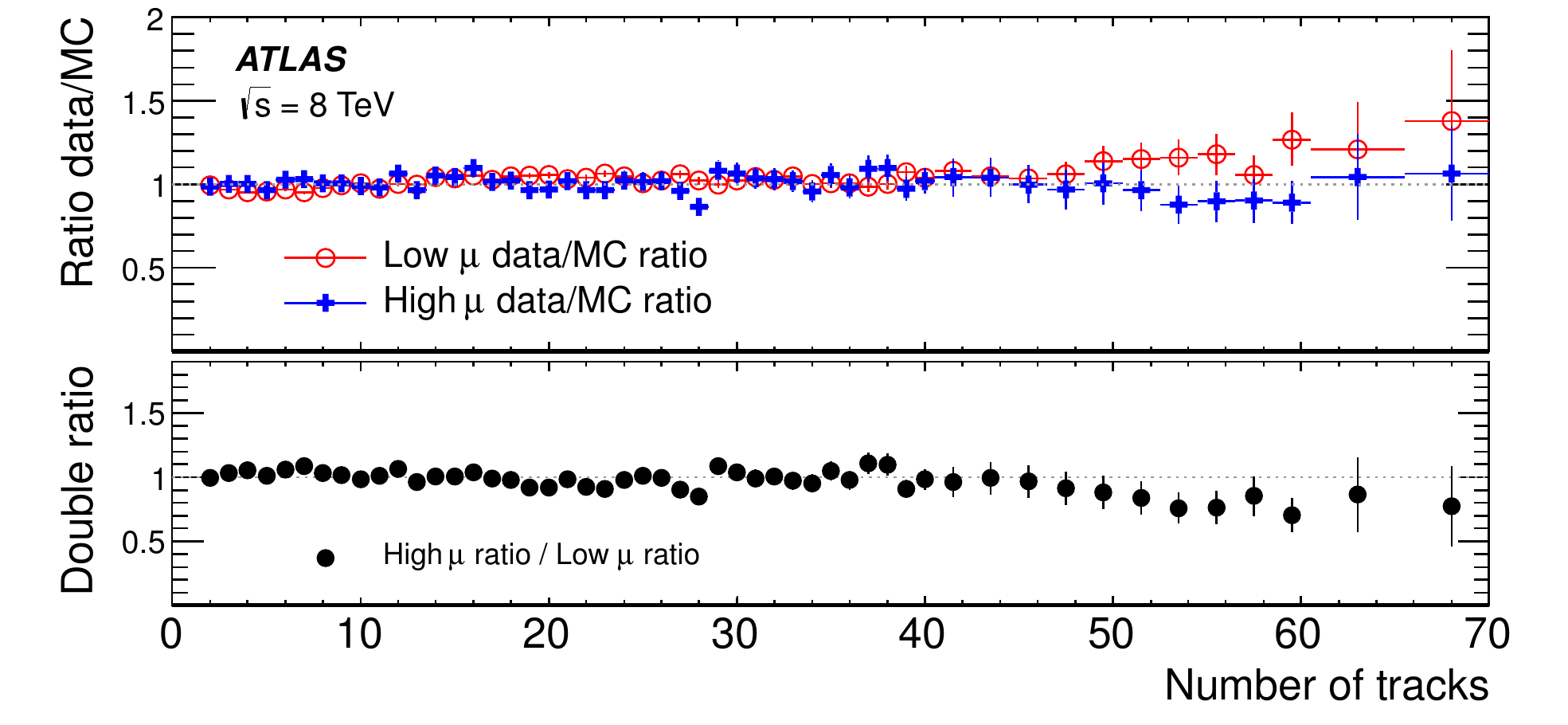}}\\
\subfloat[]{\includegraphics[width=0.75\textwidth]{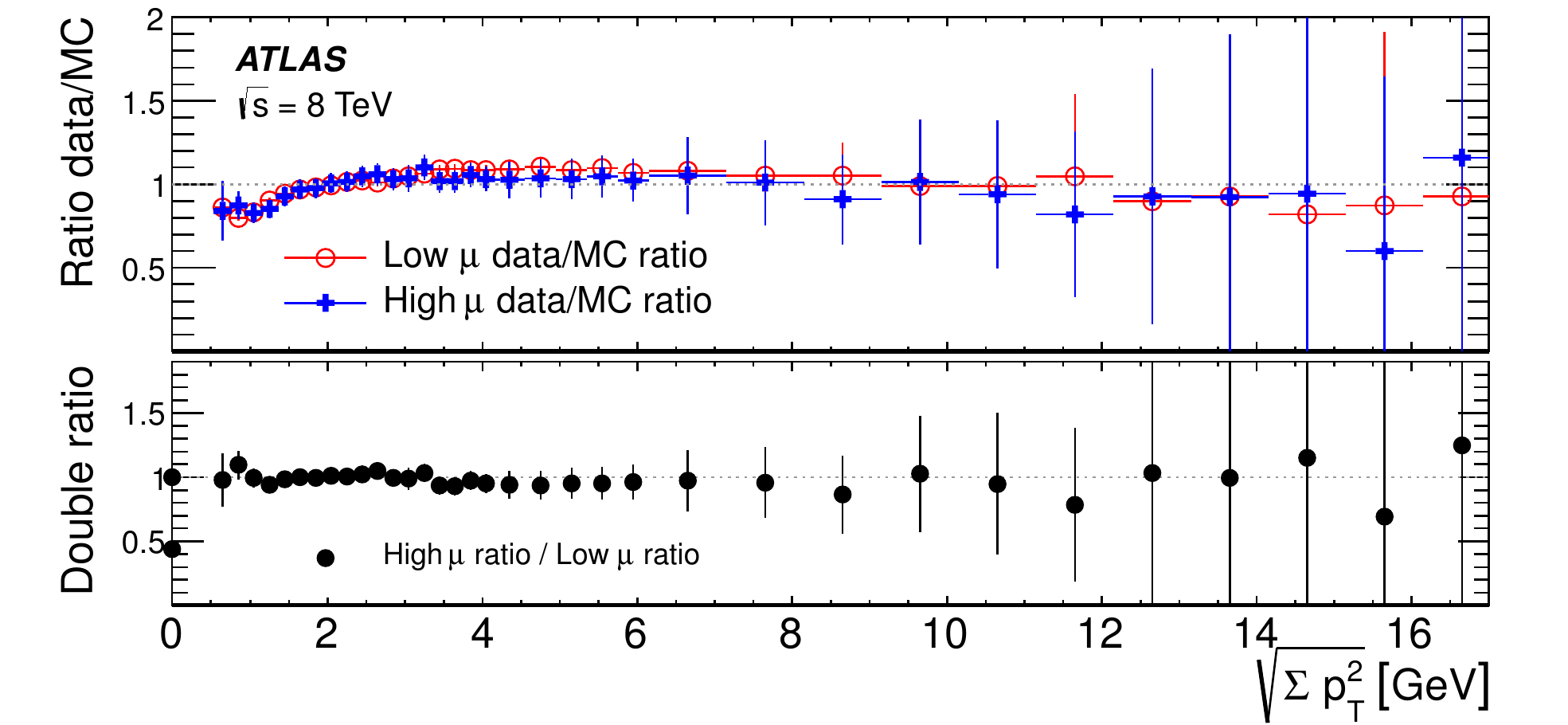}}
\caption{Ratios of data to MC simulation for observables relevant to the primary vertex reconstruction 
performance:  (a) track transverse momentum  $p_{\textrm T}$,   (b)  number of tracks per vertex,  
 (c) $\sqrt{ \Sigma p_{\textrm T}^{2}}$ of the tracks in each vertex. Error bars represent only statistical uncertainties.
The ratios are shown for low (0--1) and high (55--72) values of $\mu$. The bottom panel in each figure shows 
the double ratio of high to low $\mu$.
\label{fig:DoubleRatio}}
\end{figure}

The main observables relevant to the primary vertex reconstruction performance are in reasonable agreement 
 between data and simulation with only small discrepancies attributed to the physics modelling of soft interactions
 (see Fig.~\ref{fig:DataMC_LowMu_TrackpT_TrackEta}).   To quantify the agreement between data and simulation 
 at high values of $\mu$, the same observables  are studied and the ratios of data to simulation are compared between 
 low and high values of $\mu$. This is shown in Fig.~\ref{fig:DoubleRatio} for the track $p_{\textrm T}$, 
the number of tracks per primary vertex, and the $\sqrt{ \Sigma p_{\textrm T}^{2}}$ per primary vertex. The data to simulation
ratios are overlaid for low and high $\mu$ samples in the upper panels.  The lower panels show the 
 double ratios of data to simulation between high and low values of $\mu$.

The double ratios agree with unity, showing that there is similar agreement between data and simulation at low and high $\mu$.
In the case of track multiplicity,  the agreement  between data and simulation for high track multiplicities is somewhat better 
at high $\mu$ than at low $\mu$. This arises possibly because discrepancies in physics modelling are diluted by the contributions from merged vertices 
as $\mu$ increases.

%% file: VertexProjection.tex
An analytical model to predict  the number of reconstructed vertices as a function of
event multiplicity has been developed. This model is based on the measured primary vertex reconstruction 
efficiency and on the the probability of vertex merging.
\subsection{Modelling the number of reconstructed vertices}
In the ideal case of perfect reconstruction efficiency, the number of reconstructed vertices would scale linearly with $\mu$.
In reality there are a number of effects that cause the relation to be non-linear. 
As discussed in Section~\ref{sec::HighPileUp},   one of the most important effects is  vertex merging, when two or more vertices are merged 
and  reconstructed as one vertex. Other effects include reconstruction inefficiencies, detector 
acceptance, and, at a small level for low track multiplicities, non-collision background. As already mentioned, the impact of 
fake and split vertices is negligible.

The average number of reconstructed vertices, $\langle n_{\textrm Vertices} \rangle$, can be parameterised as a function of $\mu$ as follows:
\begin{equation}
\langle n_{\textrm Vertices} \rangle = p_{0} + \epsilon \mu - F( \epsilon \mu, p_{\textrm merge}),
\label{eq:fit_vxvsmu}
\end{equation}
where $\epsilon$ is the efficiency of the vertex reconstruction algorithm before including vertex merging effects, and $p_{0}$
accounts for any small offset arising from non-collision background. Based on the results shown in 
Sections~\ref{sec::HardScatter}, \ref{sec::Perf}, and \ref{sec::HighPileUp}, the value of $\epsilon$ is considered to be independent of $\mu$. 
The quantity $\epsilon \mu$ represents the average number of vertices that would be reconstructed 
in the absence of any pile-up induced vertex merging effects.
This quantity is referred to, hereafter, as the number of reconstructible vertices. 
In this study the parameter  $\epsilon$ is obtained from a fit to the MC simulation.
The function $F( \epsilon \mu, p_{\textrm merge})$ represents the average number of vertices lost due to merging effects, 
taking into account the number of reconstructible vertices 
and the vertex merging probability, $p_{\textrm merge}$.  These effects are primarily responsible for the non-linear dependence of the number of 
reconstructed vertices as a function of $\mu$.  The evaluation of this function is described in the next section.

The proposed model only describes the primary vertex reconstruction and does not account for pile-up effects 
in the reconstruction of tracks. The model assumes that the track reconstruction efficiency and the 
corresponding fake rate are constant for the studied range of pile-up values.

\subsection{\label{sec:vertex-merging-correction}Determination of correction for merging of primary vertices}
\begin{figure}[tp]
\centering
\includegraphics[width=0.6\textwidth]{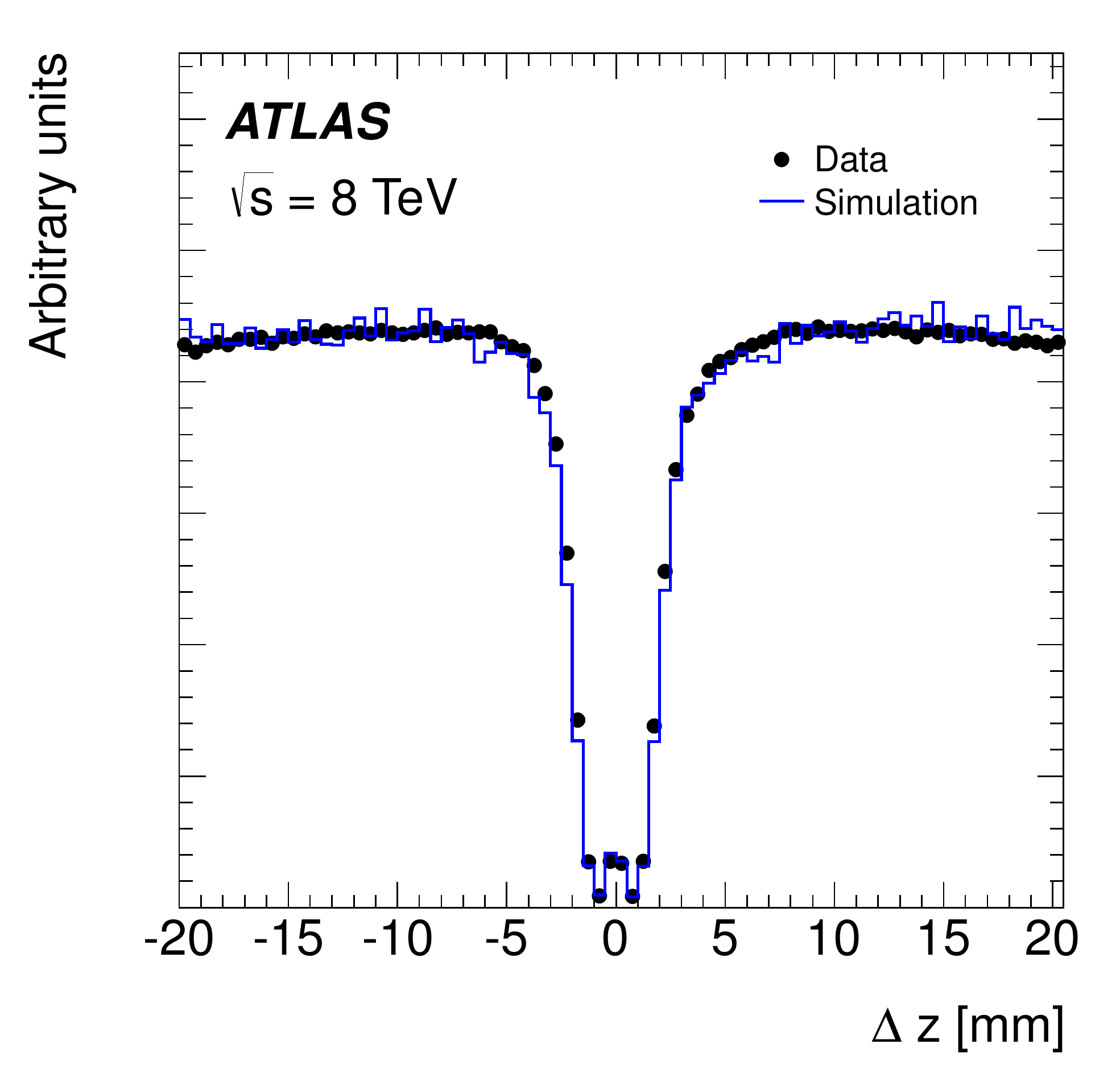}
\caption{ Distribution of the 
longitudinal separation between pairs of adjacent primary vertices in a typical Run 1 minimum-bias data sample and in MC simulation.}
\label{fig:DeltaZ}
\end{figure}

The effects of vertex merging are studied using the longitudinal separation, $\Delta z$, between
 pairs of adjacent reconstructed primary vertices. The distribution of $\Delta z$
in a typical Run 1 minimum-bias data sample is shown in Fig.~\ref{fig:DeltaZ} together with the prediction from simulation.

At low values of $\Delta z$  close-by vertices can no longer be separated and are reconstructed as a single vertex.
In Fig.~\ref{fig:DeltaZ}, this effect is visible as a steep decrease of the number of reconstructed vertices at values of $\Delta z$ 
below a few mm. The small peak around $\Delta z = 0$ is due to the  effect of splitting of primary vertices: in this case, close-by vertices are
 reconstructed with longitudinal separations well below the typical 
primary vertex resolution.  The distribution of $\Delta z$ measured in a low pile-up data sample ($\mu$ below 10) is used 
to derive a two-vertex merging probability density function $p_{\textrm merge}(\Delta z)$.
This function can then be  combined with a given beam-spot shape to derive an analytical relationship between 
the number of reconstructible vertices per event, $\epsilon\mu$, and the average number of
 reconstructed vertices, $\langle n_{\textrm Vertices}\rangle$.  
Using this approach, the effect of different beam-spot sizes on the merging probability can then
also be evaluated.

The analytical function is derived as follows:
\begin{enumerate}
  \item The $\Delta z$ distribution for pairs of adjacent vertices reconstructed in low pile-up data is fitted 
 with a Gaussian function in a range where the merging of vertices is negligible:
$|\Delta z| > 30$~mm.  The Gaussian has an expected width of $\sqrt{2} \lumsigz$,
where $\lumsigz$ is the longitudinal beam-spot RMS, assuming the beam spot has a Gaussian shape 
distribution along the $z$-axis.
\item A merging probability density function, $p_{\textrm merge}(\Delta z)$, is constructed by
taking the difference between the  distribution  of $\Delta z$ observed in data  in the 
range $|\Delta z| < 30$~mm and the prediction obtained from the Gaussian fit,  $f_{\textrm exp}(\Delta z)$.
This difference  is then normalised to the prediction  probability density function:
\begin{equation}
  \label{first_merging_eq}
                p_{\textrm merge}(\Delta z) = \frac{f_{\mathrm{exp}}(\Delta z) -  f_{\mathrm{\textrm obs}}(\Delta z)}{f_{\mathrm{\textrm exp}}(\Delta z)}.
\end{equation}
Here, $f_{\textrm obs}(\Delta z)$ represents the observed probability density function of  $\Delta z$ in the 
range $|\Delta z| < 30$~mm. An example of  the observed distribution $f_{\textrm obs}(\Delta z)$ is shown in Fig.~\ref{fig:DeltaZ}.

The $p_{\textrm merge}(\Delta z)$ PDF is  parameterised using a step function convolved with a Gaussian function with
 parameters fit to the observed distribution.    The  $p_{\textrm merge}(\Delta z)$ PDF is derived  in the low 
 pile-up regime,  where only the merging of adjacent pairs of vertices is assumed to be significant. The possible effects of
 merging  more than two $pp$ collisions into a single reconstructed primary vertex are assumed to be negligible in this low 
pile-up regime.

\item The total merging probability  $p_{\textrm{merge}}$ for two independent 
reconstructible vertices is computed from the product of the
 merging PDF and the expected $f_{\textrm{exp}}(\Delta z)$ distribution:
 \begin{equation}\label{second_merging_eq}
                p_{\textrm merge} = \int f_{\textrm exp}(\Delta z) p_{\textrm merge}(\Delta z) \textrm d(\Delta z).
 \end{equation}
It is assumed that the merging PDF for a pair of adjacent vertices $p_{\textrm merge}(\Delta z)$ is independent of the beam conditions.
 The overall probability of merging two random  reconstructible vertices depends on the particular beam-spot distribution, 
and therefore on $f_{\textrm exp}(\Delta z)$.
\item  The total number of vertices lost due to merging effects is given by:
\begin{equation}\label{third_merging_eq}
  F(\epsilon \mu, p_{\textrm merge}) = \epsilon \mu - \sum\limits_{\mathcal{N}_{\textrm Vertices}} P(\mathcal{N}_{\textrm Vertices}, \epsilon \mu) \wp_{\textrm merge}(\mathcal{N}_{\textrm Vertices}, p_{\textrm merge}),
   \end{equation}
where   $P(\mathcal{N}_{\textrm Vertices}, \epsilon \mu)$ is a 
%Poisson PDF, representing the probability of reconstructing $\mathcal{N}_{\textrm Vertices}$ vertices 
PDF, representing the probability of reconstructing $\mathcal{N}_{\textrm Vertices}$ vertices 
given $\epsilon \mu$  potentially reconstructible vertices. Since the number of visible $pp$ collisions varies according to Poisson 
with the mean of $\mu$, this function  $P(\mathcal{N}_{\textrm Vertices}, \epsilon \mu)$ is a Poisson with a mean $\epsilon\mu$.

The function $\wp_{\textrm merge}(\mathcal{N}_{\textrm Vertices}, p_{\textrm merge})$  represents
the number of reconstructed vertices after taking into account merging effects, for a number,  $\mathcal{N}_{\textrm Vertices}$, of vertices which 
would be reconstructed in the absence of any merging. This number is defined as follows:
\begin{equation}
\label{forth_merging_eq}
\wp_{\textrm merge}(\mathcal{N}_{\textrm Vertices}, p_{\textrm merge}) = \sum\limits_{i=1}^{\mathcal{N}_{\textrm Vertices}} p_{i},
   \end{equation}
where $p_i = p_{i-1} (1-p_{i-1}p_{\textrm merge})$, $i \geq 2$ and $p_1 = 1$. The $p_i$ represents the probability to reconstruct $i$ vertices in 
the presence of merging effects.
\end{enumerate}

\subsection{Comparison of data to simulation}
 To quantitatively compare data with simulation, additional effects and systematic uncertainties need to be taken into account.
 To account for the difference in visible cross section between data and simulation discussed in Section~\ref{sec:datasets}, 
the parameter $\epsilon$, extracted from the simulation fit, is scaled by a factor $1/1.11$,
 which is equivalent to a scaling of $\mu$.
A $6\%$ uncertainty is assigned to this procedure, where the dominant contribution comes from the uncertainty in the measured value of
$\mu$.

The impact of possible discrepancies in longitudinal beam-spot size between data and MC simulation was also  assessed since
the observed data values represent an average over a range
of different and non-uniform experimental values. The MC simulation samples used in this study 
were generated with a beam-spot size equal to the average observed in data.
The effect of a change in beam-spot size on the merging probability can be evaluated with
 Eq. \ref{second_merging_eq}.
A small additional uncertainty is  assigned to account for the variations of up to $\pm 2$~mm 
in beam-spot size in data. 

\begin{figure}[tp]
\centering
\subfloat[]{\includegraphics[width=0.7\textwidth]{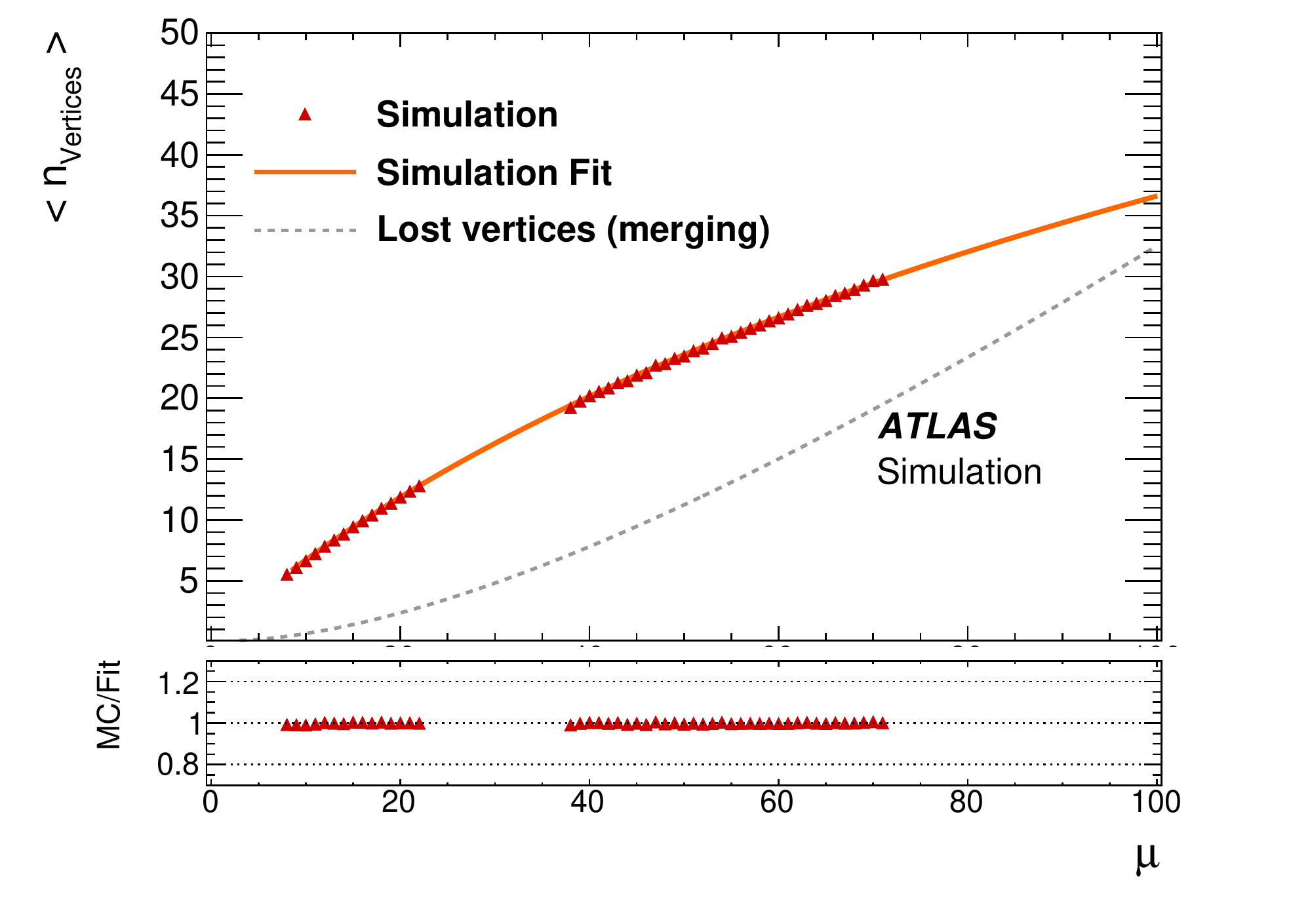}}\\
\subfloat[]{\includegraphics[width=0.7\textwidth]{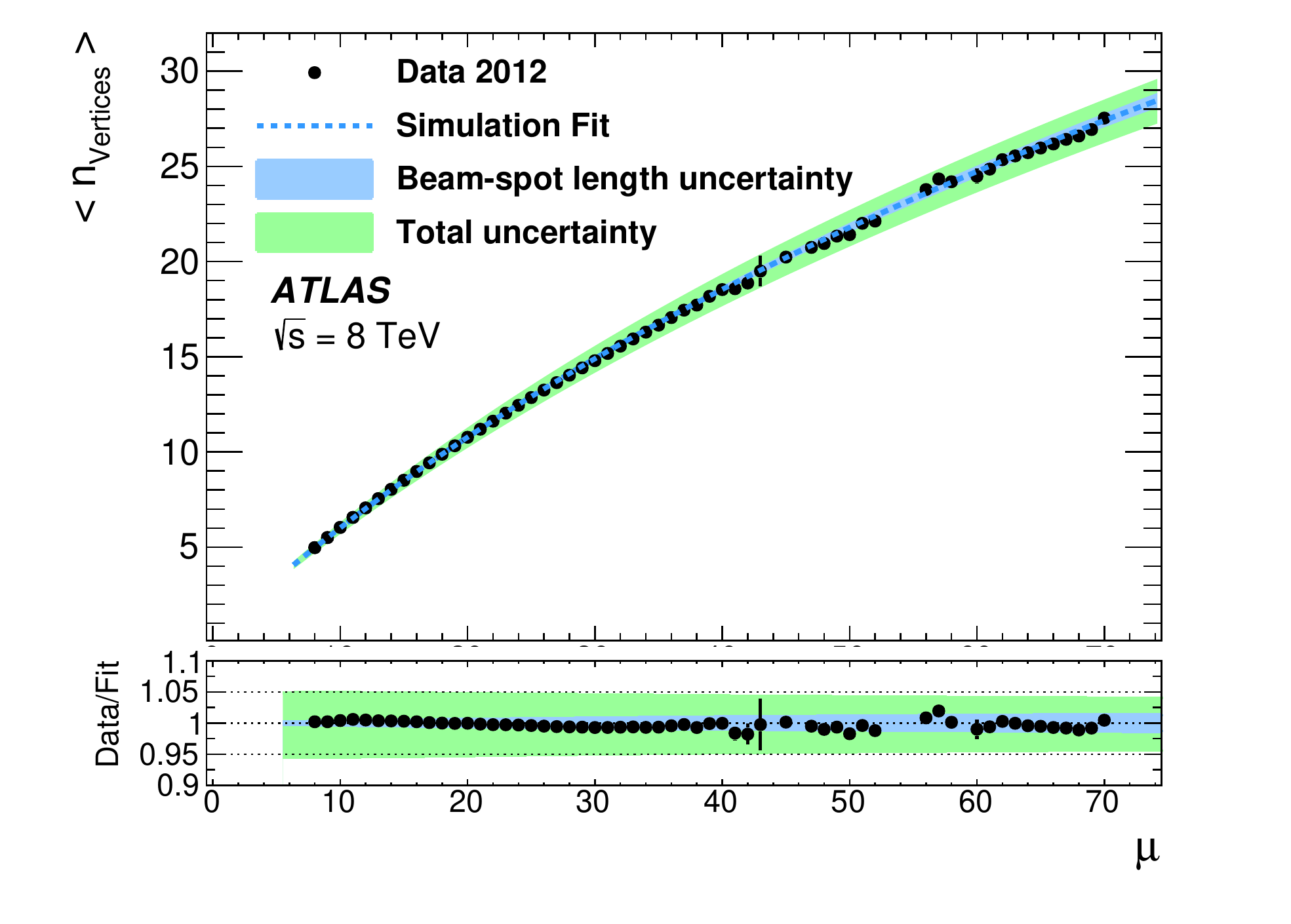}}
\caption{\label{fig:VxVsMu_MC} Distribution of the average number of reconstructed vertices as a function of the number of interactions per bunch crossing, $\mu$.
(a) MC simulation of minimum-bias events (triangles) and the analytical function in Eq.~\ref{eq:fit_vxvsmu} fit to the simulation (solid line).
The dashed curve shows the average estimated number of vertices lost to merging.
(b) Minimum-bias data (black points). The curve represents the result of the fit to the simulation in
 (a) after applying the  $\mu$-rescaling  correction described in the text.  The inner dark (blue) band shows
 the systematic uncertainty in the fit from the beam-spot length, while the outer light (green) band shows  the total uncertainty in the fit.
The panels at the bottom of each figure represent the respective ratios of simulation (a) or data (b) to the fits described in the
 text.
}
\end{figure}

A fit using Eq.~\ref{eq:fit_vxvsmu} was performed on MC simulation, allowing parameters $p_0$, $\epsilon$, and $p_{\textrm merge}$ to vary.
The efficiency, $\epsilon$,  and merging probability, $p_{\textrm merge}$, are extracted from the fit to simulation and found to be,
  $ 0.618 \pm 0.004{\textrm(stat.)}\pm 0.037 {\textrm (syst.)}$ and $0.0323 \pm 0.0002{\textrm (stat.)} \pm 0.0013{\textrm (syst.)} $ respectively,
after correcting $\epsilon$ with the $\mu$-rescaling factor and taking into account the systematic uncertainties,
as described above. The fit to MC simulation is shown in Fig.~\ref{fig:VxVsMu_MC}(a).

Data are compared to Eq.~\ref{eq:fit_vxvsmu} with the parameters $\epsilon$ and $p_{\textrm merge}$ fixed
to the values from the fit to simulation, 
 and with the small value of 
$p_0$ extracted from a fit to the data. The $p_0$ parameter is irrelevant in MC simulation, which
 does not account for the small non-collision background present in
 data at low values of $\mu$. The result  is shown in Fig.~\ref{fig:VxVsMu_MC}(b).
The uncertainty bands in Fig.~\ref{fig:VxVsMu_MC}(b) show the  beam-spot size uncertainty and the total uncertainty,
which is computed by summing in quadrature the beam-spot size and the dominant $\mu$-rescaling uncertainty terms.
The overall agreement between the data and the prediction is within $3\%$, with the largest observed discrepancies 
well within the systematic uncertainty bands.

This comparison shows that the simulation describes  the primary vertex reconstruction efficiency dependence on $\mu$ accurately.
Vertex merging is  the effect that has the largest impact on 
primary vertex reconstruction efficiency as $\mu$ increases. The analytical description 
proposed to describe this effect is validated by the measurements based on minimum-bias data. 
This  confirms that  the main factors related to the vertex reconstruction  in pile-up conditions are correctly taken into account and that the remaining 
 effects related to the presence of fake and split vertices are negligible, as expected.

The predicted average number of reconstructed vertices, as obtained from data for a given value 
of $\mu$ in Fig.~\ref{fig:VxVsMu_MC}(b), can be used to estimate the primary vertex selection efficiency for a specific hard-scatter process.
This is done by combining the prediction with the simulated distributions of track $\sqrt{\sum \pt^2}$ for this process and for minimum-bias events,
as shown in Fig.~\ref{fig:hs_mb_sumpt2}. For the highest $\mu$ value  ($\mu = 40$) studied in terms of hard-scatter primary vertex 
reconstruction and selection efficiencies in Section~\ref{sec::HardScatter}, Fig.~\ref{fig:VxVsMu_MC}(b) predicts an average number of
reconstructed vertices from pile-up interactions of~$17 \pm 1$. Of all the reconstructed vertices, the one with highest 
$\sum \pt^2$ is selected as the hard-scatter vertex with a very high efficiency for most processes.
To estimate the small probability that a pile-up vertex is selected by this procedure instead,
the simulated distribution of track $\sqrt{\sum \pt^2}$ for inelastic interactions in Fig.~\ref{fig:hs_mb_sumpt2}
is compared to the much harder one expected for the hard-scatter process of interest. For $Z \rightarrow \mu\mu$ events,
a randomly selected point on the  $\sum \pt^2$ distribution is found to be lower than the largest of the values found for 17 random
samplings of the distribution for minimum-bias events in approximately 4\% of the cases.
 This estimate, which is partially based on data but does not account for all experimental effects 
such as the distortion of the track $\sum \pt^2$ distribution of minimum-bias events due to merging of primary 
vertices, is in reasonable agreement with the estimate of 2\% obtained based on simulation in Fig.~\ref{fig:hs_select}.

%% file: Conclusion.tex
This paper presents primary vertex reconstruction and selection methods and their performance  for 
 proton--proton collision data recorded by the ATLAS experiment at the LHC during 
 Run 1. The primary vertex position resolution  measured in data is consistent 
with the predictions from simulation. A longitudinal vertex position resolution of about $30\;\mu{\textrm m}$ has been
achieved for events with high track-multiplicity.  A  significant improvement of the vertex transverse-position resolution
is obtained using the beam-spot constraint in the vertex fit, giving a resolution below $20\;\mu{\textrm m}$ for all
multiplicities.

The primary vertex reconstruction efficiency has been measured using  MC simulation.
 For minimum-bias events, the single vertex reconstruction efficiency is
 above $99\%$ for all processes, provided at least two charged particles are 
reconstructed within the ATLAS inner detector. 
For hard-scatter interactions, the reconstruction and selection efficiency has been 
 studied for a number of benchmark processes as a function of pile-up. In all cases,
the overall signal vertex reconstruction efficiency exceeds $99\%$. A  significant 
contamination from pile-up minimum-bias vertices is however observed for high values of $\mu$
in the case of hard-scatter processes with a small number of charged-particle tracks, such as 
$H\to \gamma \gamma$ and $Z\to \mu \mu$.  The efficiency to reconstruct and then correctly select
the primary vertex at $\mu =$ 40 in the case of $Z\to \mu \mu$  is predicted to remain very high, namely $98\%$, when
 both muons are reconstructed within the inner detector acceptance.

The impact of multiple $pp$ interactions in the same bunch
crossing on the reconstruction of primary vertices has been studied 
in detail.  Comparisons of the modelling of vertex input quantities
were made for low and high values of $\mu$ 
and good agreement between data and the MC simulation is observed
 for values of $\mu$ up to $70$.
The largest impact of pile-up is the merging of nearby vertices,
which has been quantified precisely by studying the relationship
between $\mu$ and the number of reconstructed vertices.
The corresponding non-linear
effects due to merging are  well modelled within the uncertainties in the MC 
 simulation for values of $\mu$ as high as $70$, confirming the validity of the proposed model.

%% file: Acknowledgements.tex
% Acknowledgements for papers with collision data
% Version 06-March-2017

% Standard acknowledgements start here
%----------------------------------------------
We thank CERN for the very successful operation of the LHC, as well as the
support staff from our institutions without whom ATLAS could not be
operated efficiently.

We acknowledge the support of ANPCyT, Argentina; YerPhI, Armenia; ARC, Australia; BMWFW and FWF, Austria; ANAS, Azerbaijan; SSTC, Belarus; CNPq and FAPESP, Brazil; NSERC, NRC and CFI, Canada; CERN; CONICYT, Chile; CAS, MOST and NSFC, China; COLCIENCIAS, Colombia; MSMT CR, MPO CR and VSC CR, Czech Republic; DNRF and DNSRC, Denmark; IN2P3-CNRS, CEA-DSM/IRFU, France; SRNSF, Georgia; BMBF, HGF, and MPG, Germany; GSRT, Greece; RGC, Hong Kong SAR, China; ISF, I-CORE and Benoziyo Center, Israel; INFN, Italy; MEXT and JSPS, Japan; CNRST, Morocco; NWO, Netherlands; RCN, Norway; MNiSW and NCN, Poland; FCT, Portugal; MNE/IFA, Romania; MES of Russia and NRC KI, Russian Federation; JINR; MESTD, Serbia; MSSR, Slovakia; ARRS and MIZ\v{S}, Slovenia; DST/NRF, South Africa; MINECO, Spain; SRC and Wallenberg Foundation, Sweden; SERI, SNSF and Cantons of Bern and Geneva, Switzerland; MOST, Taiwan; TAEK, Turkey; STFC, United Kingdom; DOE and NSF, United States of America. In addition, individual groups and members have received support from BCKDF, the Canada Council, CANARIE, CRC, Compute Canada, FQRNT, and the Ontario Innovation Trust, Canada; EPLANET, ERC, ERDF, FP7, Horizon 2020 and Marie Sk{\l}odowska-Curie Actions, European Union; Investissements d'Avenir Labex and Idex, ANR, R{\'e}gion Auvergne and Fondation Partager le Savoir, France; DFG and AvH Foundation, Germany; Herakleitos, Thales and Aristeia programmes co-financed by EU-ESF and the Greek NSRF; BSF, GIF and Minerva, Israel; BRF, Norway; CERCA Programme Generalitat de Catalunya, Generalitat Valenciana, Spain; the Royal Society and Leverhulme Trust, United Kingdom.

The crucial computing support from all WLCG partners is acknowledged gratefully, in particular from CERN, the ATLAS Tier-1 facilities at TRIUMF (Canada), NDGF (Denmark, Norway, Sweden), CC-IN2P3 (France), KIT/GridKA (Germany), INFN-CNAF (Italy), NL-T1 (Netherlands), PIC (Spain), ASGC (Taiwan), RAL (UK) and BNL (USA), the Tier-2 facilities worldwide and large non-WLCG resource providers. Major contributors of computing resources are listed in Ref.~\cite{ATL-GEN-PUB-2016-002}.
%----------------------------------------------

%% file: atlas_authlist.tex
% ATLAS Collaboration author list
% Data extracted on 15-Apr-2017 for paper reference PERF-2015-01
%\documentclass[11pt]{article}
%\usepackage{a4wide}\begin{document}
\begin{flushleft}
{\Large The ATLAS Collaboration}

\bigskip

M.~Aaboud$^\textrm{\scriptsize 136d}$,
G.~Aad$^\textrm{\scriptsize 87}$,
B.~Abbott$^\textrm{\scriptsize 114}$,
J.~Abdallah$^\textrm{\scriptsize 65}$,
O.~Abdinov$^\textrm{\scriptsize 12}$,
B.~Abeloos$^\textrm{\scriptsize 118}$,
R.~Aben$^\textrm{\scriptsize 108}$,
O.S.~AbouZeid$^\textrm{\scriptsize 138}$,
N.L.~Abraham$^\textrm{\scriptsize 152}$,
H.~Abramowicz$^\textrm{\scriptsize 156}$
% ...
\bigskip
\\
% ...
$^{12}$ Institute of Physics, Azerbaijan Academy of Sciences, Baku, Azerbaijan\\
% ...
$^{65}$ University of Iowa, Iowa City IA, United States of America\\
% ...
$^{87}$ CPPM, Aix-Marseille Universit{\'e} and CNRS/IN2P3, Marseille, France\\
% ...
$^{108}$ Nikhef National Institute for Subatomic Physics and University of Amsterdam, Amsterdam, Netherlands\\
% ...
$^{114}$ Homer L. Dodge Department of Physics and Astronomy, University of Oklahoma, Norman OK, United States of America\\
% ...
$^{118}$ LAL, Univ. Paris-Sud, CNRS/IN2P3, Universit{\'e} Paris-Saclay, Orsay, France\\
% ...
$^{136}$ $^{(a)}$ Facult{\'e} des Sciences Ain Chock, R{\'e}seau Universitaire de Physique des Hautes Energies - Universit{\'e} Hassan II, Casablanca; $^{(b)}$ Centre National de l'Energie des Sciences Techniques Nucleaires, Rabat; $^{(c)}$ Facult{\'e} des Sciences Semlalia, Universit{\'e} Cadi Ayyad, LPHEA-Marrakech; $^{(d)}$ Facult{\'e} des Sciences, Universit{\'e} Mohamed Premier and LPTPM, Oujda; $^{(e)}$ Facult{\'e} des sciences, Universit{\'e} Mohammed V, Rabat, Morocco\\
% ...
$^{138}$ Santa Cruz Institute for Particle Physics, University of California Santa Cruz, Santa Cruz CA, United States of America\\
% ...
$^{152}$ Department of Physics and Astronomy, University of Sussex, Brighton, United Kingdom\\
% ...
$^{156}$ Raymond and Beverly Sackler School of Physics and Astronomy, Tel Aviv University, Tel Aviv, Israel\\
% ...
$^{a}$ Also at Department of Physics, King's College London, London, United Kingdom\\
$^{b}$ Also at Institute of Physics, Azerbaijan Academy of Sciences, Baku, Azerbaijan\\
$^{c}$ Also at Novosibirsk State University, Novosibirsk, Russia\\
$^{d}$ Also at TRIUMF, Vancouver BC, Canada\\
$^{e}$ Also at Department of Physics {\&} Astronomy, University of Louisville, Louisville, KY, United States of America\\
% ...
\end{flushleft}